\documentclass[12pt]{iopart}
\usepackage{psfig}
\begin{document}
\jl{4}
%
%
%%%%%%%%%%%%%%%%%%%%%%%%%%%%%%%%%%%%%%%%%%%%%%%%%%%%%%%%%%%%%%%
\topical[Review of QGP Signatures]{Signatures of Quark-Gluon-Plasma formation in
high energy heavy-ion collisions: A critical review}

\author{S A Bass\dag\ddag\footnote[5]{Feodor Lynen Fellow 
	                          of the A. v. Humboldt Foundation},
        M Gyulassy\S{}\ddag, H St\"ocker$\parallel$\ddag{} 
	and W Greiner$\parallel$}

\address{\dag\  Department of Physics \\
	       Duke University \\
               Durham, NC, 27708-0305, USA\\}

\address{\ddag Institute of Nuclear Theory \\
		University of Washington \\
	        Seattle, WA, 98195-1550, USA\\}

\address{\S\ Physics Department \\ 
	       Columbia University \\
	       550 West 120th, New York, NY 10027, USA\\}

\address{$\parallel$\  Institut f\"ur Theoretische Physik\\
             Johann Wolfgang Goethe Universit\"at\\
             D-60054 Frankfurt am Main, Germany\\}
%%%%%%%%%%%%%%%%%%%%%%%%%%%%%%%%%%%%%%%%%%%%%%%%%%%%%%%%%%%%%%

\submitted
\noindent{INT preprint DOE/ER/40561-11-INT98}

\maketitle

\tableofcontents

\pagebreak

%%%%%%%%%%%%%%%%%%%%%%%%%%%%%%%%%%%%%%%%%%%%%%%%%%%%%%%%%%%%%%%%%%%%%%%%%%%%%
	\section{Probing dense matter of elementary particles}
%%%%%%%%%%%%%%%%%%%%%%%%%%%%%%%%%%%%%%%%%%%%%%%%%%%%%%%%%%%%%%%%%%%%%%%%

Ultra-relativistic heavy ion collisions 
offer the unique opportunity to probe
highly excited dense nuclear matter under controlled laboratory
conditions. 
%One of the objectives of this research field at the interface of
%high--energy and nuclear physics is the creation and study
%of super-dense matter.
The compelling driving force for such studies is the expectation that an
entirely new form of matter may be created from such reactions. That
form of matter, called the Quark Gluon Plasma (QGP), is the QCD
analogue of the plasma phase of ordinary atomic matter. However, unlike such
ordinary plasmas, the deconfined quanta of a QGP are not directly
observable because of the fundamental confining property of the
physical QCD vacuum.
What is observable are hadronic and
leptonic residues of the transient QGP state.  There is a large
variety of such individual probes. Leptonic probes, 
$\gamma,\, e^+ e^-,\, \mu^+ \mu^-$ carry information about the spectrum
of electromagnetic current fluctuations in the QGP state; the
abundance of quarkonia $\Psi,\, \Psi',\, \Upsilon,\,\Upsilon'$ (also
observed via $l^+ l^-$) carry information about the chromoelectric
field fluctuations in the QGP. The arsenal of hadronic probes,
$\pi,\, K,\, p,\, \bar p ,\, \Lambda ,\, \Xi ,\, \Omega ,\, 
\phi ,\, \rho,$ \ldots provide
information on the quark flavor chemistry and baryon number transport.
Theory suggests that with decays such as $\rho \to e^+ e^-$ the properties
of the hadronization and chiral symmetry breaking can be indirectly
studied. Quantum statistical interference patterns in $\pi \pi$,
$K K$, $p p$, $\Lambda \Lambda$ correlations provide somewhat cloudy
lenses with which the space-time geometry of hadronic ashes of the
QGP can be viewed.
The detailed rapidity and transverse momentum spectra of hadrons provide
barometric information of pressure gradients during the explosive
expansion of the QGP drop. 

The central problem with all the above probes is precisely that they
are all indirect messengers. If we could see free quarks and gluons
(as in ordinary plasmas) it would be trivial to verify the QCD
prediction of the QGP state. However, nature choses to hide those
constituents within  the confines of color neutral composite many
body systems -- hadrons.

The QGP state formed in nuclear collisions is a
transient rearrangement of the correlations among quarks and 
gluons contained in the incident baryons  into a larger but globally still
color neutral system with however remarkable theoretical properties.
The task with heavy ion reactions is to provide experimental information
on that fundamental prediction of the Standard Model.
 
This topical review covers 
current (1998) theoretical and experimental attempts to disentangle
popular scenarios on QGP signatures from the complex, off-equilibrium physics.
The start of the RHIC experiment program is only a year away. 
This will be a dedicated machine to study the QGP. Nevertheless
a very large effort has been made at the AGS and SPS over the last
12 years and has resulted in an impressive amount of exciting new
findings.
The search for the QGP can be traced via the proceedings of the
High Energy Heavy Ion Studies 
\cite{hehis1,hehis2,hehis3,hehis4,hehis4a,hehis5,hehis6,hehis7,hehis8,hehis9}
and Schools \cite{peniscola,bodrum,wilderness}
and the ``Quark Matter'', ``Nucleus-Nucleus'' and ``Strange Quark Matter'' 
conferences
%quark matter refs.
\cite{qm79,qm80,qm81,qm82,qm83,qm84,qm86,qm87,qm88,qm90,qm91,qm93,qm95,qm96,qm97,
nn1,nn2,sqm95,sqm96,sqm97}.
Some textbooks
%buecher
\cite{mueller85book,hwabook1,hwabook2,csernai94book,wong94book},
and a vast number of review articles have been published -- reference samples of
early
%alte reviews
\cite{shuryak80a,mclerran86a,csernai86a,stock86a,stoecker86a,clare86a,schuermann86a,kajantie87a}
and of the latest
%neue reviews
\cite{singh93a,mueller95a,harris96a,greiner96a,alam96a}
review papers are given here.
The list of more than 500 references given in this topical review is by no means
complete. Apologies are offered to those whose contributions
could not be included into the references.

%%%%%%%%%%%%%%%%%%%%%%%%%%%%%%%%%%%%%%%%%%%%%%%%%%%%%%%%%%%%%%%%%%%%%%%%
	\section{QCD matter and relativistic 
	heavy-ion collisions}

	\subsection{Infinite stationary systems in equilibrium}

	\paragraph*{The deconfinement phase transition and chiral
	symmetry restoration\\}
	\label{qcdphase}
%%%%%%%%%%%%%%%%%%%%%%%%%%%%%%%%%%%%%%%%%%%%%%%%%%%%%%%%%%%%%%%%%%%%%%%%

Phase transitions are among the most dramatic many body effects in physics.
Examples for 
restored symmetry via a phase transition at high temperatures, $T_C$, are
ferro-magnetism, super-conductivity and the solid - liquid phase transition.
In nuclear physics evidence for a liquid gas 
phase transition
of nuclear matter has been claimed 
for temperatures of $T \approx 5$ MeV \cite{pochodzalla95a}.
Phase transitions to abnormal nuclear matter states at high densities
have also been predicted early on \cite{lee74a,scheid74a}.

QCD is a non-abelian gauge theory,
it's basic constituents are quarks and anti-quarks interacting
through the exchange of color-charged gluons. 
At short space-time-intervals -- large  momentum transfers --
the effective coupling constant decreases logarithmically 
(``asymptotic freedom'' -- meaning weak coupling
of quarks and gluons) whereas it becomes strong for large distances and
small relative momenta. 
This results in the phenomena of chiral symmetry breaking
and quark-gluon confinement.

At very high temperatures and densities, in the domain of weak 
coupling between quarks and gluons, long range interactions are dynamically 
screened \cite{collins75a,polyakov77a}.
Quarks and gluons are then no longer confined to bound hadronic states
(``deconfinement''). Furthermore, chiral symmetry is restored -- for
baryon-free matter -- apparently at the same temperature $T_C$.
This novel phase of nuclear
matter is called the {\em quark gluon plasma} \cite{collins75a}.

A transition from the deconfined quark-gluon phase to confined color
singlet states has (probably) occured during the rapid expansion of
the early universe. Temperatures were very high then, but the net baryon
density was small. Therefore one often assumes zero baryon chemical
potentials in calculating the thermodynamic properties of strongly
interacting matter in the early universe. It is sought to 
re-establish these conditions and thus enable a study of quark
deconfinement in the laboratory via heavy ion collisions  
\cite{scheid68a,chapline73a}. 
At the highest in the near future obtainable energies (LHC) the initial
net baryon density may be around nuclear matter ground state
density. The entropy per baryon ratio, however, is estimated to be
in the order of  $10^3$ to $10^4$ \cite{eskola96a}
(early universe: $10^9$) and thus
a vanishing  baryon chemical potential is considered a viable approximation.

%However, zero baryon chemical
%potential may not actually be attainable, 
%even at the highest energies at the LHC
% (c.f. section \ref{stopping}).

The energy densities currently 
achievable in ultra-relativistic heavy ion collisions
at the AGS and SPS are on the order of 0.5--10 GeV/fm$^3$ \cite{bjorken83a} 
and temperatures are in the 
range of 100-200 MeV. At temperatures $T_C \sim 150 - 200$ MeV 
the effective coupling constant
of QCD, however, is still on the order of 1. Therefore, perturbative techniques
of QCD \cite{kapusta83a,braaten90a,thoma95b} are not applicable.
%Frankfurt mod #1
The situation may change at extremely high energies (LHC?) where QCD
predicts a large cross section for minijet production. Then there may
exist a possibility to reach a new regime of large parton densities
at a small coupling constant. The QCD interaction in this regime
would be highly nonlinear \cite{baier96a}.
% end of mod #1
Recent calculations of the Debye screening mass using 
{\em lattice gauge simulations}, however, indicate that perturbative QCD
techniques may not even be applicable at such energies \cite{kajantie97a}.

%%%%%%%%%%%%%%%%%%%%%%%%%%%%%%%%%%%%%%%%%%%%%%%%%%%%%%%%%%%%%%%%%%%
		\paragraph*{Perturbation theory\\}
%%%%%%%%%%%%%%%%%%%%%%%%%%%%%%%%%%%%%%%%%%%%%%%%%%%%%%%%%%%%%%%%%%

\begin{figure}[tbh]
\centerline{\psfig{figure=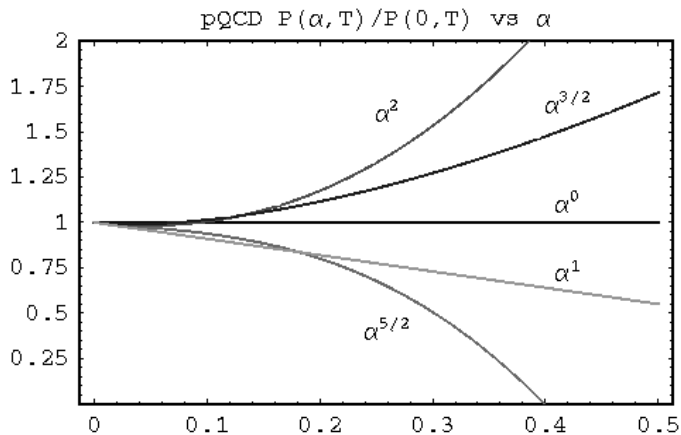,width=9cm}}
\caption{ \label{braatenfig} 
Perturbative contributions up to $O(g^5)$
\protect\cite{braaten96b} to the pressure vs. the coupling constant.}
\end{figure}

The fundamental asymptotic property of QCD leads to the naive
expectation that the properties of a  QGP can be calculated via 
perturbation theory.
%Perturbation theory has been widely applied to study the importance of strong
%interactions in the QGP. 
However, due to infrared divergences, 
especially in the chromo-magnetic sector,
perturbation theory may not even be
applicable for temperatures far above $T_C$.
QCD perturbation theory has been improved \cite{braaten90a} by resummation
to screen color-electric divergences. 
These ``hard--thermal--loop'' methods, originally developed
for zero baryon density, have now been extended to finite densities
\cite{vija95a}. Nevertheless, its applicability at temperatures and densities
{\em accessible to experiment} is severely limited. At energy densities
on the order of 20 GeV/fm$^{3}$ the temperature is in the order of
600 MeV and the coupling constant g is still of the order of 2, 
which invalidates the necessary
assumption $g T \ll T$ for the hard--thermal--loop resummation
scheme. 
%Frankfurt mod #6
Only at the Planck scale is QCD a weakly interacting
theory  as QED \cite{kajantie97a}. 
%end of mod #6
Recent calculations of the perturbative contributions up to $O(g^5)$
\cite{braaten96b}
to the pressure are shown in fig.\ref{braatenfig}. The oscillation
of the results suggest a zero radius of convergence of thermal
pQCD.

%%%%%%%%%%%%%%%%%%%%%%%%%%%%%%%%%%%%%%%%%%%%%%%%%%%%%%%%%%%%%%%%%%%%%
		\paragraph*{Lattice Gauge Theory\\}
		\label{lattice}
%%%%%%%%%%%%%%%%%%%%%%%%%%%%%%%%%%%%%%%%%%%%%%%%%%%%%%%%%%%%%%%%%%%%

{\em Lattice gauge simulations of QCD} \cite{wilson74a,creutz83a}, 
provide therefore the only rigorous method to compute
the {\em equation of state} of strongly
interacting elementary particle matter. In principle
 both, the non-perturbative hadronic matter 
and the non-perturbative QGP phases of QCD can be investigated.
The main disadvantage of lattice simulations is the practical restriction to
finite, periodic, baryon free systems in global equilibrium, a scenario far
from the highly inhomogeneous off-equilibrium
situation found in complex heavy-ion reactions.
Technically, the strong dependence of the results
on the lattice spacing and periodic box size is presently a problem.
Nevertheless, lattice data provide the most compelling theoretical
evidence of a rapid transition region from the confined to the 
QGP state.

Lattice calculations allow at least for the computation of thermodynamic
averages of different quantities related to hadron masses and 
to the phase transition
(in the infinite volume and zero baryon number limit). There have been
considerable improvements in algorithms \cite{laermann96a} and in computing
power in recent years. For finite temperature full QCD
simulations lattices with spatial sizes of $24^3$ and 48 points in 
Euclidean time direction have been used \cite{blum95a,bernard98a}, 
while for pure gauge theory
(without quarks) lattices of $32^3 \times 12$ \cite{boyd95a,boyd95b} 
are in use.

\begin{figure}[tbh]
\centerline{
\psfig{figure=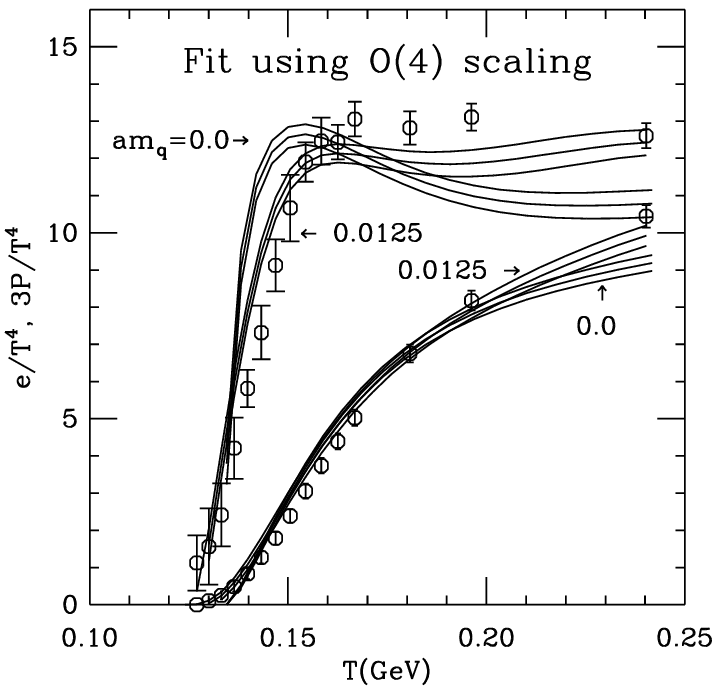,width=7cm}
\psfig{figure=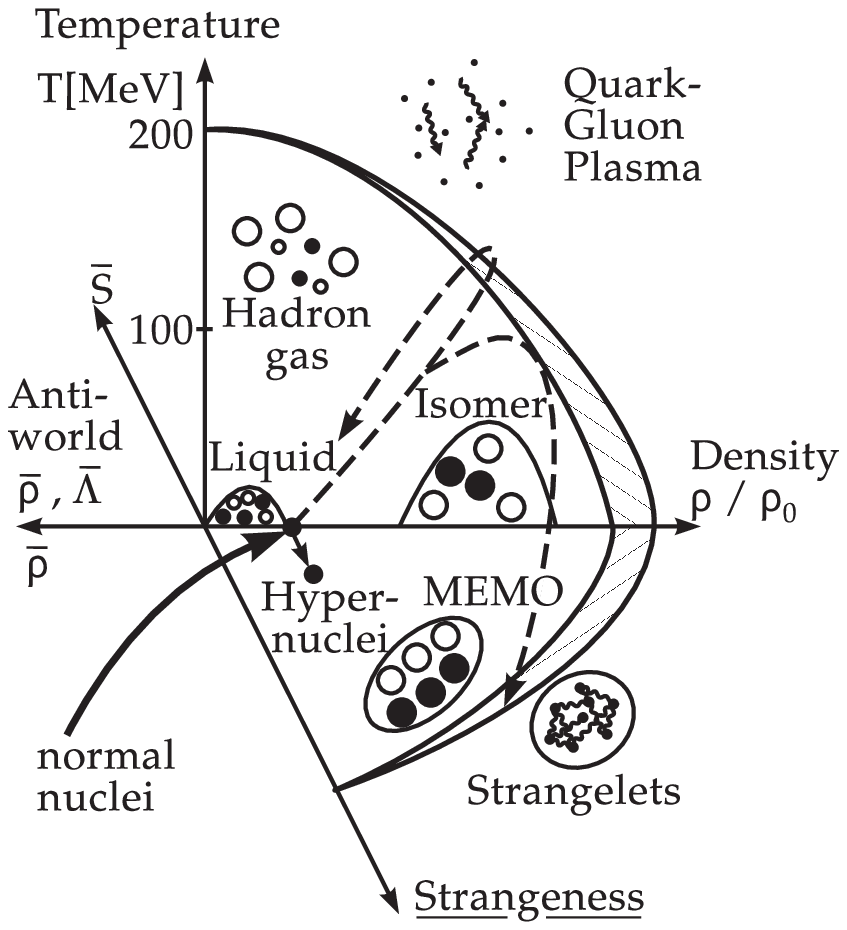,width=7cm}
}
\caption{ \label{latticefig1} 
Left: Lattice calculation (circles) of the equation of state for two flavor QCD
\protect \cite{bernard96a}. The lines show
fits using O(4) scaling and extrapolations to zero quark mass. The critical
temperature is in the order of 140 MeV. 
%Center: equation of state in
%the linear sigma model \protect\cite{papazoglou97a}.
Right: schematic overview of theoretical phases of QCD matter. }
\end{figure}

Lattice calculations
yield a critical temperature of $T_C = 265^{+10}_{-5}$ MeV
in the quenched approximation \cite{laermann96a}
-- where neither dynamical quarks, nor a
chiral phase transition exist. Simulations including
dynamical quarks at $\mu_B=0$ indicate a critical temperature in the order of
$T_C = 140$ MeV (see figure \ref{latticefig1}). 
However, in this case finite size effects of the lattice have
not yet been fully overcome and the precision is not as high as in the
quenched case \cite{laermann96a}. 
The inclusion of the second most important thermodynamic variable, the chemical
potential $\mu_B$ into a full fletched lQCD calculation is, presently, 
still out of reach. This raises a practical question, whether 
conclusions based on $\mu_B=0$ estimates, might misguide physical
argumentation for observables in nuclear collisions.
This warning is particularly appropriate for those QGP-signals, where a 50\%
quantitative change of an observable is used to differentiate 
QGP production scenarios from ordinary hadronic transport ones.

The behavior of order parameters as a function of temperature, 
such as the quark condensate,
%($\frac{1}{3}\langle \psi \bar{\psi} \rangle \approx -235$ MeV)
indicate that
the transition between the confined and deconfined state of QCD
may show a discontinuity of some thermodynamic derivative 
-- i.e. a phase transition.
The order of this phase transition is crucial for some
proposed signatures of the QGP. Many striking signatures 
depend heavily on the assumption of
a first order phase transition and the existence of a mixed phase of 
QCD matter.
For {\em pure } SU(3) gauge theory and for {\em full}  QCD with
four massless flavors of dynamical quarks, lattice QCD results and universality
class arguments \cite{svetitsky82a,pisarski84a} indicate a 
first order phase transition.
For two (massless) flavors, however, these arguments predict 
\cite{pisarski84a,rajagopal93a} that {\em if} the phase transition
is of second order, it should have the critical exponents of the
O(4) Heisenberg anti-ferromagnetic spin--model in three dimensions.
Numerical evidence for this was recently obtained \cite{karsch94a},
suggesting that the transition in this case is indeed of second order.
%Frankfurt mod #2
With three light flavors the chiral phase transition corresponds
to a change in a continuous symmetry (i.e. chiral symmetry) and
is automatically of second order. 
%end of mod#2

For the most realistic case of QCD with two flavors of light quarks 
with masses between 5 and 10 MeV and one flavor with a mass around
200 MeV,
the situation remains unclear: 
the order of the phase transition seems to depend on the numerical 
values for the masses of the light and heavy quarks \cite{brownfr90a}. If the
latter is too heavy, the transition might be smeared out to a mere
rapid increase of the energy density over a small temperature interval.
In this case the use of simple deconfinement scenarios 
may lead to wrong expectations for observables.
The elementary excitations
in such a phase transition region 
ought not be described by quarks and gluons but could 
physically resemble more 
hadronic excitations with strongly modified ``in--medium'' properties
\cite{detar88a}.

In any case the QCD phase transition observed on the lattice is
-- when dynamical quarks are included -- only very weakly first order,
if there is a discontinuity at all. The latent heat across the
discontinuity is at most a small fraction of the total jump in
the normalized entropy density $s/T^3$ between the hadronic phase
and the asymptotic QGP. A real phase-coexistence region between
hadrons and (possibly strongly interacting, but deconfined) quarks
and gluons, as often discussed in the past, seems therefore no longer
a realistic possibility. However, nothing is known from first principles
about the phase transition at finite baryon density and therefore a
strong first order transition (with a phase coexistence region) is
still a possibility at large baryon 
densities \cite{rajagopal98a,shuryak98a}.
An additional complication is that for systems of finite volume
(V$\le$ 100 fm$^3$) the deconfinement cannot be complete. Fluctuations
lead to a finite probability of the hadronic phase above $T_C$.
The sharp discontinuity (e.g. $\varepsilon/T^4$) is thus smeared out
\cite{spieles97b}.

Purely hadronic models, such as the $\sigma - \omega$ Model or the
linear $\sigma$-model exhibit a similar phase transition (from
normal to abnormal nuclear matter), 
but are not constrained to $\mu_B=0$. The equation of state for 
nuclear matter does not only depend on temperature and density but 
will also depend on the net-strangeness content, which may be
non-zero in subsystems (e.g. individual phases) present in
heavy ion collisions.
A schematic view of the resulting complex multidimensional 
phase-diagram is illustrated on the right in figure~\ref{latticefig1}.

%%%%%%%%%%%%%%%%%%%%%%%%%%%%%%%%%%%%%%%%%%%%%%%%%%%%%%%%%%%%%%%%%%
\subsection{Non-equilibrium models}
%%%%%%%%%%%%%%%%%%%%%%%%%%%%%%%%%%%%%%%%%%%%%%%%%%%%%%%%%%%%%%%%%%

In order to connect the theoretical thermodynamic properties of a 
QGP with experimental data on finite nuclear collisions, many 
non-equilibrium dynamical effects must also be estimated. Transport
theory is the basic tool to address such problems.
Non-equilibrium effects are certain to arise from 
the rapid time-dependence of the system
(even the use of the term ``state'' seems questionable), finite
size effects, inhomogeneity, $N$-body phase space, particle/resonance
production and freeze-out and collective dynamics.
Such 
microscopic and macroscopic (hydrodynamical) models attempt to describe 
the full time-evolution
from an assumed initial state of the heavy ion reaction (i.e. the two
colliding nuclei) up to the freeze-out of all initial and produced
particles after the reaction.
Hydrodynamical models 
neglect most of these effects by 
making the assumption that the initial condition can be assumed to be
in local thermal equilibrium and that local equilibrium is maintained
during evolution. Fireball models simply parameterize final spectra
and abundances via freeze-out parameters, e.g. $T, \mu_B, \vec{v}_f$.
However, the initial condition in nuclear collisions is a coherent
state $| AB \rangle$ of two quantal ($T=0$) nuclear systems.
A non-equilibrium quantum evolution of $| AB \rangle$ introduces complex
high order Fock-State components. A key dynamical assumption is that
decoherence occurs rapidly during the early phase of the collision
yielding a mixed state density matrix (with $S={\rm Tr}\rho \ln \rho >0$).
There is no theorem to insure that $\rho$ evolves to a local
equilibrium form $\exp(-u_\mu p^\mu /T)$ at any time during the reaction.
That can only be tested via a transport theory approximation
to the evolution equations. The question of the form of the initial state
$\rho(\tau_0)$ must still be addressed, but once that is specified,
transport theory can reveal if local equilibrium is achieved and what 
observables are least sensitive to uncertainties in $\rho(\tau_0)$.

Depending on the most convenient basis for expanding $\rho(\tau_0)$,
transport theory assumes different forms. At low energies the initial
ensemble is most conveniently described in terms of mesons and baryons.
Here hadronic transport theory is appropriate. At collider energies, 
pQCD minijet processes are expected to produce a high density mostly
gluonic gas. In that regime parton cascade models are more appropriate.

\paragraph*{Parton cascades\\}
Parton cascade models 
\cite{wang91a,geiger92a,wang92a,wang94a,geiger95a,ellis95a} 
evolve partonic degrees of freedom.
They are therefore mostly applied to study the initial compressional 
and the high density phase of ultra-relativistic heavy ion collisions
(collider energies, $\sqrt{s}\ge 200$~GeV). 
These models all contain the general structure \cite{geiger92a}:
\begin{enumerate}
\item Initialization: the nucleons of the colliding nuclei are resolved
into their parton substructure according to the measured nucleon
structure functions and yield the initial parton distributions.
\item Interaction: parton interactions as described by
perturbative QCD are used to model the evolution
of the ensemble of partons during the course of the collision. This includes
multiple scatterings together with associated space-like and time-like
parton emission processes before and after each scattering. The
sequence of scatterings is, however, incoherent 
and the neglect of quantum interference effects is questionable.
\item Hadronization: partons are recombined or converted
via string fragmentation into final hadron states.
\end{enumerate}
The propagation is performed on straight lines -- 
soft non-perturbative collective field effects have so far been
neglected. On the other hand,
hadronization has to be modeled by brute force to mock up 
confinement in the final reaction
stage.

%%%% miklos fig
\begin{figure}[tbh]
\centerline{\psfig{figure=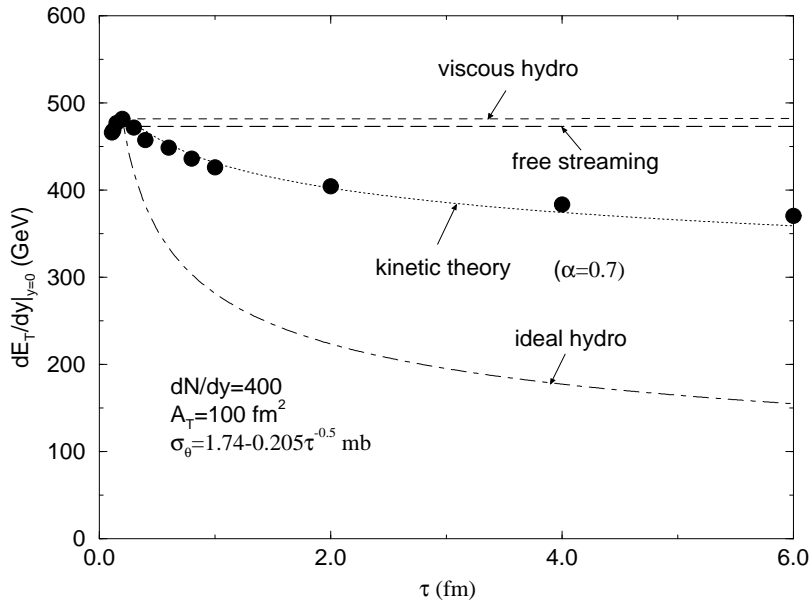,width=11cm,angle=-90}}
\caption{ \label{miklosfig1} 
Comparison of time evolution of transverse energy per rapidity in 
analytic kinetic theory results with numerical parton cascade calculations
\protect \cite{gyulassy97a}. Strong deviations from hydrodynamic
behavior are visible. }
\end{figure}
%%%%

One of the central issues addressed by parton cascades is the question
of energy deposition processes in space--time as well as momentum space.
Partonic cascades predict that roughly 50\% of the expected energy
deposition at RHIC and a larger fraction at LHC takes place at the
partonic level \cite{eskola94a}.
Rapid thermalization is caused by radiative energy degradation and 
spatial separation
of partons with widely different rapidities due to free streaming;
transverse momentum distributions of initially scattered partons are
almost exponential if radiative corrections are taken into account
\cite{harris96a}.
For RHIC energies thermalization is predicted on a proper time scale
of 0.3 -- 0.5 fm/c \cite{eskola94a}.
A recent analysis of parton cascade evolution \cite{gyulassy97a} shows
that local equilibrium is not maintained due to rapid expansion.
Very large dissipative corrections to hydrodynamics appear.
The thermalized QGP is initially gluon rich and depleted of quarks due
to the larger cross section and higher branching ratios for gluons
\cite{shuryak92a}. Chemical equilibrium is achieved over a time of 
several fm/c \cite{biro93a,geiger93a}. This may be reduced 
if higher order pQCD processes
are taken into account \cite{xiong94a}.

\paragraph*{Hadronic transport models\\}
Hadronic transport models treat relativistic heavy-ion collisions as 
sequences of binary/$N$-body collisions of mesons, baryons, strings and 
their constituents, diquarks and quarks. The real part of the 
interaction can be  obtained in principle from G-Matrix calculations, 
with the in-medium self-energy and the imaginary part is modeled via hard
scattering cross sections
\cite{yariv79a,cugnon80a,pang92a,
kruse85a,molitoris85a,aichelin85a,aichelin87b,peilert89a,aichelin91a,
sorge89a,liba95a,weber90a,ehehalt95a,bass98a}.
%Three main elements form the general structure of these models
%\begin{enumerate}
%\item Initialization: either with a Fermi-gas ansatz or within a 
%self-consistent approach of minimizing the respective Hamiltonian.
%\item Propagation: propagate the constituents (hadrons) according to
%the equations of motion of the respective model (straight lines in the 
%CASCADE case).
%\item Collision term: parameterized or tabulated
%cross sections and decay widths for all the baryons and mesons included.
%\end{enumerate}
For high beam energies most models include particle production via string
formation -- either using the Lund \cite{anderson87a,anderson87b,sjoestrand94a} or 
a pomeron exchange scheme \cite{werner93a}.
Partonic degrees of freedom are not treated explicitly and therefore
these models do not include a phase transition.
However, some models contain further speculative scenarios such as
color-ropes \cite{biro84a,sorge92a},
breaking of multiple-strings \cite{werner93b} or decay of multi-quark 
droplets \cite{aichelin93a} which clearly go beyond hadronic physics.

Hadronic transport models are critical for assessing the influence of 
ordinary or exotic hadronic phenomena on the observables proposed
to search  for a QGP. They therefore provide a background basis to 
evaluate whether an observable shows evidence for non-hadron physics.

\paragraph*{Nuclear fluid dynamics\\}
NFD 
is so far the only dynamical model in which a phase transition can
explicitly be incorporated (see e.g. 
\cite{bjorken83a,clare86a,amelin91a,brachmann97a} for details). 
This is possible since the equation
of state (including a phase transition) is a direct input for the
calculations.
However, NFD is an idealized continuum description based on local equilibrium
and energy--momentum conservation. 
Therefore it is very well suited to
study kinematic observables such as collective flow.
Since NFD is a macroscopic kinetic theory it is not directly applicable
to the study of hadron abundances and particle production. However,
NFD calculations predict (local) temperatures and chemical potentials
which can be used, e.g. by chemical equilibrium calculations of hadron
abundances, to study particle production. 
Different observables predicted by nuclear fluid dynamics 
will be discussed in section \ref{flow}.

In the ideal fluid approximation (i.e. neglecting off-equilibrium effects), 
the EoS is the {\em only}
input to the equations of motion that relates directly to properties
of the matter under consideration. The EoS influences the
dynamical evolution of the system, and final results are uniquely
determined. 
The initial condition can be chosen from two colliding nuclei (in
a full 3D calculation with up to three fluids) or
an equilibrated QGP or hadronic matter with prescribed temperature
and chemical potential and velocity/flow profiles 
(for simpler, more schematic calculations). The time-evolution
is then studied until hadronic freeze-out for which 
a decoupling (freeze--out) hyper-surface needs to be specified.

However, the ideal fluid ansatz is only a rough approximation. In the
parton cascade study \cite{gyulassy97a} for example, large deviations
from even the Navier Stokes fluid approach were found.

%%%%%%%%%%%%%%%%%%%%%%%%%%%%%%%%%%%%%%%%%%%%%%%%%%%%%%%%%%%%%%%%%%
	\section{Observables: prospects and limitations}
%%%%%%%%%%%%%%%%%%%%%%%%%%%%%%%%%%%%%%%%%%%%%%%%%%%%%%%%%%%%%%%%%

As we have seen in the previous section it is obviously difficult
to find a robust theoretical description of relativistic heavy ion 
collisions involving the QCD phase-transition to predict observables. 
Not only is even the
order of the phase-transition from $\mu=0$ not known from the
ab-initio lQCD calculations, but also has the physical situation
of present or near future relativistic heavy ion collisions,
namely finite $\mu_B$, not been addressed yet in this theory.
However, even if this would be the case, one would only know the
behavior for static infinite systems. The second major unknown is the
influence of the non-equilibrium evolution on the (small) many-body
system. 
The very nature of even the thermodynamic limit of a QGP is not
completely understood. Real time response has only been studied
via pQCD, which however may have zero radius of convergence
 in $g$ in the thermodynamic limit.
Theory in this situation can thus serve mainly to motivate particular
experimental studies and provide overall consistency checks in the 
interpretation of data.
Data are needed to fix the uncertain phenomenological parameters
of the transport models, while such model calculations with
plausible parameters are essential to motivate the taking of the
data in the first place. This symbiotic relation between theory
and experiment in this field is very important as emphasized also
for example by Van~Hove \cite{hove_sym} and Kajantie \cite{kajantie_sym}.

A strategy for the detection of quark matter -- in our opinion --
must collect at least circumstantial evidence from several ``signals''
or anomalies. 
In the following we discuss each of the individual signals.
The strategy does then consist in a systematic variation of an external
parameter (system size, impact parameter and -- in particular -- 
bombarding energy); i.e. the measurement of excitation functions
of several signals which in the case of a phase-transition show
{\em simultaneously} the predicted anomalous behavior.

%%%%%%%%%%%%%%%%%%%%%%%%%%%%%%%%%%%%%%%%%%%%%%%%%%%%%%%%%%%%%%%%%%%%%%%%%%%%%%%%
	\subsection{Creation of high baryon density matter: 
	nuclear stopping power}
	\label{stopping}
%%%%%%%%%%%%%%%%%%%%%%%%%%%%%%%%%%%%%%%%%%%%%%%%%%%%%%%%%%%%%%%%%%%%%%%%%%%%%%%%

\paragraph*{Theoretical concepts\\}

It has been proposed more than two decades ago that
head-on collision of two nuclei can be used to create highly excited
nuclear matter \cite{scheid68a,chapline73a}. 
The longitudinal momentum 
is converted via multiple collisions 
into transverse momentum and secondary particles,
causing the creation of a zone of high energy density. Nuclear shock
waves have been suggested as a primary mechanism of creating high
energy densities in collisions with $\sqrt{s} \le 20$~GeV.
\cite{scheid68a,chapline73a,scheid74a}. This is analogous to the
well known Rankine-Hugoniot analysis of ordinary dense matter up to
$\sim 1$~Mbar pressures. In the nuclear shock wave case, the 
Rankine-Hugoniot analysis predicts that pressures up to $10^{23}$~Mbar
$\sim 100$~MeV/fm$^3$ may be reached. 

The term {\em nuclear stopping power} \cite{busza84a} 
characterizes the degree of stopping which
an incident nucleon suffers when it collides with 
another nucleus. 
For A+A collisions stopping manifests itself in a shift of the
rapidity-distributions of the incident nucleons towards mid-rapidity.
The heaviest
systems available, such as Pb+Pb or Au+Au, are best suited
for the creation of high baryon density matter.

The shape of the baryon rapidity distribution can give clear indications
on the onset of critical phenomena: Due to the strong dependence
of the baryon rapidity distribution on the baryon--baryon cross section
\cite{hartnack89a,berenguer92a,schmidt93a},
a rapid change in the shape of the scaled $dN/d(y/y_p)$
distribution with varying incident
beam energy is a clear signal for new degrees of freedom which show up
during
the reaction (i.e. deconfinement), e.g. due to phenomena such as critical
scattering \cite{gyulassy77a}. The width of the $dN/d(y/y_p)$
distribution for baryons is inversely proportional to their cross section. 

\begin{figure}[thb]
\begin{minipage}[t]{7cm}
\centerline{\psfig{figure=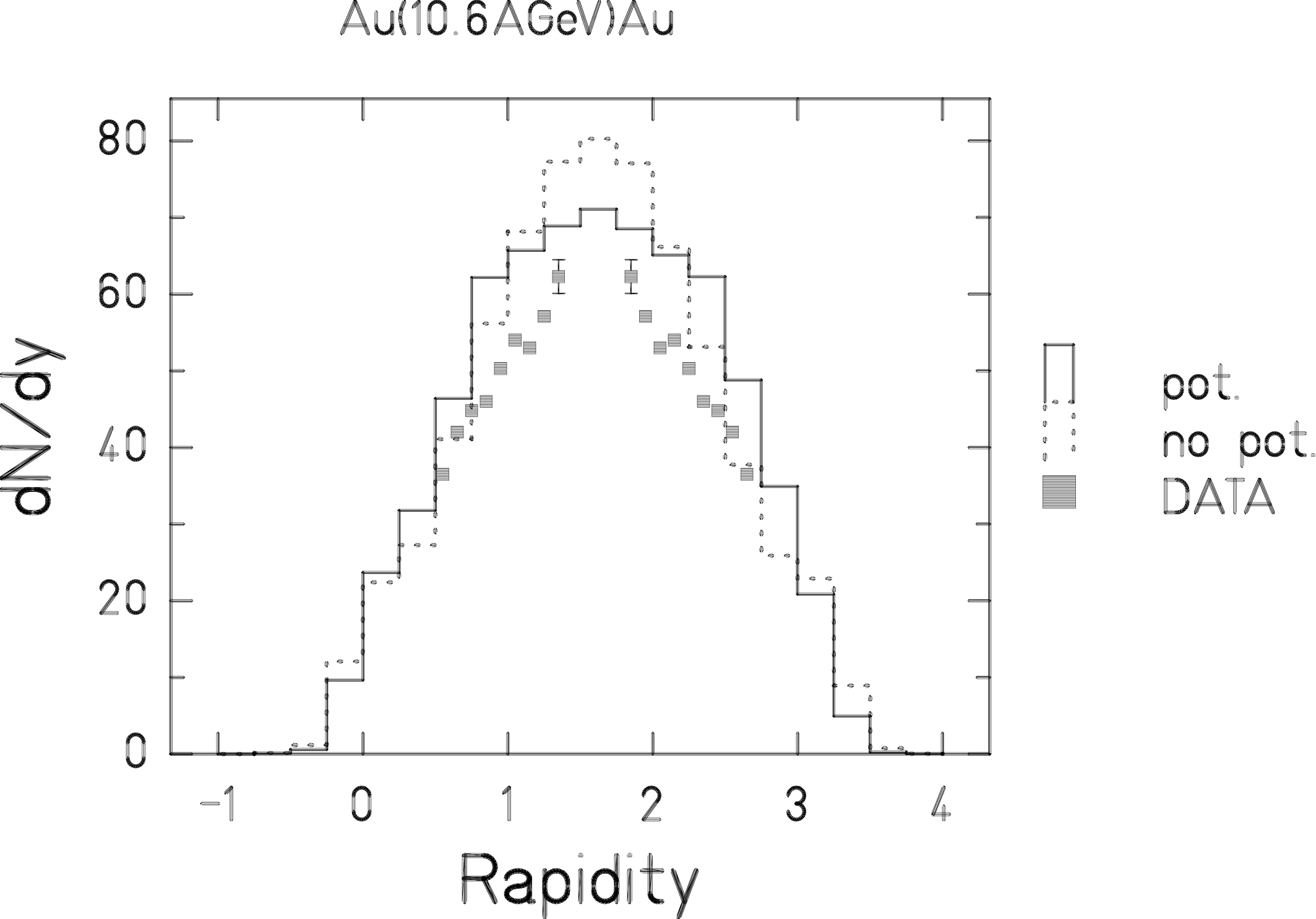,width=8.5cm}}
\end{minipage}
\hfill
\begin{minipage}[t]{6.5cm}
\centerline{\psfig{figure=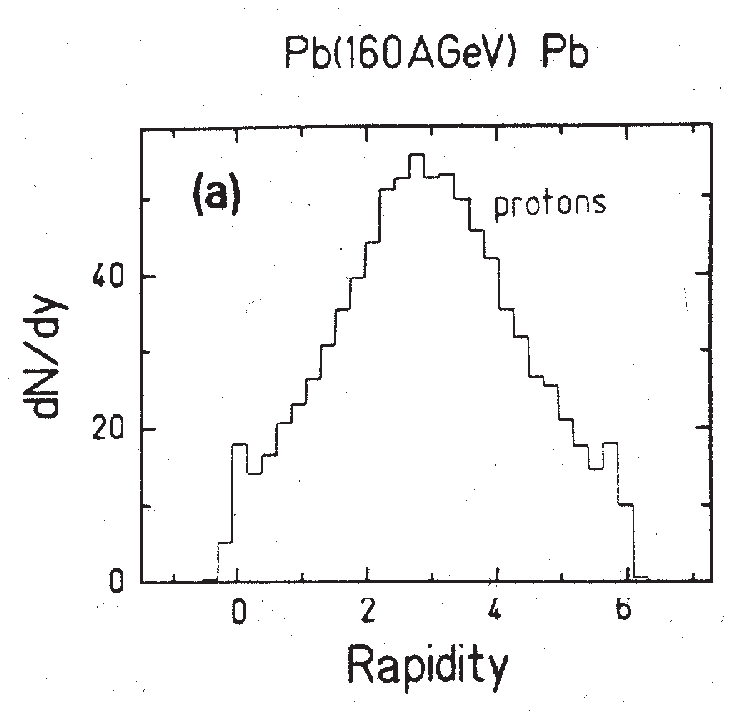,width=6.5cm}}
\end{minipage}
\caption{ \label{stopp-rqmd} Left: 
RQMD prediction \protect \cite{mattiello95b}
of stopping in central Au+Au collisions at 10.6 GeV/nucleon. The
preliminary data are from the E866 collaboration \protect \cite{gonin93a}.
Note that the current status of data analysis indicates a flatter
shape for the experimental distribution \protect \cite{videbaek95a}.
Right: RQMD prediction \protect \cite{keitz91a}
of stopping in central Pb+Pb collisions at 160 GeV/nucleon.}
\end{figure}

\begin{figure}[tbh]
\centerline{\psfig{figure=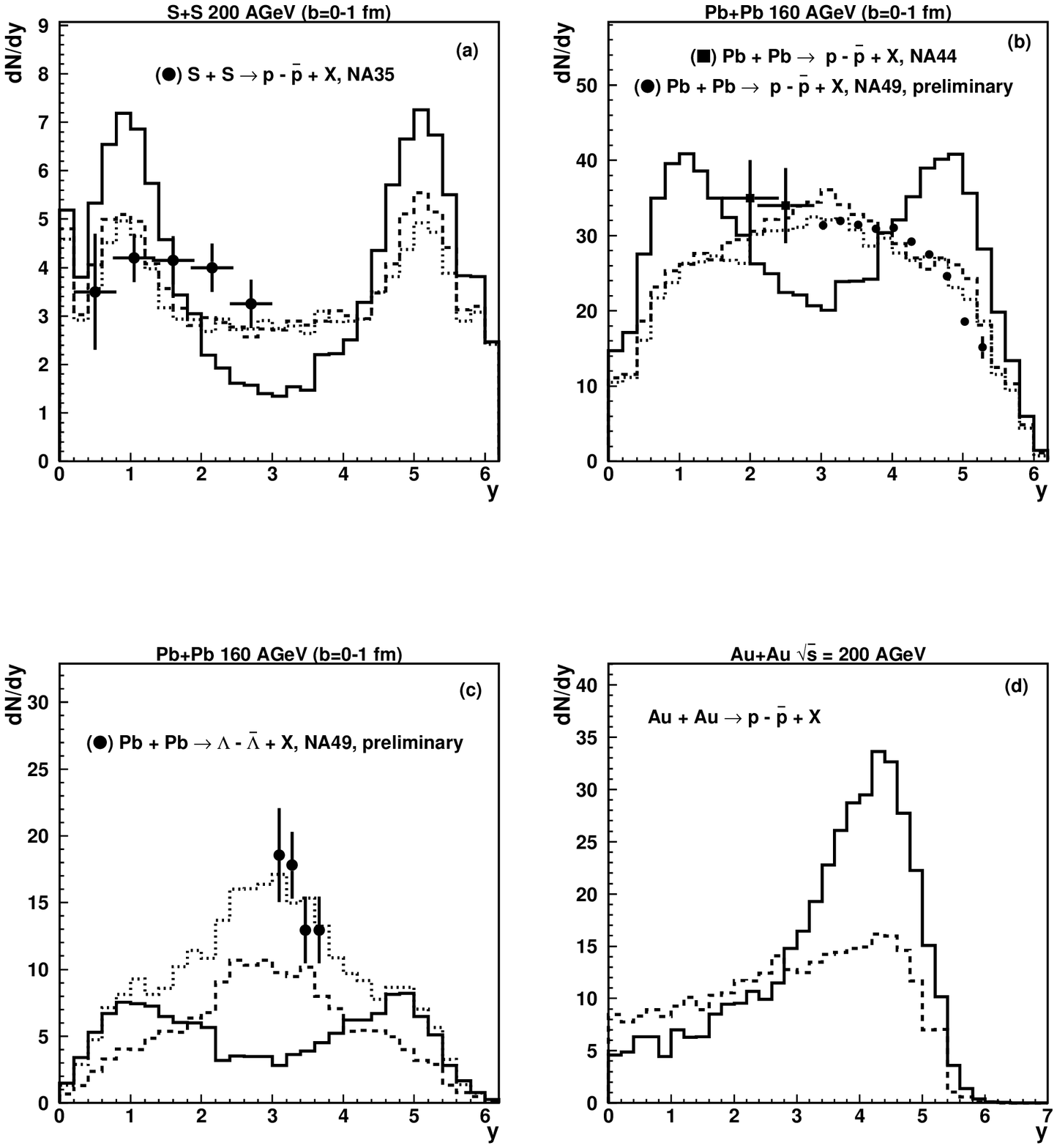,width=11cm}}
\caption{ \label{hijingfig} 
Comparison of baryon stopping in HIJING (solid), HIJING/B (dashed) and
HIJING/B with ``ropes'' with various data
\protect \cite{vance98a}.  }
\end{figure}

Hadronic transport model calculations have predicted stopping for
heavy collision systems at CERN/SPS energies \cite{sorge90a,keitz91a}
(see figure~\ref{stopp-rqmd}). Even for RHIC energies the central rapidity
zone is not expected to be net-baryon free.
RQMD has predicted a net-baryonnumber density of $>10$ 
at mid-rapidity \cite{schoenfeld93a} and HIJING/B yields similar 
estimates \cite{vance98a} (see figure~\ref{hijingfig}).

%Frankfurt mod #7
%Another interesting question is whether the leading particle effect
%observed in hadron-hadron and hadron-nucleus collisions disappears in
%nucleus-nucleus collisions or not. So far no measurements have been
%performed in the nucleus fragmentation region at CERN/SPS-energies.
%
%At LHC-energies a change in the shape of the baryon rapidity distribution
%would mark the transition to the hard regime, if the minijet production
%cross sections are as high as predicted by QCD \cite{}.
%end of mod #7

%%%% miklos fig
%%%%

\begin{figure}[thb]
\centerline{\psfig{figure=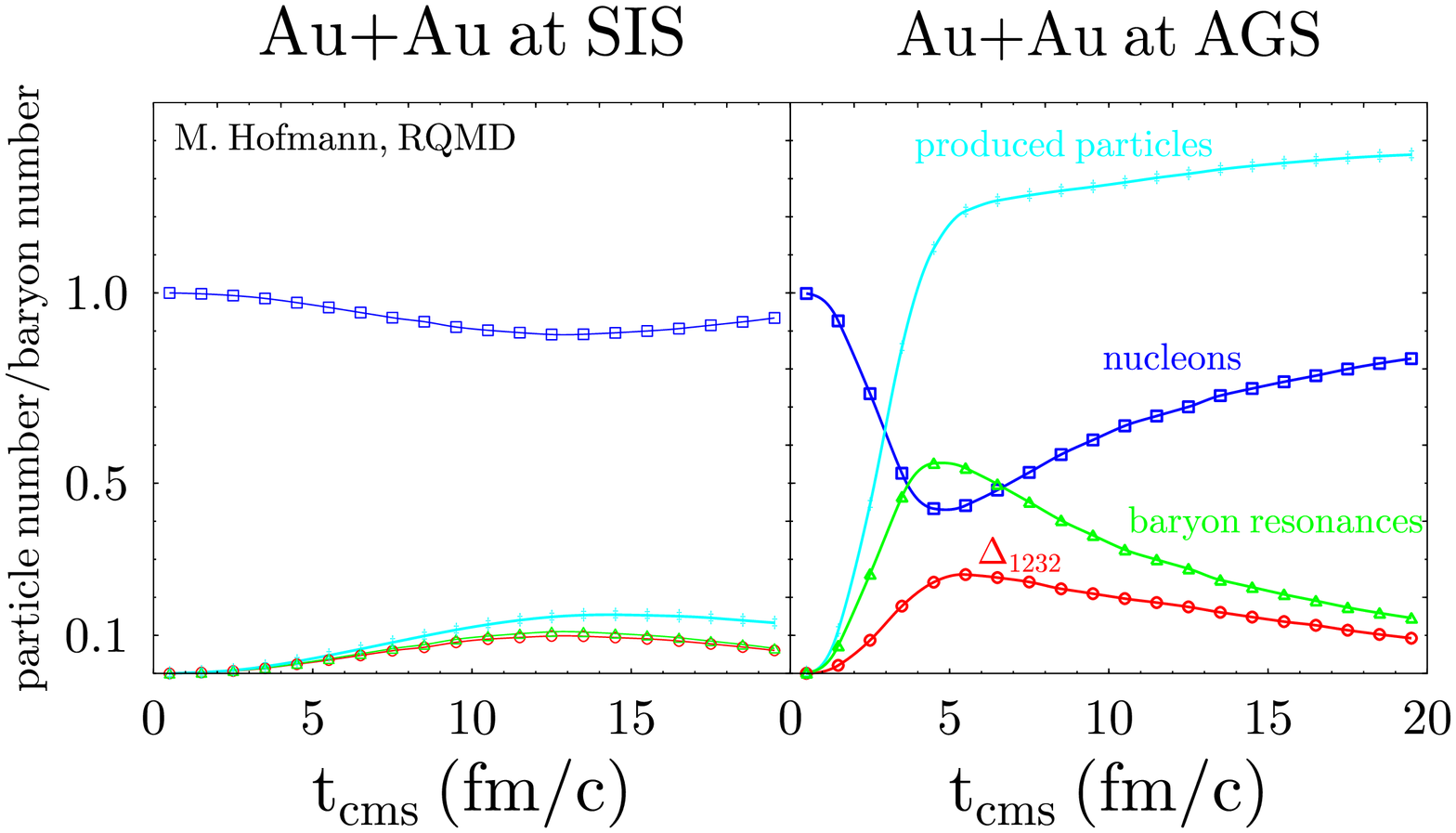,width=13cm}}
\caption{ \label{deltamatter}  Time evolution of particle multiplicities
(scaled with the number of incident nucleons) for central Au+Au collisions
at 1 GeV/nucleon (SIS) and at 10.6 GeV/nucleon (AGS). At SIS energies,
only about 10\% of the nucleons are excited to resonances whereas at
AGS energies the degree of excitation exceeds 50\%. For a time-span
of up to 10 fm/c the baryons are in a state of $\Delta-$matter.
The figure has been taken from \protect \cite{hofmann95a}.
}
\end{figure}

The creation of a zone of high baryon and energy density around mid-rapidity
results in massive excitation of the incident nucleons. A state of
high density resonance matter may be formed \cite{chapline73a,boguta81a,liZ97a}.
Transport model calculations indicate that this excited state of baryonic
matter is dominated by the $\Delta_{1232}$ resonance. They predict
a long apparent lifetime ($>$ 10 fm/c) and a rather large volume
(several hundred fm$^3$) for this $\Delta-$matter state in central 
Au+Au collisions at the AGS \cite{hofmann95a} (see figure~\ref{deltamatter}).

The degree of  stopping can furthermore be used to estimate the 
achieved energy density in the course of the collision within the
Bjorken scenario of scaling hydrodynamics \cite{bjorken83a}. 
For such an estimate not the rapidity distribution of the incident, leading
particles is required, but that of secondary particles,
those produced during the course of the reaction.
One often assumes that particles produced at $y=y_{CM}$ originate
from the central reaction zone at $z=0$ and the initial
proper time $\tau_0$.
The rapidity distribution of these produced particles could then be used
to estimate the initial energy density in the central reaction
zone:
\begin{equation}
\epsilon_0 \,=\, 
\frac{m_T}{\tau_0 A} \,\left. \frac{dN}{dy} \right|_{y=y_{CM}}\quad. 
\end{equation}
Here $A$ is the transverse overlapping region area in the collision and
$m_T$ the transverse mass of the produced particles.
The proper production time $\tau_0$ is very uncertain and estimates are on 
the order of 0.5 - 1 fm/c.
Estimates for the CERN/SPS energy density at proper time
$\tau_0 \sim$ 1 fm/c are on the order
of $\epsilon \approx 1$ to 5 GeV/fm$^3$ \cite{bjorken83a}, with
baryon densities up to $\rho \le 1$ fm$^{-3}$. In ref.~\cite{vance98a}
extrapolations to RHIC suggest that energy densities up to 
20~GeV/fm$^3$ at $\rho \sim 2 \rho_0$ may be reached 
(see figure~\ref{hijingfig}).

\paragraph*{Experimental status\\}

\begin{figure}[thb]
\begin{minipage}[t]{7.5cm}
\centerline{\psfig{figure=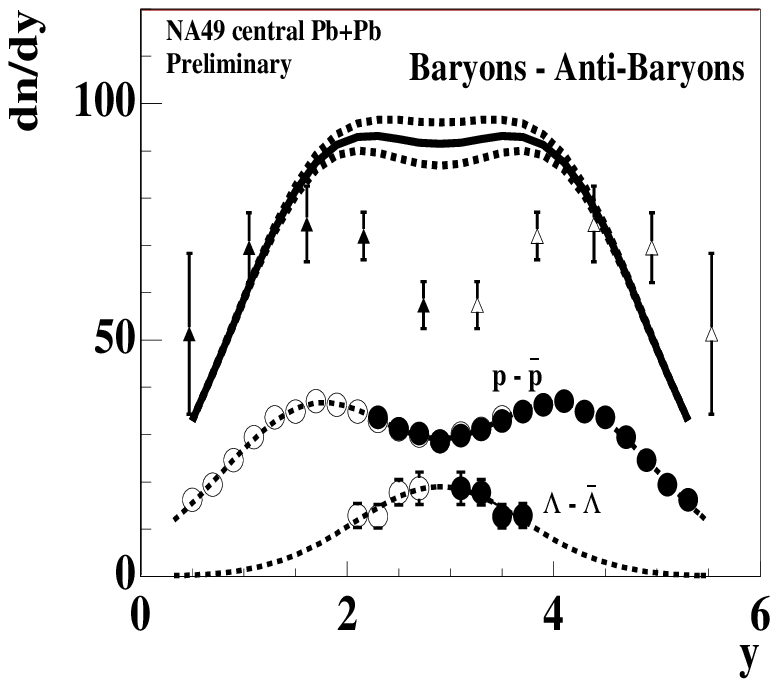,width=7cm}}
\end{minipage}
\hfill
\begin{minipage}[t]{7.5cm}
\centerline{\psfig{figure=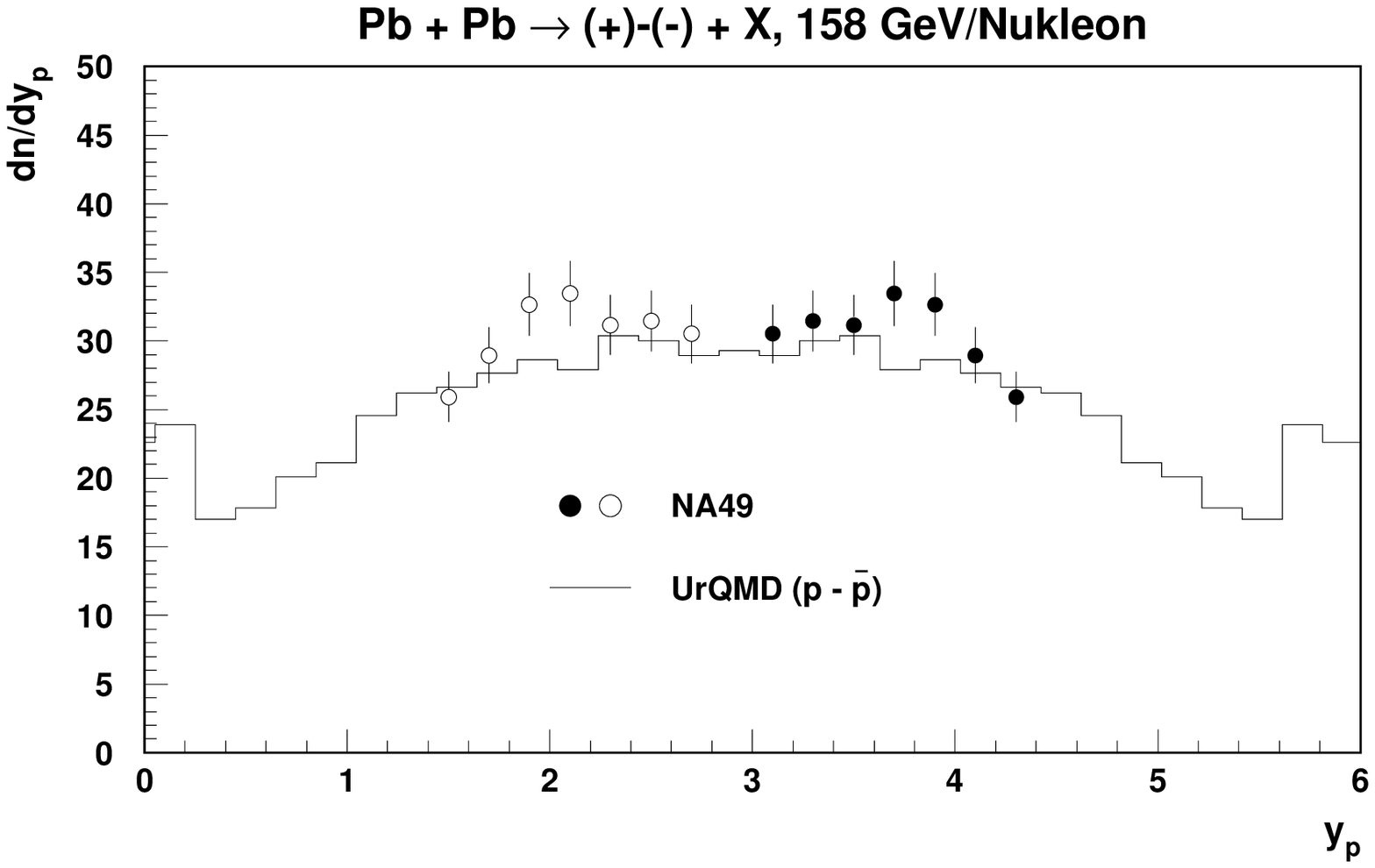,width=7cm}}
\end{minipage}
\caption{ \label{stoppfig1} Baryon stopping in central Pb+Pb collisions
at 160 GeV/nucleon. Left: data by the NA49 collaboration (preliminary,
figure taken from \protect\cite{roland98a}) The solid line shows the
rapidity distribution for net-baryons which can be decomposed into
contributions from net-protons and net-$\Lambda$. For comparison the
net-baryon distribution for central S+S collisions is also plotted
(triangles). Right: UrQMD prediction compared to the same data 
(figure taken from \protect\cite{bass98a}).
}
\end{figure}

At AGS and SPS an extensive investigation of the nuclear stopping
power is near completion. Proton-proton \cite{blobel74a} and peripheral 
nucleus-nucleus  interactions at AGS \cite{abbott94a,barrette94a} 
and SPS \cite{baechler94a}  energies yield a forward--backward 
peaked $dN/dy$ distribution in the C.M. frame, 
and a low degree of baryon stopping.

A higher degree of stopping is observed 
for central collisions of intermediate mass nuclei 
(Si+Si at AGS, S+S at SPS):
The rapidity distribution is flat at C.M. rapidities, 
two broad bumps are observed between 
projectile/target and C.M. rapidities respectively
\cite{abbott94a,barrette94a,baechler94a}. 
The heaviest collision systems (gold and lead respectively) 
exhibit the largest stopping
power and thus correspond to the creation of the highest baryon densities:
At AGS energies, the baryon rapidity distribution
exhibits a  pile-up at mid-rapidity \cite{videbaek95a,wienold96a}
(see figure~\ref{stopp-rqmd}). 
Whether the shape of the $dN/dy$ distribution at SPS
energies is flat or shows two
bumps is currently 
not fully resolved (the SPS-data are preliminary). 
There are indications, however, that with rising
beam energy the scaled $dN/d(y/y_p)$ distribution stretches over
the increasing rapidity gap between projectile and target;
this can be seen in figure \ref{stoppfig1}.
Recently the NA49 collaboration \cite{bormann97a} reported a  
$\Lambda$ rapidity distribution which may be peaked strongly at mid-rapidity 
for Pb+Pb at 160 GeV/nucleon. This finding, however, is preliminary and
in disagreement with equally preliminary results by the WA97 
collaboration \cite{wa97sqm}, which indicate rapidity densities for
$\Lambda$'s lower by a factor of $2-3$.

Transverse energy measurements at the AGS \cite{barrette93a} indicate that
the transverse energy
$E_T$ increases by 50\% faster than 
predicted by an independent nucleon--nucleon interaction model
when going from a light system (Si+Al) to a
heavy system (Au+Au). In terms of a microscopic hadronic model this can be
understood as a strong increase in baryonic density in the initial 
reaction phase and a corresponding large increase in the volume of
high density matter \cite{pang92a,barrette94b}.

The $\Delta(1232)$ abundance has been measured via $\pi^+-p$ correlations 
at the AGS by the E814 and E877
collaborations \cite{hemmick94a,barrette95a}. The pion spectra 
can be decomposed into
a thermal contribution and a contribution due to $\Delta-$resonance decays. 
The $\Delta(1232)$-to-nucleon ratio at freeze-out was determined to be
$\approx 35$\% for central silicon nucleus collisions.  
Hence, one can conclude that a large fraction of the system resides
in hadron resonances, which produce most of the observed hadrons by
their decay (``feeding''), after the resonances have decoupled. The dense
state before this decay can therefore be called ``resonance-matter''.
It exists due to the inertial confinement of energy 
and baryon number in the early phase of the reaction 
(see figure~\ref{deltamatter}).

At CERN/SPS measurements of $E_T$ have been used to estimate the created
energy density: For 200 GeV/u S+Au central collisions \cite{baechler91a} 
$\epsilon$ reaches $\approx$ 
3 GeV/fm$^3$. For the Pb+Pb
experiment at 160 GeV/u similar values were extracted \cite{alber95a}, but over
a much larger volume.
The reader is reminded here of the sensitivity of these
extracted values on the hadron production time $\tau_0$, which is 
uncertain to at least a factor of two.

\begin{figure}[thb]
\centerline{\psfig{figure=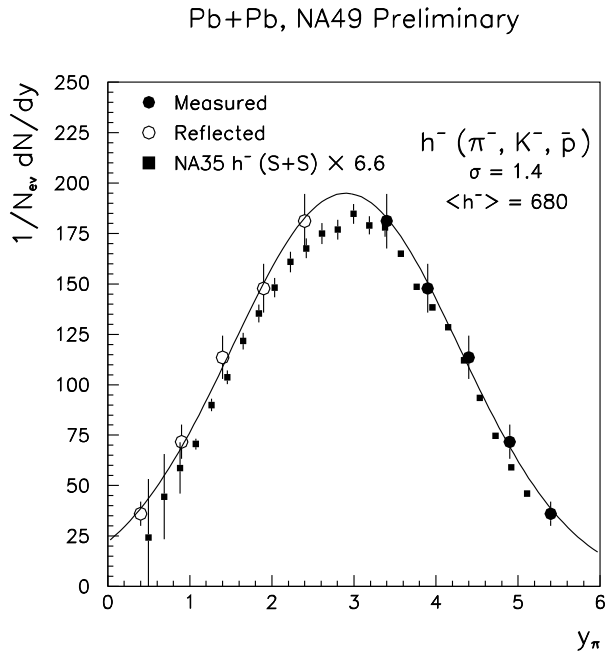,width=8cm}}
\caption{ \label{stoppfig2} Rapidity density of negative hadrons
for central collisions. The circles represent the preliminary Pb+Pb 
measurement by the NA49 collaboration \protect \cite{jones96a}, whereas
the squares are from the NA35 S+S experiment \protect \cite{alber94a}.
The latter are scaled by a factor of 6.6 , which corresponds to the relative
number of participants. This scaled sulfur distribution agrees well
with the lead distribution. The figure has been taken from 
\protect \cite{jacobs96a}.}
\end{figure}

The mass-dependence of the rapidity distribution of produced particles,
i.e. pions or kaons, can also used to search for scaling violations
which could signal the onset of new physics phenomena.
A comparison of the 
negative hadron rapidity  distributions for S+S \cite{alber94a} 
with those for the
lead on lead run \cite{jones96a}  shows that the preliminary lead
data can be matched by scaling the sulfur data with 
a factor 6.6 , close to 
the relative number of 
participant nucleons in central lead-lead collisions
($A_{Pb}/A_S = 208/32 \approx 6.5$) \cite{jones96a}
(Figure \ref{stoppfig2}).

\paragraph*{Discussion\\}

The form of the measured baryon rapidity distributions 
shows experimentally that the central rapidity region 
up to $E_{lab} \sim 200$~GeV/nucleon is not net-baryon free,
in contrast to what
had been expected in most early papers. Rather 
strong stopping as assumed first in hydrodynamic model studies
\cite{stoecker86a,clare86a}
is observed. Therefore, results of
theoretical analyses, which rely heavily on a net-baryon 
free mid-rapidity region
with zero baryo-chemical potential have to be taken with
care. The quantitative measurements of the $A$-dependent stopping
of baryons is one of the most important results
of the AGS and SPS measurements.

If the preliminary findings of a strongly peaked $\Lambda$ rapidity 
distribution \cite{bormann97a}
and a rather broad Gaussian or flat baryon rapidity distribution
\cite{jones96a} by the NA49 collaboration are both confirmed, then 
this would be a hard obstacle for models which rely on global thermal
equilibrium (plus flow) for the description of the final state of the 
the reaction \cite{cleymans93a,letessier94a,braun-munzinger96a}.

Simple ``first collision models'' without rescattering 
\cite{werner93a,anderson87a,anderson87b,sjoestrand94a} do not suffice
to reproduce the data, whereas transport theory has correctly
predicted the observed degree of baryon-stopping
\cite{sorge90a,keitz91a,pang92a,schmidt93a,mattiello95b,bass98a}.
An alternative mechanism of baryon stopping based 
on diquark breaking 
\cite{capella96a}, is also able to describe 
the corresponding experimental data, in contrast to 
the simple first collision approach. These models extrapolated
to RHIC energies imply that even at $\sqrt{s} = 200$~GeV/nucleon
the dense matter is created with baryon density $\sim 2 \rho_0$ at 
$\tau \sim 1$~fm/c. In \cite{vance98a} the beam energy dependence
of the initial baryon density is estimated to vary as 1/$s^{1/4}$.

The energy densities of $\epsilon \approx$ 3 GeV/fm$^3$ 
estimated (with a factor of $\approx 3$ uncertainty) from 
rapidity distributions of produced particles  indicate that part of 
the system may have entered the predicted state of 
deconfinement \cite{harris96a}.
Hadronic transport models, however, predict or reproduce the
measured rapidity distributions, if baryon and meson
rescattering and particle production via string decay 
\cite{sorge90a,keitz91a,sorge92a,pang92a,bass98a} are included.
Also hadronic models which include 
multi--quark droplets \cite{aichelin93a} above
$\epsilon_{\rm crit}$ seem to give similar results. 

\begin{figure}[htb]
\begin{minipage}[t]{5.6cm}
\centerline{\psfig{figure=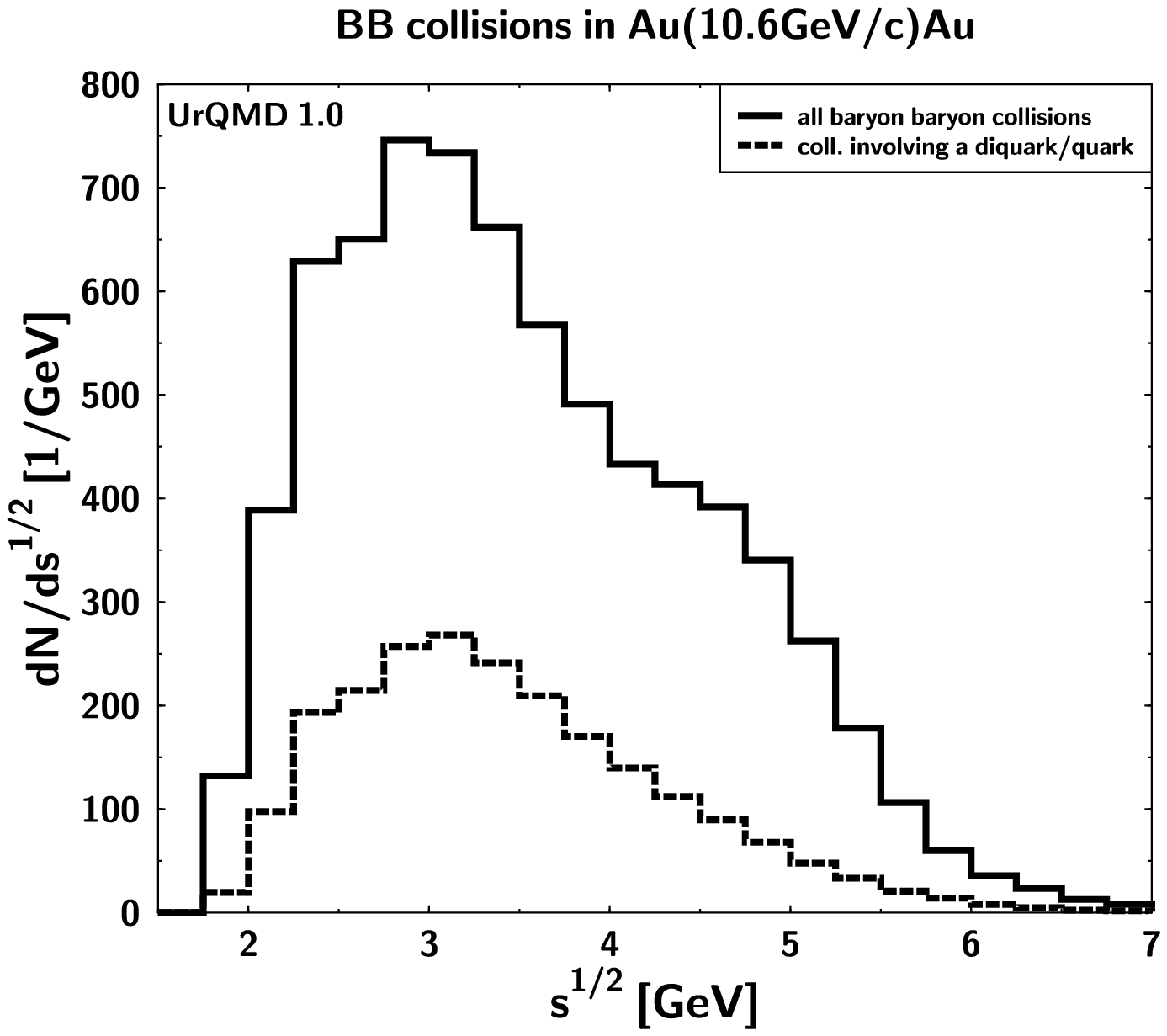,width=7.0cm}}
\end{minipage}
\hfill
\begin{minipage}[t]{5.6cm}
\centerline{\psfig{figure=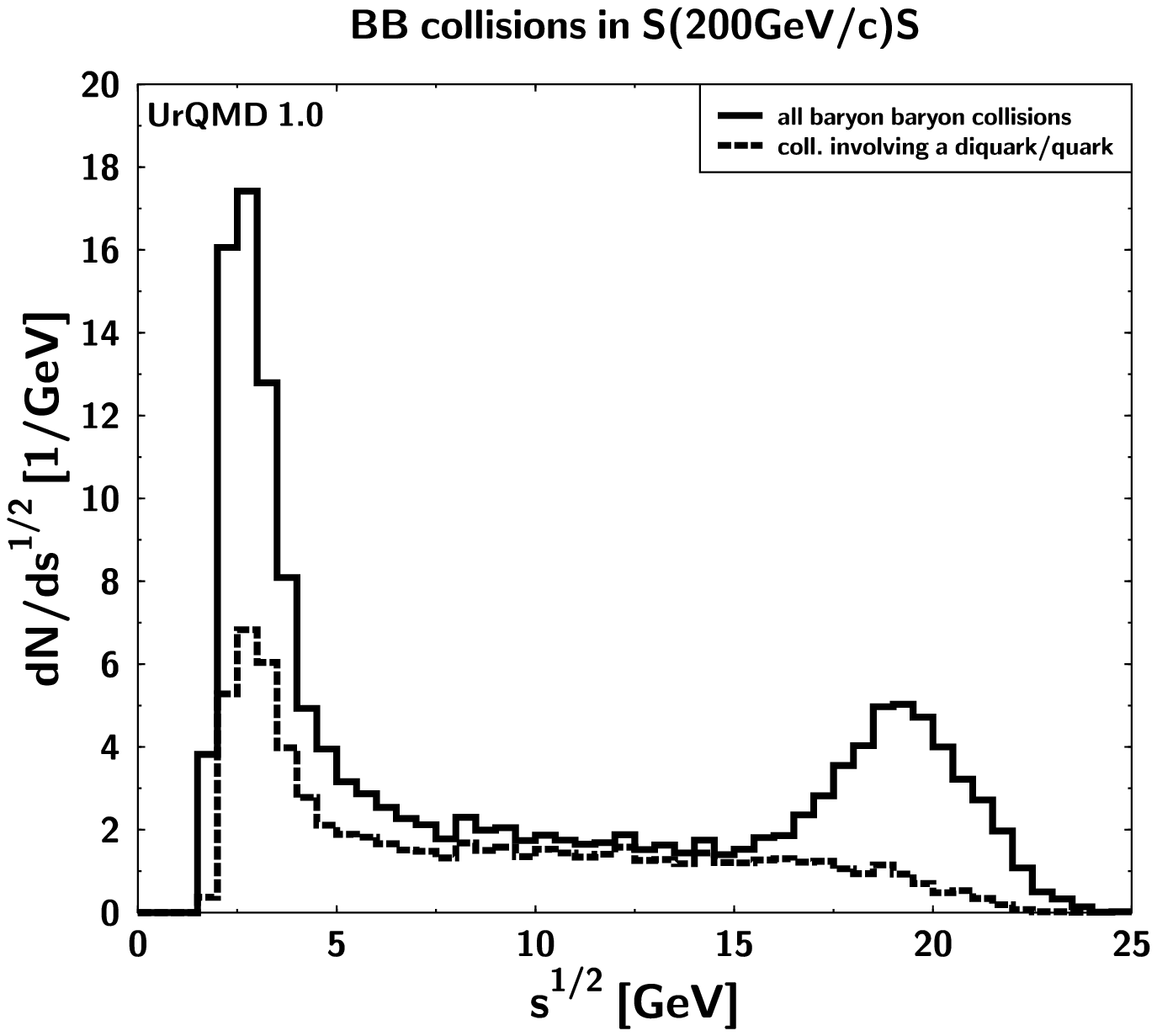,width=7.0cm}}
\end{minipage}
\caption{\label{uqmdsrt}  $E^{coll}_{CM}$ distribution for baryon baryon 
collisions in central Au+Au reactions at the AGS (left) and in central
S+S reactions at the SPS, calculated with the UrQMD transport model
\protect \cite{bass98a}. }
\end{figure}

The inclusion of string excitations, collisions and decays are a
first step towards modeling the parton/quark substructure of hadrons.
In this sense these models go beyond what one would term purely 
hadronic model:
Figure~\ref{uqmdsrt} shows $\sqrt{s}$ distributions of baryon-baryon 
interactions for Au+Au collisions at AGS and S+S collisions at SPS
energies \cite{bass98a}. At the AGS, the collision spectrum is
dominated by collisions of fully formed baryons. It exhibits a
maximum at low energies, $\sqrt{s} \approx 3$~GeV. 
Approximately 20\% of the collisions
involve a diquark, i.e. a leading baryon originating from a string decay.
In contrast to the heavy system at AGS, 
the collision spectrum for S+S at SPS 
exhibits two pronounced peaks. They are dominated by full BB collisions, 
one peak at the  energy of initial projectile-target collisions, 
and one peak in the low (``thermal'') energy  range.
Approximately 50\% of the BB collisions, most of them represented by the
bump at intermediate
$\sqrt{s}$ values, involve  diquark- or constituent 
quark-collisions with baryons.

The linear scaling behavior of the mass dependence of the negative
hadron rapidity distributions 
precludes a strong pQCD minijet component at these
energies. We note that the agreement of the VNI parton cascade 
model with the $E_T$ systematics at SPS \cite{geiger97a} 
may be due to adjusting
a strongly model dependent soft beam jet component to fit proton
proton data. This issue is 
important because in ref.~\cite{geiger97b} it was claimed
that at SPS already a partonic energy density
$\ge 5$~GeV/fm$^3$ was created. 
The $A^1$ scaling of $E_T$ therefore constrains
very strongly against hard scattering models used for example
in \cite{kharzeev96c} to argue for a QGP interpretation of $J/\Psi$
suppression. We return to this point later.

The experimental results demonstrate that highly
excited dense matter is formed at mid-rapidity. They prove that a new
state of elementary matter has been created. 
However, the inclusive central distributions do not
give a clear and decisive answer to the question of whether this matter
is predominantly of hadronic or quark nature.

%%%%%%%%%%%%%%%%%%%%%%%%%%%%%%%%%%%%%%%%%%%%%%%%%%%%%%%%%%%%%%%%%%%%%%%%%%%%%%%%%
	\subsection{Creation of high temperatures: particle spectra}
	\label{spectra}
%%%%%%%%%%%%%%%%%%%%%%%%%%%%%%%%%%%%%%%%%%%%%%%%%%%%%%%%%%%%%%%%%%%%%%%%%%%%%%%%%

\paragraph*{Theoretical concepts\\}

The hot, dense reaction zone consists of slowed down incident nucleons and 
produced particles. The {\em fireball} model considers these hadrons
as a mixture of ideal gases in thermodynamic equilibrium.
For temperatures above 50 MeV and moderate densities, 
the Fermi- and Bose-Einstein-distribution 
functions for baryons and mesons (except for the pions) may
be approximated by a Maxwell-Boltzmann distribution 
\cite{gosset77a,hagedorn80a}
with the temperature $T$  and the chemical potentials $\mu_i$ 
(connected to conserved quantum numbers $i$) as only free 
parameters.

Kinetic equilibration is thought to be visible predominantly in the transverse
degrees of freedom;
therefore, transverse momentum or transverse mass distributions are 
used to extract temperatures from the spectral slopes.
 
It has been suggested that 
abnormal nuclear matter, e.g. a QGP, may be observed via a secondary,
high temperature component in the particle spectra or via a shoulder
in the pion multiplicity distributions \cite{stoecker81b}.

It has also  been suggested that
the equation of state, that is the energy density $\epsilon$
vs. temperature $T$, can be probed experimentally by 
plotting the mean transverse momentum $\langle p_t \rangle$ vs.
the rapidity density $dN/dy$ or the transverse energy density
$dN/dE_T$. If a phase transition occurs (i.e. a rapid change in the number of 
degrees of freedom) one expects a monotonously rising curve interrupted by
a plateau: This plateau is 
is caused by the saturation of $\langle p_t \rangle$ 
during the mixed phase. 
After the phase transition from e.g.  color singlet states
to colored constituents has been completed \cite{vanhove82a} the mean
transverse momentum rises again.
However, detailed hydrodynamical studies \cite{kapusta85,gersdorf87}
showed that the plateau is washed out due to collective flow.

Collective (radial) flow \cite{scheid74a,siemens79a,stoecker81b} as well as 
feeding from  resonances strongly influence
the shape of the particle spectra 
\cite{stoecker81b,sollfrank90a,leeKS90a,schnedermann93a,
schnedermann94a,mattiello95a,mattiello96a}.
For light composite particles, such as deuterons, 
the influence of collective flow is
visible in a shoulder-arm shape of the transverse momentum spectra
\cite{stoecker81b}.
This can be seen in figure \ref{mattfig1}.
In order to account for flow effects, the spectra can be fitted with
a thermal distribution including collective flow. The temperature $T$ and
the transverse flow velocity $\beta_t$ are the fit-parameters.
The shapes of the velocity profile and density profile at freeze-out 
should enter as additional degrees of freedom in the analysis.
Usually a box shaped density profile and a linearly increasing transverse
velocity profile are assumed \cite{stoecker81b,sollfrank90a,letessier94a,
braun-munzinger95a}. This results in severe distortions into the
analysis, as discussed in the following \cite{konopka95a}:

\begin{figure}[thb]
\centerline{\psfig{figure=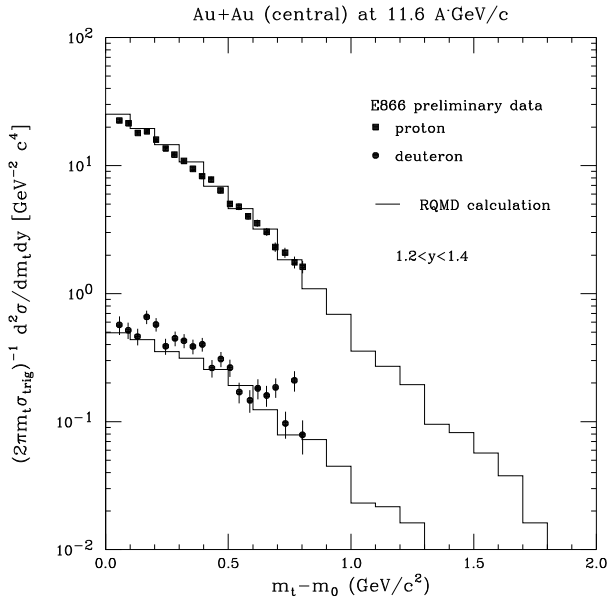,width=8cm}}
\caption{ \label{mattfig1} RQMD prediction of transverse mass spectra
for protons and deuterons in central Au+Au collisions at the AGS 
compared with preliminary data by the E866 collaboration.
For deuterons a shoulder is visible in the low $m_t$ range of the spectrum.
This structure is due to collective flow.
%spectra at mid-rapidity for protons
%and light composite particles in central Si and Au collisions
%at the AGS. The bold solid histogram is a calculation with potentials,
%the thin histogram the respective calculation without potentials and
%the smooth solid line depicts  a Boltzmann parameterization adjusted
%to the high momentum part of the spectra. The strong shoulder arm structure
%is due to collective flow.
The figure
has been taken from \protect \cite{mattiello96a}.}
\end{figure}

When extracting temperatures and flow velocities 
from microscopic calculations, 
the system is divided into cells and the {\em local} 
transverse and longitudinal velocity distributions are
analyzed \cite{berenguer92a,mattiello96a,konopka96a,hofmann95a}. 
The temperatures extracted via a global  two parameter fit are more than 
a factor of two higher than the
temperatures gained from such a microscopic analysis at beam
energies in the 100 MeV/nucleon to 10 GeV/nucleon regime \cite{konopka96a}.
The reason for this discrepancy lies mostly in the assumed shape of the 
freeze-out density profiles.

Whereas a linearly increasing transverse freeze-out velocity
profile seems a tolerable assumption, the shape of the freeze-out density
profile has -- due to collective flow --
a Gaussian shape (centered at $r_t=0$), rather than the
usually assumed box-shape distribution.
When realistic density and velocity profiles are used, one finds that
the high $m_t$
components of the particle spectra reflect contributions of 
large collective flow effects
(i.e. the high expansion velocity). This analysis yields substantially 
lower values for  the temperatures $T$.
Such microscopic analyses of the spectra of protons, mesons 
and light composite particles
at AGS energies show also that $\beta_t$ and $T$ depend on the mass of the
particle \cite{mattiello96a,sorge96a}.

\paragraph*{Experimental status\\}

\begin{figure}[thb]
\centerline{\psfig{figure=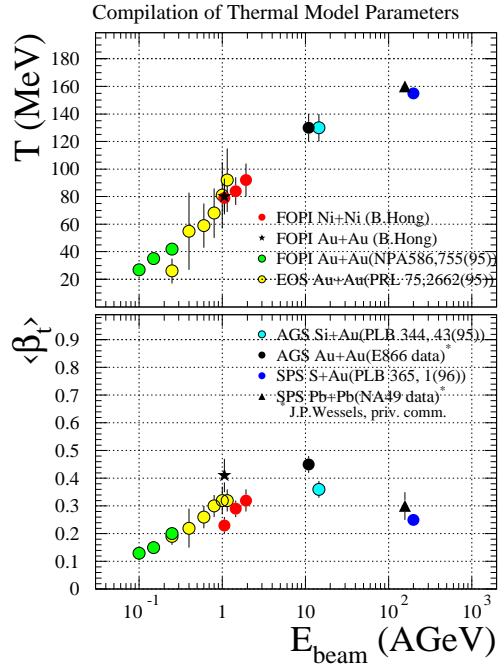,width=8cm}}
\caption{ \label{spekfig3} Excitation function of temperature T and
average transverse expansion velocity $\beta_t$. The figure
has been taken from \protect \cite{herrmann96a}.}
\end{figure}

Data taken at the AGS with Si beams \cite{abbott94a} seem on first sight 
to be consistent with an expanding, 
hadro-chemically and thermally equilibrated system with a temperature
of $130 \pm 10$ MeV and a transverse flow velocity of $\beta_t \approx 0.36$
\cite{letessier94a,braun-munzinger95a}. CERN SPS data with S beams 
have been fitted
in the same fashion, with apparent temperatures around 150 MeV and flow 
velocities between 0.35 and 0.41 \cite{schnedermann93a,braun-munzinger96a}.
Figure \ref{spekfig3} shows the extracted 
excitation function for the temperature
$T$ and the average transverse expansion velocity $\beta_t$ \cite{herrmann96a},
including also SIS and BEVALAC data.

\begin{figure}[htb]
\begin{minipage}[t]{7.5cm}
\centerline{\psfig{figure=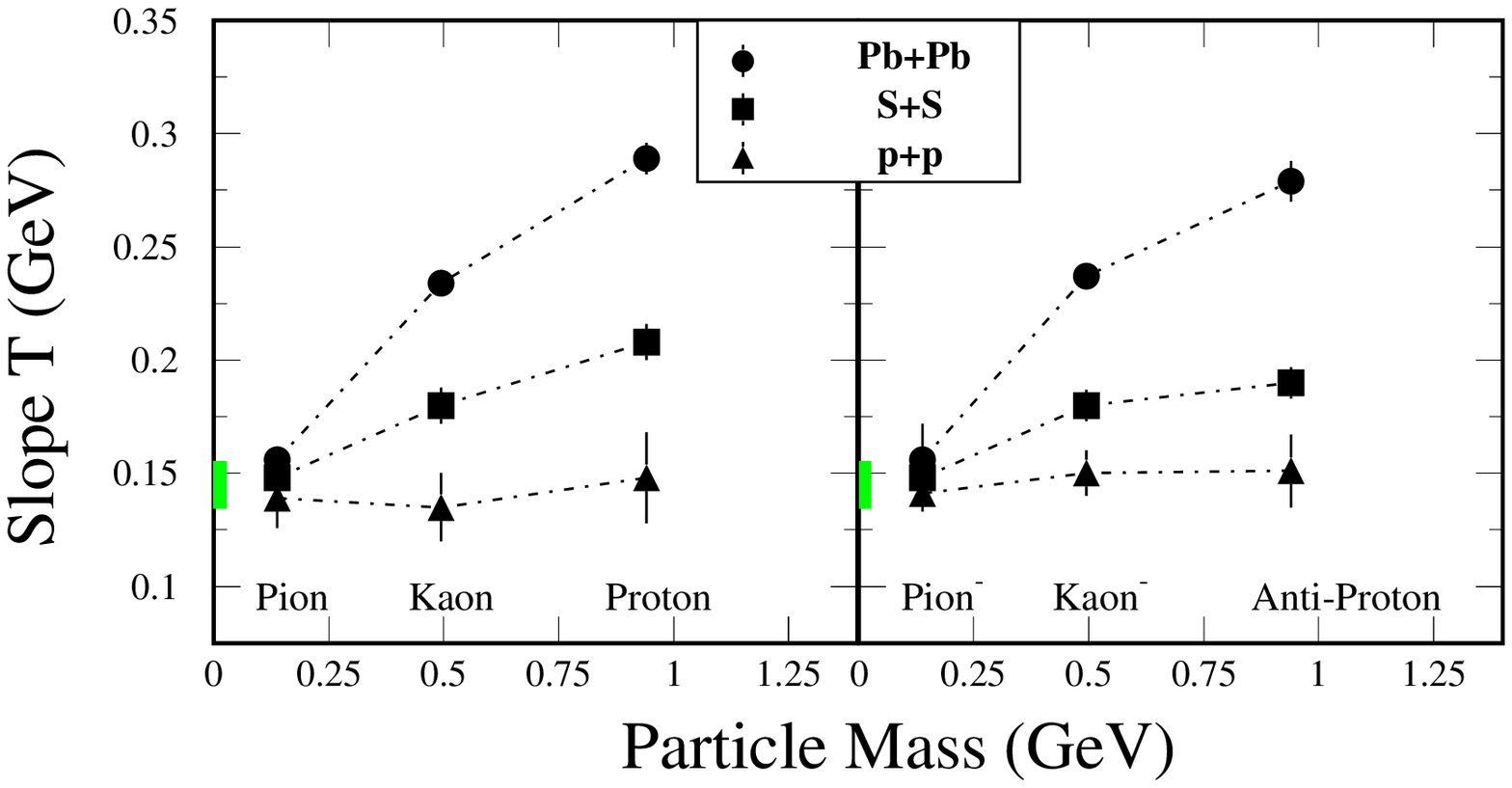,width=10cm}}
\end{minipage}
\hfill
\begin{minipage}[t]{5.5cm}
\centerline{\psfig{figure=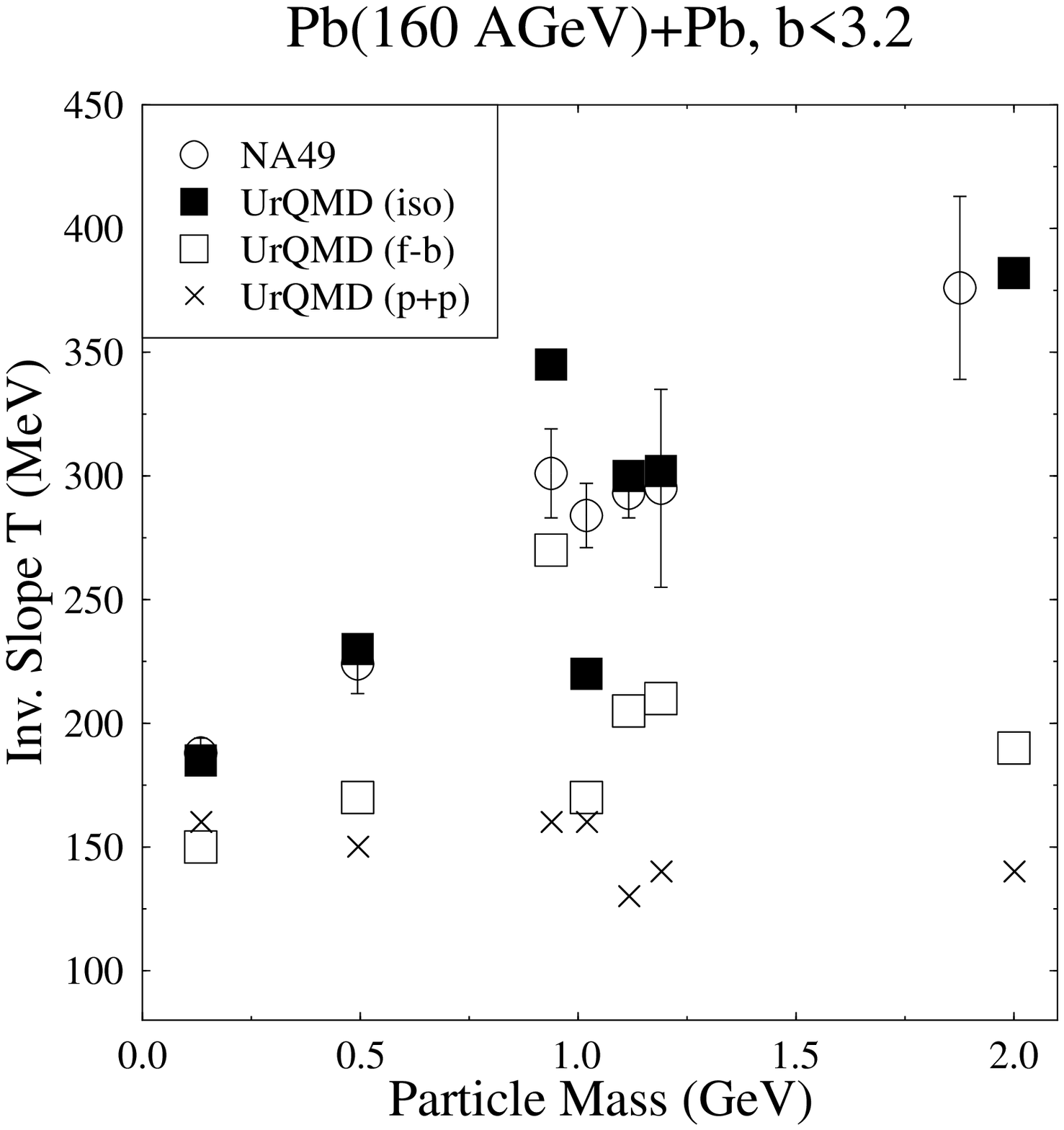,width=6cm}}
\end{minipage}
\caption{\label{xufig} Mass and collision system dependence of the 
inverse slope parameter $T$ measured by the NA44 collaboration 
\protect \cite{xun96a} (left) and calculated by the UrQMD model (right)
\protect\cite{bleicher98a}.}
\end{figure}

In order to disentangle collective flow contributions from thermal motion,
the dependence of the slope parameter $T_{sl}$ (which includes collective
flow effects) on the collision system mass and the particle
mass has been studied by the NA44 and NA49 collaborations 
\cite{xun96a,jones96a} at the SPS. 
Results can be seen in figure \ref{xufig}.
In proton-proton collisions obviously no collective effects are visible and an
inverse slope parameter of $T_{sl,pp}=145$ MeV is extracted for all analyzed
particle species ($\pi, K$ and $p$). When going to heavier collision systems,
collective flow effects become obvious: the inverse slope parameter
$T_{sl}$ increases with the mass of the emitted particle 
(see figure~\ref{xufig}). 
Empirically one finds 
$T_{sl} = T_{sl,pp} + m \cdot \langle \beta_t \rangle^2 $. $\beta_t$ is the
mean expansion velocity which depends on the mass of the collision system
and $m$ is the mass of the particle analyzed. 

The constant $T_{sl,pp}$ in the empirical result therefore
hints at the predicted {\em limiting temperature}
\cite{hagedorn65a,stoecker81b} of $140 \le T \le 200$ MeV. The observation of the
increase of the flow effects for massive particles and heavy collision systems
had been predicted with early hydrodynamical and microscopic calculations
\cite{siemens79a,stoecker81b,stoecker86a,hung95a,sorge95a,bleicher98a}.

\paragraph*{Discussion\\}

The data at AGS and CERN seem compatible with a hadro-chemically
and thermally equilibrated system. 
However, this does not mean that
the system necessarily evolved through  thermal and chemical equilibrium states
\cite{konopka96a,bleicher97a,weber97a}. 
The fits to the spectra, with the temperature $T$ and the
transverse expansion velocity $\beta_t$ as parameters, have to be
performed with great care. 
There is a broad range of $T$ and
$\beta_t$ values which are compatible with the same 
spectrum \cite{stoecker81b,schnedermann93a,schnedermann94a,konopka96a}, 
where the temperature $T$ depends crucially on
the freeze-out density and
velocity profiles -- at least in the case of composite
particles such as deuterons and 
tritons \cite{konopka95a,mattiello96a,polleri97a,bleicher98a}.

The finding of one global freeze-out temperature $T$ and velocity $\beta_t$
\cite{letessier94a,braun-munzinger95a} is to be contrasted with the independent
analysis based on RQMD and UrQMD calculations of spectra of light composite 
particles \cite{konopka95a,mattiello96a,bleicher97a} and on spectra of 
mesons \cite{sorge95a,bleicher97a,bass98a}. 
These models are well able to reproduce the data
and the analysis indicates different values (with a variation of $\sim $20\%) 
for $\beta_t$ and $T$, depending
on the mass of the particle. 
The simplified thermal plus flow model 
\cite{stoecker81b,schnedermann93a,letessier94a,braun-munzinger95a} should 
not be taken literally. In reality we expect a complicated space-time
dependent non-equilibrium freeze-out, details depending on inelastic 
production and absorption cross sections.
In particular, the anti-baryon annihilation cross sections play an important
role, as will be discussed below (section~\ref{flow}). Furthermore,
flow of mesons vs. baryons \cite{bass93a,jahns94a} in  opposite
directions clearly indicate strong deviations from the single
source fits as discussed in the following section.

Recently, the WA98 collaboration reported $\pi^0$ spectra in
Pb+Pb reactions for $p_t$ up to 4 GeV/c \cite{WA98_pi0}. The data could be fit
well by hydrodynamical models \cite{dumitru98a}.
However, it was found in \cite{wangXN98a} that the data were well reproduced
by the QCD parton model. In this sense (the non-equilibrium)
quark plasma is seen in the high $p_t$ spectra. However, as 
emphasized in \cite{gyulassy98a}, at SPS the parton model
is hypersensitive to models for soft multiple collisions. Hydrodynamics
just happens to be one of the soft multiple collision models that
can account for the data.

%If thermal equilibrium exists at some reaction stage, the 
%large interaction cross sections of the hadrons would have
%destroyed the correlations from  the early
%hot and dense reaction phase (i.e. entropy or energy density) -- the 
%extracted temperature would reflect the
%status of the system at the time of freeze--out.

%Nevertheless some abrupt change in the monotonic behavior
%of the $T$ excitation function would hint at the occurance of some
%exotic physics phenomenon -- so far nothing in that direction has
%been observed.

%%%%%%%%%%%%%%%%%%%%%%%%%%%%%%%%%%%%%%%%%%%%%%%%%%%%%%%%%%%%%%%%%%%%%%%%%%%%%%%%%
	\subsection{Transverse collective radial and directed flow}
	\label{flow}
%%%%%%%%%%%%%%%%%%%%%%%%%%%%%%%%%%%%%%%%%%%%%%%%%%%%%%%%%%%%%%%%%%%%%%%%%%%%%%%%%

\paragraph*{Theoretical concepts\\}

The excitation function of transverse collective flow is 
the earliest predicted signature for probing 
compressed nuclear matter
\cite{scheid68a,scheid74a}.
It has been shown that the excitation function of flow is sensitive
to the EoS and can be used to search for abnormal matter states 
and phase transitions \cite{hofmann76a,stoecker86a,hartnack_iso}.

In the fluid dynamical approach, the transverse collective flow is directly
linked to the pressure of the matter  in the reaction
zone: 

With the pressure $P(\rho, S)$  (depending on the density $\rho$
and the entropy $S$), one can get a physical feeling for 
the generated collective transverse momentum $\vec{p}_x$ by writing
it as an integral of the pressure acting on a surface and over 
time \cite{stoecker81a}:
\begin{equation}
\label{pxeqn}
\vec{p}_x \,=\, \int_t \int_A P(\rho,S) \, {\rm d}A \, {\rm d}t
\end{equation}
where d$A$ represents the surface element between the participant and
spectator matters and the total pressure is the sum of the
potential pressure and the kinetic pressure:
The transverse collective flow depends directly on the equation of state,
$P(\rho,S)$.

\begin{figure}[thb]
\begin{minipage}[t]{8.5cm}
\centerline{\psfig{figure=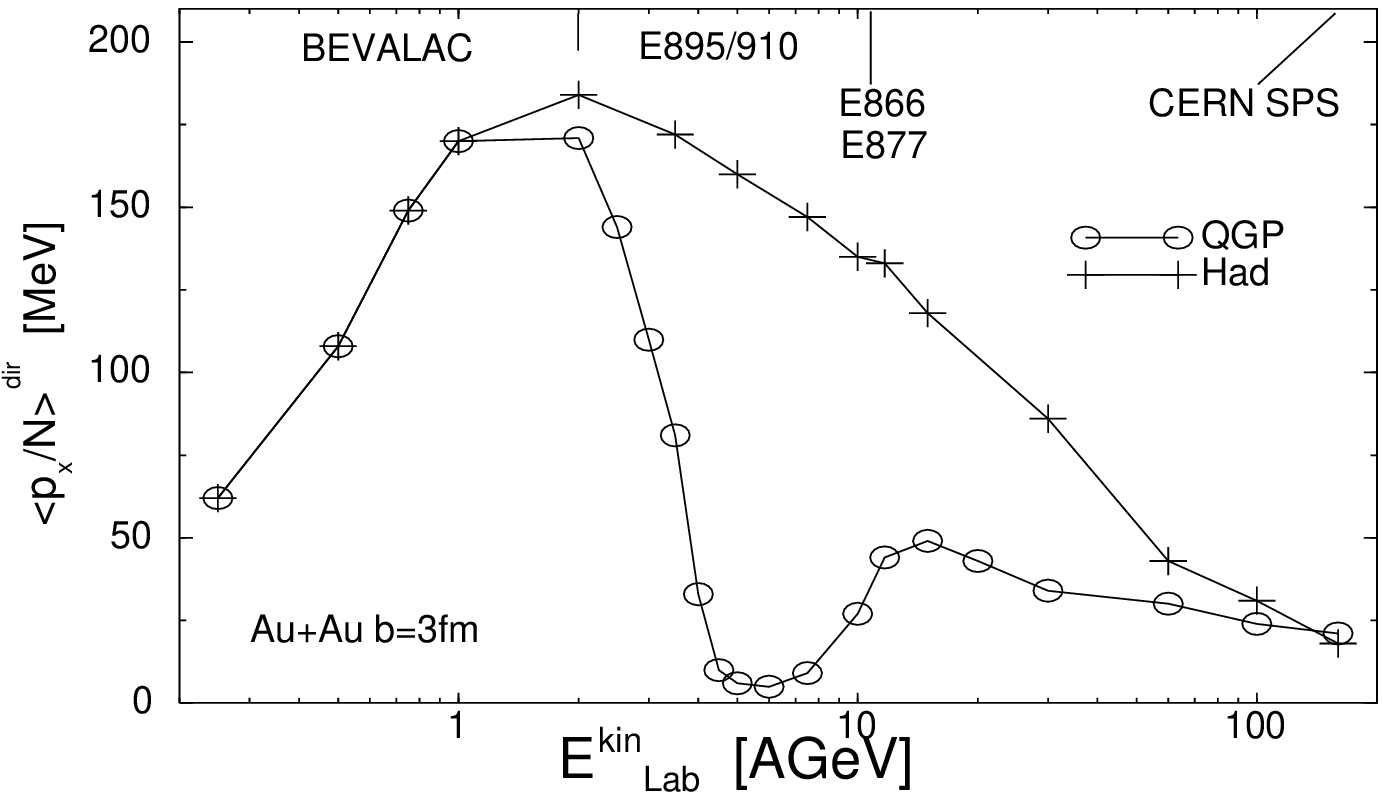,width=9cm}}
\end{minipage}
\hfill
\begin{minipage}[t]{7.5cm}
\centerline{\psfig{figure=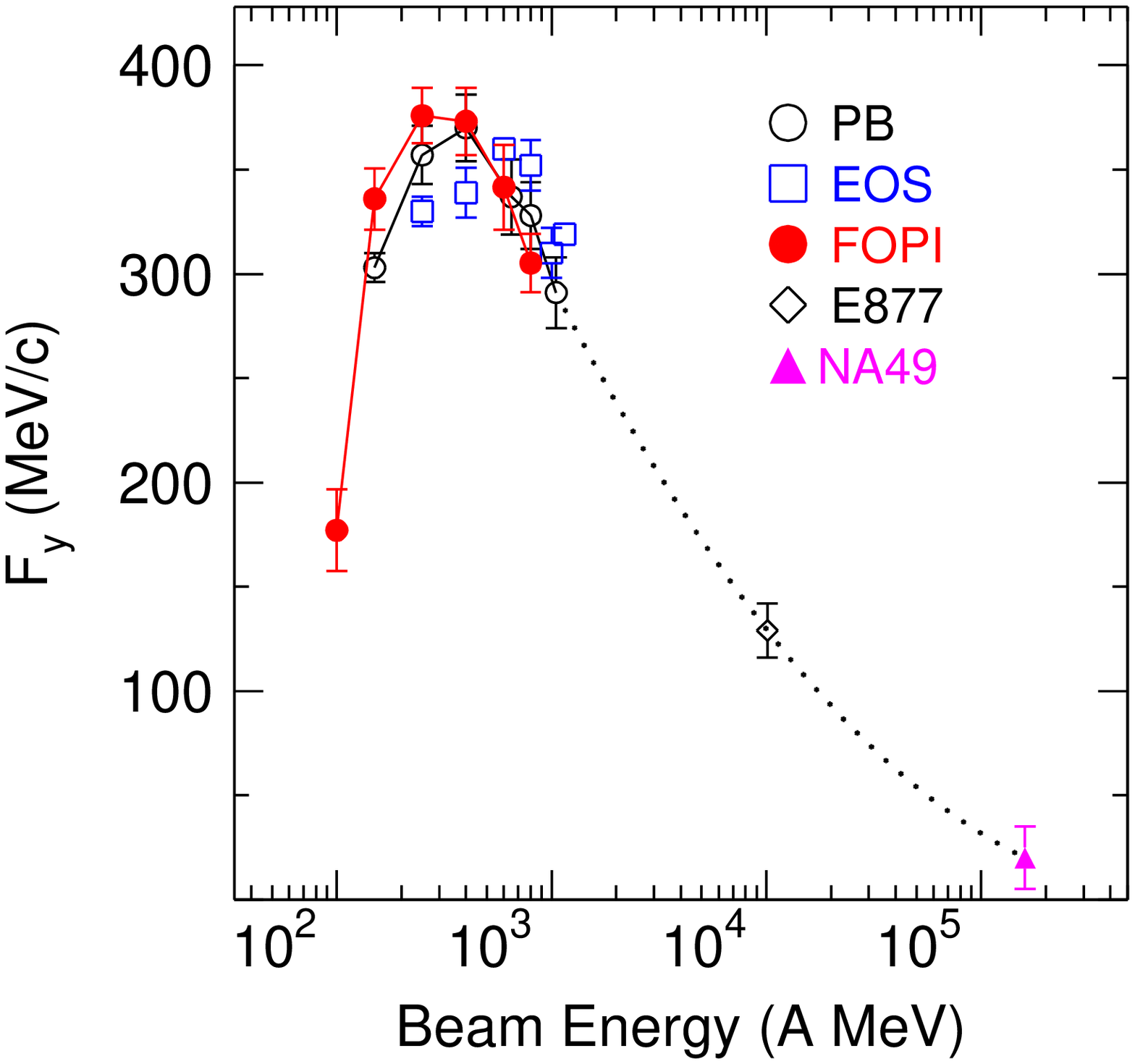,width=6cm}}
\end{minipage}

\caption{ \label{flowfig1} Excitation 
function of directed transverse flow. Left: prediction in the
framework of  nuclear hydrodynamics 
\protect \cite{rischke95b,rischke96a}, 
with and
without deconfinement phase transition. In the case of a phase 
transition a minimum in the excitation function is clearly visible. 
Right: Data compilation taken from \protect\cite{braun-munzinger98a}}
\end{figure}

Directed collective flow has been  predicted by nuclear fluid dynamics (NFD) 
\cite{scheid68a,scheid74a,stoecker80a,stoecker82a,buchwald84a}. 
Microscopic models such as VUU
(Vlasov Uehling Uhlenbeck), and QMD (Quantum Molecular Dynamics) have
predicted smaller flow than ideal NFD, these models show good 
agreement with viscous NFD \cite{schmidt93a} and
with the experimental findings 
\cite{moli84,moli85,hartnack89a,ch92}.
It has been discovered initially
at the
the BEVALAC \cite{gus84,do86,gut89} for charged particles by the Plastic-Ball
and Steamer Chamber  collaborations \cite{renfordt84a}, at SATURNE by the
DIOGENE collaboration \cite{gosset90a}
and has been studied extensively at GSI by the FOPI 
\cite{ramillien95a,herrmann96a}, LAND \cite{leifels93a}, TAPS 
\cite{kugler94a} and KaoS \cite{brill96a} collaborations.

One has to distinguish two different signatures of directed collective flow:
\begin{itemize}
\item[a)] The {\em bounce--off} \cite{stoecker80a} of compressed matter 
{\em in the reaction plane} and 
\item[b)] the 
{\em squeeze--out} \cite{stoecker82a} of the participant matter 
{\em out of the reaction plane}. 
\end{itemize}
The most strongly stopped, compressed matter
around mid-rapidity is seen directly in the {\em squeeze--out} \cite{ha90}.
A strong dependence of these collective effects  
on the nuclear equation of state
is predicted \cite{ch92}. For higher beam energies, however,
projectile
and target spectator decouple quickly from the reaction zone, giving
way to a preferential emission of matter in the reaction plane, 
even at mid-rapidity \cite{ollitrault93a}.
An excitation function of the {\em squeeze--out} at midrapidity, possibly
showing the transition from out of plane enhancement to preferential 
in-plane emission
has been predicted to enhance the sensitivity to the nuclear equation of state
\cite{danielewicz98a,bass98a}.

Apart from the above discussed {\em directed} flow, 
the so-called ``radial'', i.e.
undirected, flow component can be used for simplicity (azimuthal symmetry)
\cite{siemens79a,stoecker81b}. 
It has to be taken into account for
the  interpretation of particle spectra used for temperature extraction
which may drop by as much as a factor of 2.

Due to it's direct dependence on the EoS, $P(\rho,T)$, flow excitation
functions can provide unique information about phase transitions:
The formation of abnormal nuclear matter, e.g., yields a reduction 
of the collective flow \cite{hofmann76a}.
A directed flow excitation function as
signature of the phase transition into the QGP has been proposed 
by several authors \cite{stoecker86a,amelin91a}.
A microscopic analysis showed that the existence of
a first order phase transition 
can show up as a reduction in the directed transverse
flow \cite{ha90}.

\begin{figure}[tbh]
\begin{minipage}[t]{5.6cm}
\centerline{\psfig{figure=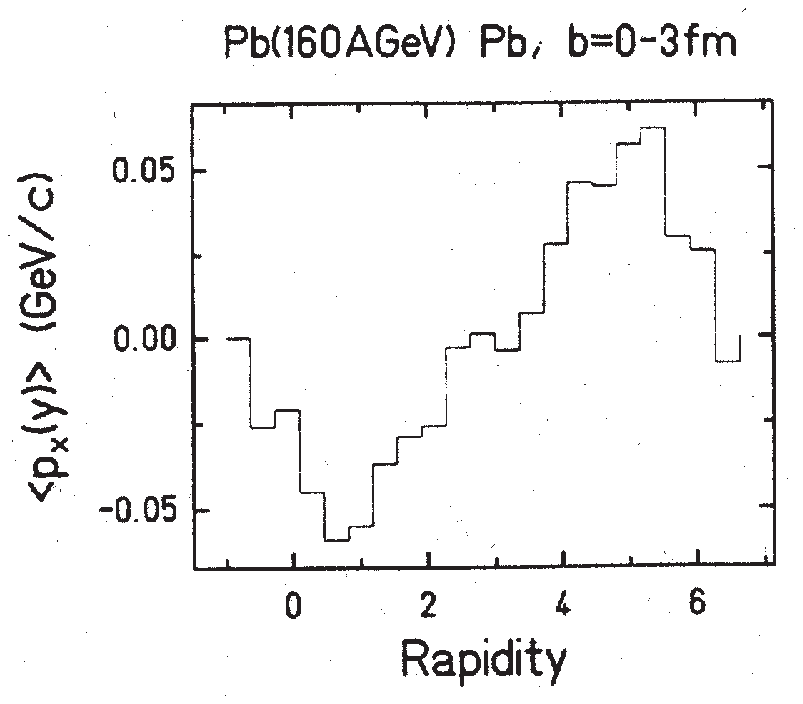,width=7cm}}
\end{minipage}
\hfill
\begin{minipage}[t]{5.6cm}
\centerline{\psfig{figure=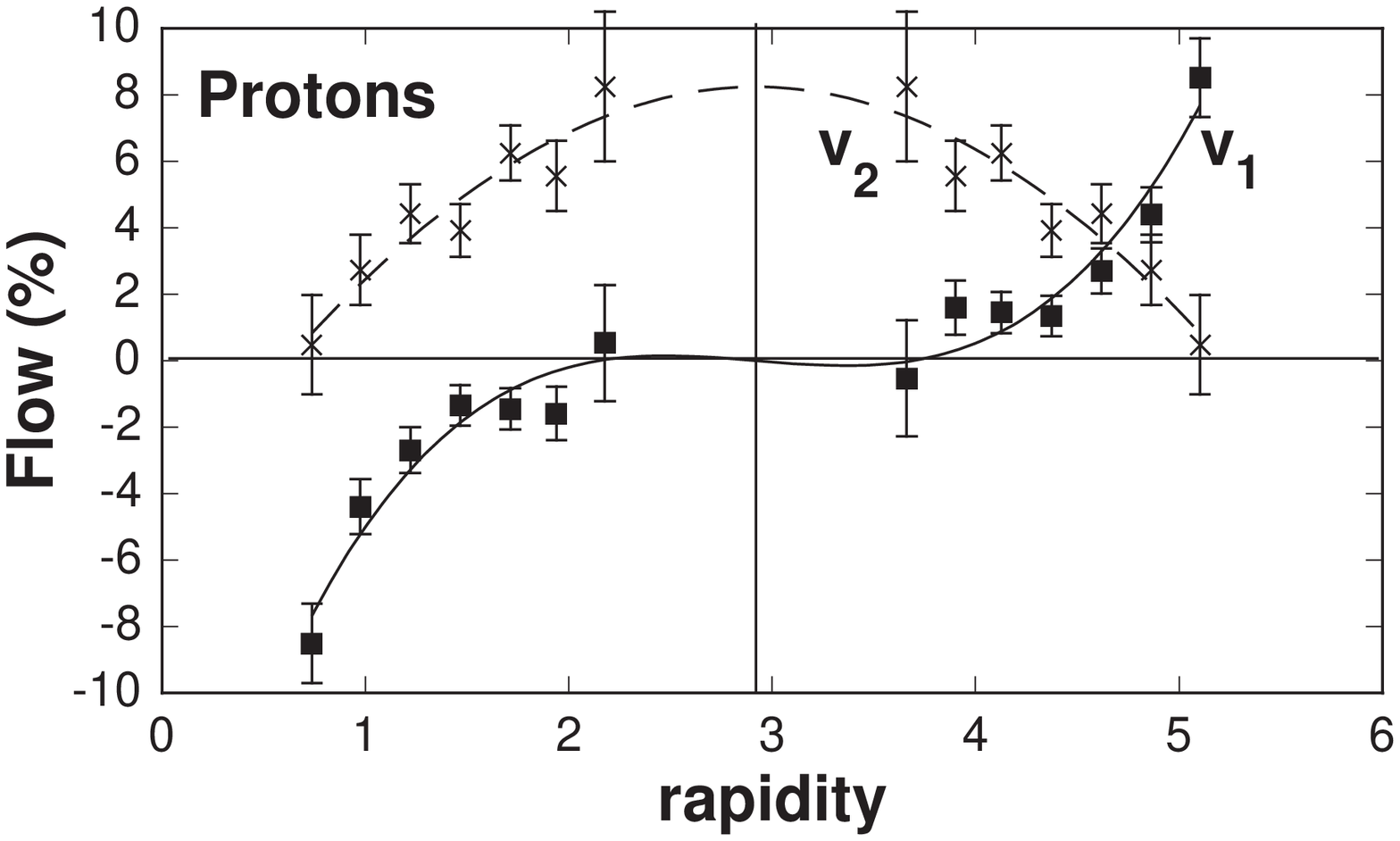,width=8cm}}
\end{minipage}
\caption{ \label{spsflow} RQMD 1.05 prediction of collective 
sideward flow for the system Pb+Pb at 160 GeV (left, figure taken from
\protect \cite{keitz91a}). The rhs shows data on directed and elliptic
flow vs. rapidity by the NA49 collaboration (figure taken from 
\protect\cite{appelshaeuser98a}).}
\end{figure}

For first order phase transitions, the pressure remains constant
in the region of the phase coexistence. This results in a vanishing 
velocity of sound $c_s= \sqrt{\partial p/\partial \varepsilon}$.

The expansion of the system is driven by the pressure gradients, therefore
expansion depends crucially on $c_s^2$. Matter in the  mixed phase
expands less rapidly than a hadron gas or a QGP at the same energy density
and entropy.
In case of rapid changes in the EoS without phase transition, the
pressure gradients are finite, but still smaller than for an ideal gas
EoS, and therefore the system expands more slowly \cite{kapusta85,gersdorf87}.

This reduction of $c_s^2$ in the transition region is commonly referred to
as {\em softening} of the EoS. The respective region of energy densities
has been called the 
{\em soft region} \cite{hung95a,rischke95a,rischke96a,rischke96b}.
Here the
flow will temporarily slow down (or possibly even stall).
Consequently a {\em time delay} is expected in the expansion
of the system.
This prevents the deflection of spectator matter
(the {\em bounce--off}) and, therefore, causes 
a reduction of the directed transverse 
flow \cite{bravina94a,rischke95b} 
in semi-peripheral collisions. 
The softening of the EoS should be observable in the excitation function of the
transverse directed flow of baryons (see figure \ref{flowfig1}).

The overall decrease of $\vec{p}_x$ seen in Fig. \ref{flowfig1} 
for $E_{lab} > 10$ GeV both for
the hadronic and the QGP equation of
state demonstrates that faster spectators are less easily
deflected (because $A$ and $t$ in equation~\ref{pxeqn} are decreasing
with $E_{lab}$) by the hot, expanding participant matter. 
For the QGP equation of state, however, these one-fluid calculations
show a {\em local minimum} in the excitation function, at about
6~GeV/nucleon. 
This can be related to  the
QGP phase transition, i.e. to the existence of the 
{\em soft region} in the EoS.

The limitation of one-fluid hydrodynamic calculations is that they
assume instantaneous
thermalization. This becomes unrealistic for increasing beam energies since
due to the average rapidity loss of only one unit per proton-proton 
collision, nucleons require several collisions for thermalization.
A more realistic three-fluid calculation, in which the third fluid
represents the {\em fireball} of produced particles 
and only local thermal equilibrium
is assumed, yields
much lower flow values -- even without a first order phase-transition
\cite{brachmann97a}. The position of the minimum (the magnitude of
the overall effect) therefore strongly depends on the degree of stopping
(i.e. which type of fluid-dynamical model is employed)
and on the details of the chosen EoS and phase transition parameters.

Taking the finite volume of the reaction zone into account, one finds
that fluctuations hinder a sharp separation between the QGP-phase
and the hadronic phase and lead to a {\em rounding} of the 
phase transition \cite{spieles97b}. For realistic reaction volumes
the softening of the equation of state is reduced considerably and
thus the minimum-signal in the flow excitation function is washed out.

A second order phase transition may not exhibit
this minimum in the flow excitation function:
The existence of a minimum in $p_{x,dir}(E_{lab})$ is rather a
{\em qualitative} signal for a strong first order transition.
If such a drop of $p_{x,dir}(E_{lab})$ is observed, it remains
to be seen which phase transition caused this behavior: a 
hadron--quark-gluon phase transition or, e.g., 
a resonance matter -- ground state matter phase transition
in confined nuclear matter \cite{boguta81a,waldhauser87a}.

\paragraph*{Experimental status\\}

\begin{figure}[thb]
\begin{minipage}[t]{7cm}
\centerline{\psfig{figure=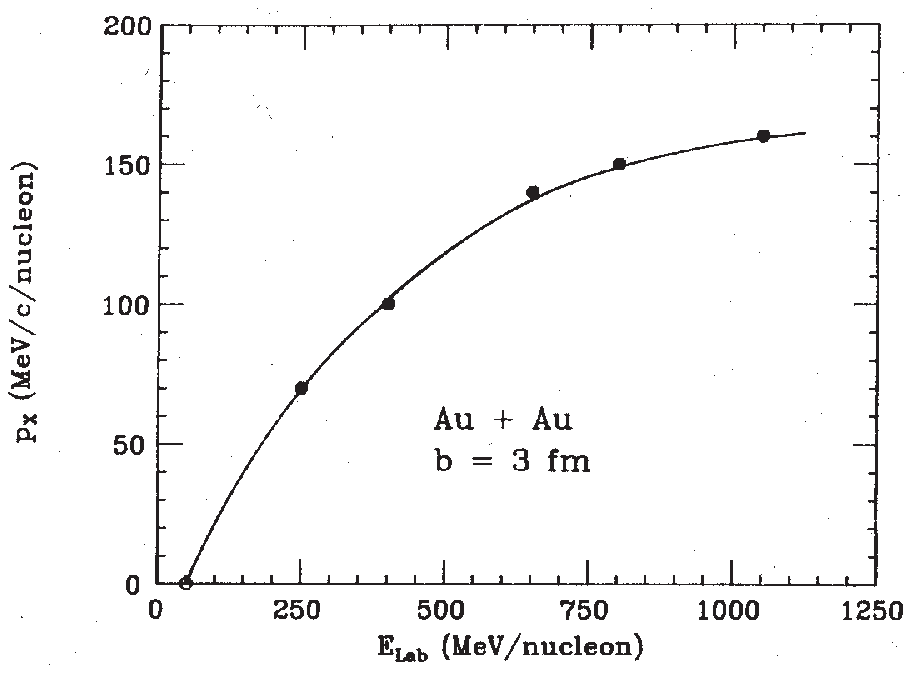,width=7.25cm}}
\end{minipage}
\hfill
\begin{minipage}[t]{6.5cm}
\centerline{\psfig{figure=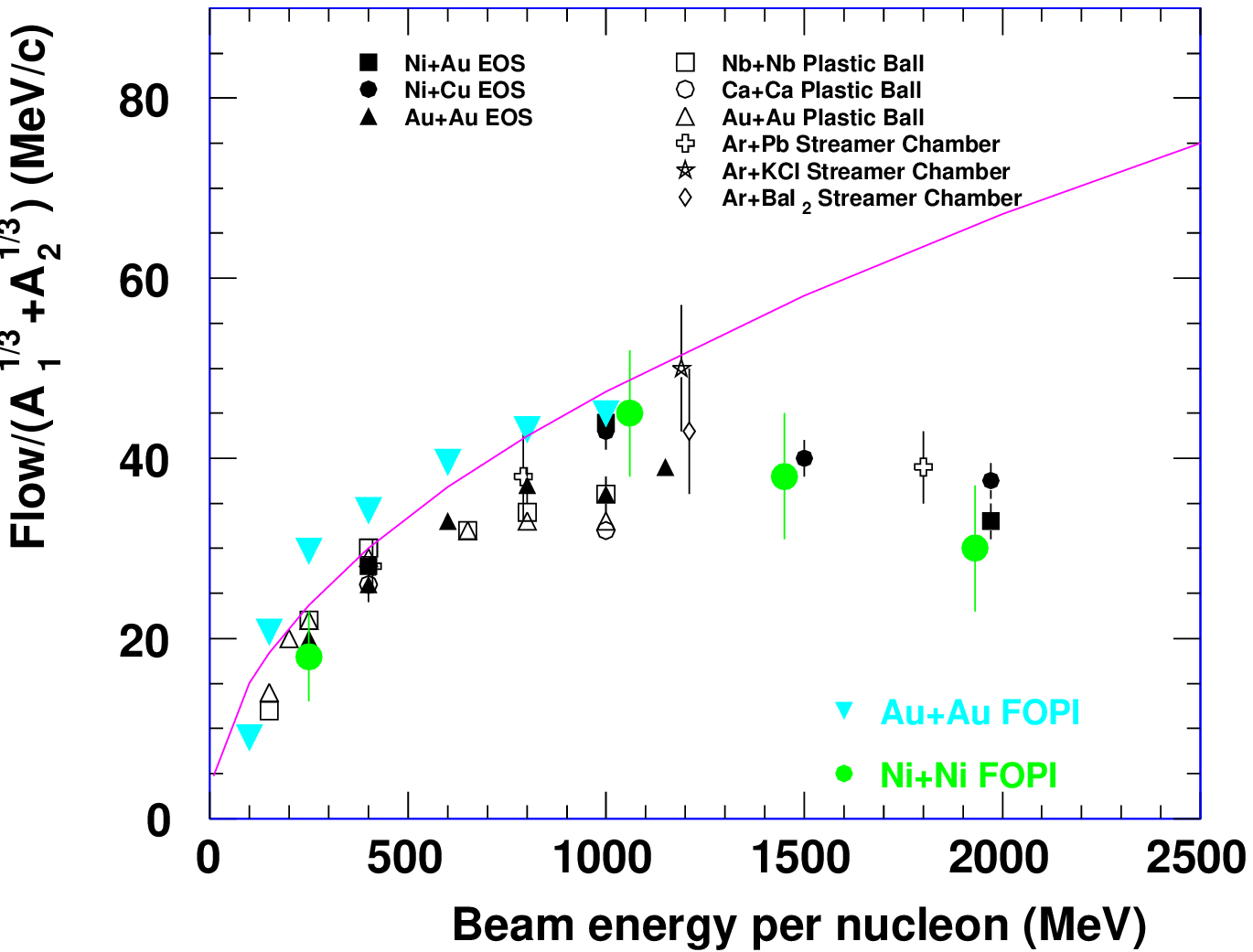,width=6.75cm}}
\end{minipage}
\caption{ \label{flowfig2} Excitation 
function of directed transverse flow for different collision systems 
from 100 MeV/nucleon up to 2 GeV/nucleon. 
Left: VUU prediction \protect\cite{molitoris85a,stoecker86a} 
for the system Au+Au. 
Right: data from the Streamer Chamber, Plastic
Ball and EOS experiments at the BEVALAC and from the FOPI experiment
at SIS \protect\cite{gus84,do86,gut89,chance96a,herrmann96a} (the
figure has been taken from \protect\cite{herrmann96a}). The data
has been scaled by $(A_1^{\frac{1}{3}} + A_2^{\frac{1}{3}})$ to account
for the different collision systems.}
\end{figure}

Collective flow measurements have first been performed at
the BEVALAC \cite{gus84,do86,gut89} for charged particles by the Plastic-Ball
and Streamer Chamber collaborations. A more detailed investigation of 
the excitation function between  0.1 to 1.2 GeV/nucleon for Au+Au 
has been performed by the FOPI, KaoS, LAND and TAPS 
collaborations at GSI \cite{herrmann96a,leifels93a,kugler94a,brill96a} 
and the EOS-TPC collaboration at LBNL \cite{chance96a}
(see figure~\ref{flowfig2}).

\begin{figure}[thb]
\centerline{\psfig{figure=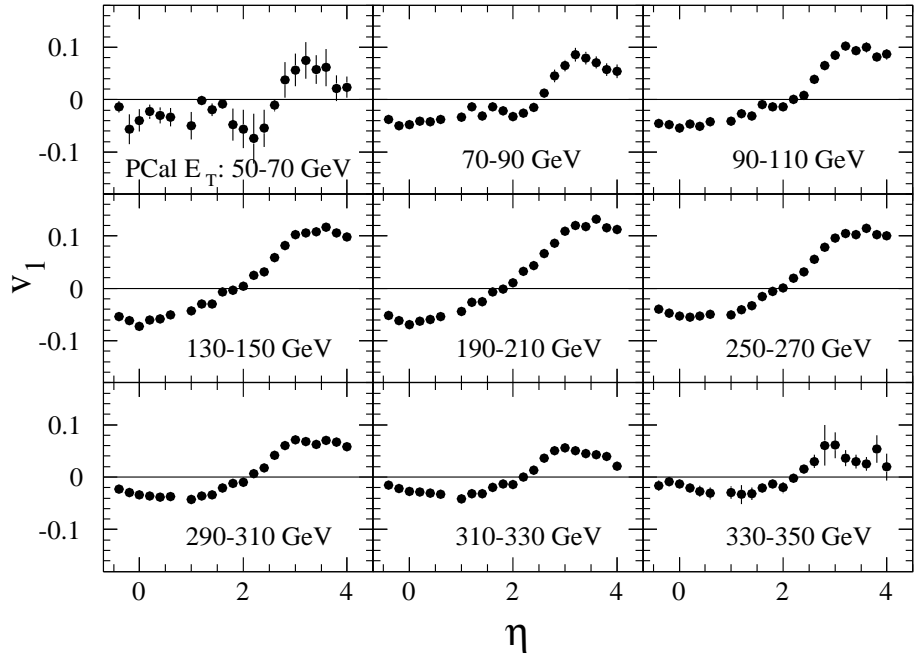,width=12cm}}
\caption{ \label{flowfig3} Transverse collective flow measured
at the AGS by the E877 collaboration \protect \cite{barrette96a}.
Plotted is d$v_1/{\rm d}\eta = {\rm d}(\langle E_x\rangle/\langle E_T\rangle)/
{\rm d}\eta$ which is a similar quantity as d$(\langle p_x\rangle/\langle p_t
\rangle){\rm d}y$, used at lower beam energies. }
\end{figure}

At 10.6 GeV collective flow has recently been discovered
by the E877 collaboration \cite{barrette94c,barrette96a}. Figure \ref{flowfig3}
shows d$v_1/{\rm d}\eta = 
{\rm d}(\langle E_x\rangle/\langle E_T\rangle)/{\rm d}\eta$ for different
centrality bins.
The E895 group has  
measured the flow excitation function for Au+Au  at the AGS in the 
energy range between 2.0 and 10.6 GeV/nucleon \cite{liu98a}. Their data
show a smooth decrease in $\langle p_x \rangle$ from 2 to 8 GeV/nucleon
and are corroborated by measurements of the E917 collaboration at
8 and 10.6 GeV/nucleon \cite{ogilvie98a}.

The E895 collaboration has also measured an elliptic flow excitation
function 
indicating a transition from out-of-plane enhancement (i.e. {\em squeeze-out})
to in-plane enhancement around
5 GeV/nucleon \cite{pinkenburg98a}.

At CERN/SPS, the existence of undirected flow has been deduced 
from a combined analysis of
particle spectra \cite{schnedermann92a,braun-munzinger96a}
and HBT correlations \cite{appelshaeuser98b}
(see also sections~\ref{spectra} and~\ref{hbtsection}).

First observations of a directed transverse flow
component have been reported by the WA98 collaboration 
\cite{peitzmann97a,nishimura98a}
using the Plastic Ball detector located at target rapidity for event plane
reconstruction. They show a strong directed flow signal for protons
and ``antiflow'' for pions, both enhanced for particles with high transverse
momenta. The same findings have been reported from the NA49 collaboration,
which due to its larger acceptance allows for an even more detailed
investigation. They report a quite strong elliptic
flow signal near mid-rapidity at 160~GeV/nucleon
\cite{appelshaeuser98a} (see figure~\ref{spsflow}).

\paragraph*{Discussion\\}

An observation of the predicted
local minimum in the excitation function of the directed transverse flow
\cite{rischke95b,rischke96a}
would be an important discovery, and an unambiguous signal for a 
strong phase transition in dense matter.
It's experimental measurement would serve as strong evidence for
a QGP and a strongly first order deconfinement transition at non-zero
baryon density.

A strong experimental effort at the AGS  and SPS has led to the
discovery of flow even at these ultra-relativistic energies.
The search for the minimum-signal in the excitation function is 
under way.

The absolute values for the ideal NFD prediction of directed flow 
shown in figure \ref{flowfig1}
overestimate the experimental values considerably 
\cite{rischke96a,rischke96b} due to lack of viscosity \cite{schmidt93a}.
The position of the minimum in $p_{x,dir}(E_{lab})$  
depends on the EoS -- therefore it
is by no means clear where (in $E_{lab}$)
the deconfinement phase transition will occur. 
Furthermore, finite volume
corrections reduce the {\em softening} of the equation of state and 
might reduce the minimum-signal considerably \cite{spieles97b}.

The combined efforts of the FOPI and EOS/E895 collaborations will allow to
map experimentally the region from 0.1 GeV/nucleon to 10 GeV/nucleon.
However, the current data show a smooth decrease in the flow
from 2 to 10 GeV. This seems to favor a hadronic scenario without
a phase transition.
An experimental search for this outset of flow in the
energy range between 10 GeV/nucleon and 200 GeV/nucleon seems necessary.

The recent measurement of the {\em squeeze-out} excitation function
between 2 and 8 GeV/nucleon may offer a new approach for studying
the nuclear equation of state \cite{pinkenburg98a,danielewicz98a,bass98a}. 

The comparison of proton spectra with $\phi$-meson spectra may help to 
disentangle ``early'', QGP-related flow components from ``late'', hadronic
contributions. Transport model calculations have shown that the 
$\phi$-meson decouples much earlier from the system ($\approx 12$ fm/c)
than the nucleons \cite{sorge97a}.
Since both particles have approximately the same mass, their ``thermal''
motion and undirected flow components should be identical and any differences
in the spectra should arise only through the additional interaction
the nucleons suffer in the later reaction stages \cite{sorge97a}.

%%%%%%%%%%%%%%%%%%%%%%%%%%%%%%%%%%%%%%%%%%%%%%%%%%%%%%%%%%%%%%%%%%%%%%%%%%%%%%%
		\subsection{Space time pictures of the reaction: HBT source radii}
	\label{hbtsection}
%%%%%%%%%%%%%%%%%%%%%%%%%%%%%%%%%%%%%%%%%%%%%%%%%%%%%%%%%%%%%%%%%%%%%%%%%%%%%%%

\paragraph*{Theoretical concepts\\}

Intensity interferometry of identical particle pairs, such as $\pi \pi, K K$ 
or $p p$
pairs, can be used to extract information about the space-time 
dynamics, freeze-out volume and reaction geometry of heavy-ion collisions.
The method was originally devised by Hanbury-Brown and Twiss
to measure the angular diameter of a star using the correlation 
of two photons \cite{HBT}: 

The probability of detecting two photons 
in coincidence in two different detectors is correlated 
to the relative separation between the two detectors.
This correlation is connected to the angular diameter of the
emitting source. This effect is commonly known as the 
{\em Hanbury-Brown-Twiss (HBT) effect}. It has been also observed
in proton-antiproton annihilations \cite{goldhaber60a}.

By applying the HBT-measurements 
to particles emitted in heavy ion reactions, such
as protons, pions or kaons, 
the two particle correlation function yields
the longitudinal and transverse radii  as well as the lifetime and flow pattern
of the emitting source at the moment of freeze out 
\cite{koonin77a,yano78a,pratt84a,pratt86a,heinz91a,heinz96a}.
The inverse
widths $R_{out}$ of the ``out'' correlation function 
and $R_{side}$ of the ``side'' correlation  function   
can be used to extract a measure for
the duration of particle emission ($R^2_{out} - R^2_{side}$)
and the transverse size of
the source ($R_{side}$) \cite{pratt86a,heinz91a}.

The prolonged life-time of the collision system in the mixed phase,
which has already been
discussed in section \ref{flow}, can be observed through 
an enhancement of the ratio of inverse widths ($R_{out}/R_{side}$) of the
two particle correlation function in out- and side-direction 
\cite{pratt86a,bertsch89a,pratt94a,rischke96b}
For energy densities estimated to be reached in Pb+Pb collisions at the CERN/SPS
one expects $R_{out}/R_{side} \sim 1.5 - 2$ \cite{rischke96b}.
Inclusion of the decays of long-lived resonances may however reduce
the $R_{out}/R_{side}$ ratio 
\cite{schlei92a,bolz93a,bolz93b,schlei96a,wiedemann97a}.

Final state interactions between non-identical particles can provide
information not only about the duration of the emission but also about
its time-ordering. It has recently been shown that an anisotropy in
the space-time distribution of emitted particles reflects 
in the directional dependence of unlike-particle correlations 
(e.g. $p-K$) and
can thus directly be used to measure the sequence of the emission
of particles of different types \cite{lednicky96a}.
Applying this technique to the correlation between a {\em strange} and an
{\em anti-strange} particle (e.g. $K^+ K^-$ interferometry) \cite{soffS97a} 
may result in the direct observation of the strangeness distillation
process \cite{greinerc87a} 
(see section~\ref{strangelets}). 
That process -- which is instrumental
to the formation of so-called {\em strangelets} -- predicts an enrichment 
of $s$ quarks in the quark phase while the $\bar s$ quarks drift into
the hadronic phase. The resulting time-ordering of the freeze-out for
{\em strange} and {\em anti-strange} particles is to be compared
to the (different) emission times due to the different mean free paths
in a purely hadronic scenario \cite{soffS97a}.

A combined analysis of single- and two-particle spectra can yield a rather
complex reconstruction of the geometry and dynamical state of the 
source at freeze-out \cite{heinz97a}. This information can be used
as a powerful test for dynamical simulations of the collision process.

\paragraph*{Experimental status\\}

$K^+K^+$ and $K^-K^-$ measurements at the SPS \cite{beker94a} show similar 
radii around 2.7 ($\pm 10$\%) fm for the system S+Pb. 
Since the $K^-$-nucleon interaction cross section is far larger than the
$K^+$-nucleon cross section, this result indicates that the dominant interaction
for kaons in the later reaction stages (close to freeze--out) at SPS energies 
are $K$-$\pi$ interactions \cite{murray94a}. 
At AGS energies, the situation might be different: the baryon to
meson multiplicity ratio is approximately one, there. A detailed analysis
has yet to be performed. 
Radii extracted from $\pi \pi$ correlations
are larger than those from $K K$, both at AGS and SPS energies
\cite{beker94a,sullivan93a}. 
The differences  are caused by different interaction cross sections 
and resonance decays \cite{gyulassy88a,schlei96b}, plus the effect of 
collective expansion \cite{heinz96a,heinz96b}. 
More theoretical work is needed to separate these
effects.

\begin{figure}[thb]
\centerline{\psfig{figure=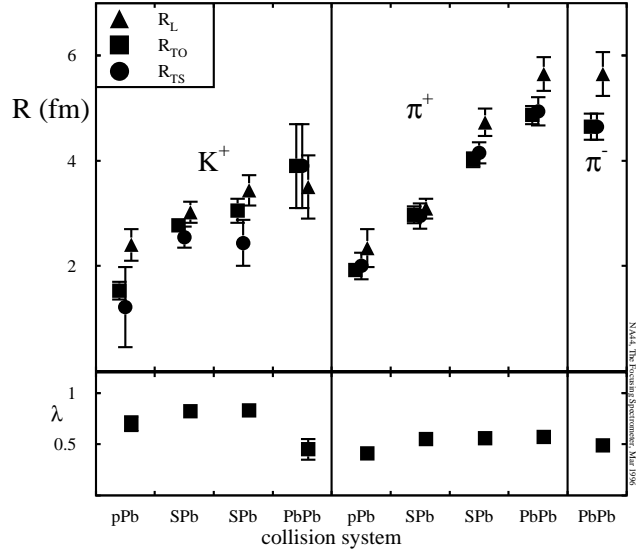,width=9cm}}
\caption{ \label{hbtfig1} Systematics of HBT radius parameters, compiled
from data by the NA44.
The figure has been taken from \protect \cite{franz96a}.}
\end{figure}

For central collisions of heavy systems the extracted transverse
radii are on the order of 5 to 7 fm for the $\pi\pi$ (both, AGS and SPS) 
and 3 fm (AGS) to 4 fm (SPS) for the $K K$ system
\cite{miskowiec96a,franz96a,kadija96a,rosselet96a,roehrich97a}.

The longitudinal and transverse radii measured at the SPS are larger
than the respective radii of both, the projectile and the target, indicating 
an expansion of the system prior to freeze--out \cite{jacak95a,heinz97a}.
Data with the sulfur beam at the SPS show that the longitudinal
radii measured as a function of rapidity \cite{alber95b} could be fitted
by a boost invariant longitudinal expansion \cite{makhlin88a}. Recent
data taken with the lead beam by the NA49 collaboration confirm this
finding for the Pb+Pb 
system \cite{kadija96a,appelshaeuser96a,roehrich97a,appelshaeuser98b}. 

Transverse source radii for $\pi \pi$  show a decrease from 4 to 2.5 fm 
for S+S and from 6 to 4 fm for Pb+Pb, respectively, 
with increasing
transverse momentum of the pions (see also Figure \ref{hbtfig1})
\cite{alber95b,kadija96a,rosselet96a,roehrich97a};
this behavior is to be expected in the presence of transverse flow
\cite{pratt84a,pratt86a,heinz91a}. 
Alternatively, it can also be explained by microscopic models which
predict that high $p_t$ particles are
emitted in the early reaction stages (by heavy resonances or strings) 
and low $p_t$ particles (which have rescattered more often) have
late freeze-out times \cite{bass94a,hofmann95a}.
Measurements of $R_{out}/R_{side}$ indicate values on the 
order of 1 \cite{jacak95a}.
For the Pb+Pb system, a duration of emission of about 3-4 fm/c has
been reported with the lifetime of the source being $\tau \approx 8$ fm/c
\cite{roehrich97a}.

Source radii cannot only be studied as a function of transverse momentum 
or beam energy,
but also as a function of impact parameter related quantities, such as
the number of participant nucleons.
The latter analysis  can be performed  either by comparing central events 
of different systems or by comparing
different centralities in very heavy systems. 

Exciting
preliminary results have been reported at QM '96 
for the system Au+Au at 10.6 GeV/nucleon by the E866 collaboration
\cite{baker96a}, 
showing a dramatic increase of 40\% in the source radius over 
the last 7\% of highest centrality.
More recent data-sets of the same collaboration
suggest a more gradual increase with centrality \cite{ogilvie96a}. 
The limited statistics, however, do not permit a final assessment, yet.

\paragraph*{Discussion\\}

At AGS energies hadronic transport models are well able to reproduce
the measured source radii \cite{sullivan93a,miskowiec96a} -- at SPS
energies a full analysis has not yet been performed, but early comparisons
showed at least qualitative agreement \cite{sullivan93a,beker94a}.

A strong first order phase transition \cite{pratt86a,bertsch89a}
and even an infinite order but rapid cross over transition \cite{rischke96b}
should result in a lower pressure,
slower expansion and perhaps a long-lived evaporating droplet of QGP.
The rather short lifetime of $\tau \approx 8$ fm/c
\cite{roehrich97a} for Pb+Pb at CERN/SPS suggests either the non-existence
of such a low-pressure system or perhaps that the initial energy-density
that is needed to create a QGP is much higher
\cite{pratt86a,rischke96b}.

HBT-interferometry shows thusfar no evidence for the characteristic 
time delay of QGP formation up to SPS energies.
The main complication of HBT analysis in nuclear collisions 
is the existence of strong collective flow that (Doppler) distorts
the interference pattern.
The $p_t$ dependence of the HBT radii has become
a useful tool to   probe this aspect of the reaction dynamics.
It will be important to search for time signatures at RHIC and LHC.

%%%%%%%%%%%%%%%%%%%%%%%%%%%%%%%%%%%%%%%%%%%%%%%%%%%%%%%%%%%%%%%%%%%%%%%%%%%%%%%
		\subsection{Remnants of hadronization: strangeness enhancement}
%%%%%%%%%%%%%%%%%%%%%%%%%%%%%%%%%%%%%%%%%%%%%%%%%%%%%%%%%%%%%%%%%%%%%%%%%%%%%%%

\paragraph*{Theoretical concepts\\}

In proton proton collisions, 
the production of particles containing strange quarks
is strongly suppressed  as compared to the 
production of particles with $u$ and $d$ quarks \cite{bailly87a,malhotra83a}. 
It has been argued that this suppression is 
due to the higher mass of the $s \bar{s}$ quark pair.
The suppression increases with the strangeness content of the particles 
produced in proton proton collisions. 

In the case of QGP formation, 
$s \bar{s}$ pairs can either be produced via the interactions of two gluons
or of $q \bar{q}$ pairs.
Leading order $\alpha_s$ pQCD calculations suggest
that the second process dominates only for 
$\sqrt{s} \le 0.6$ GeV \cite{glueck78a}.
The time-scale of chemical equilibration of (anti-) strangeness due to
gluon gluon interaction is estimated -- also based on first
order pQCD calculations -- to be about 3 to 6 fm/c,
depending on the temperature of the plasma \cite{rafelski82a}.

Following this line of argument, 
the yield of strange and multi-strange
mesons and (anti-) baryons has been 
predicted to be strongly enhanced in the presence 
of a QGP as compared to a purely hadronic scenario at the same temperature
\cite{rafelski82b,kochp86a}.
However, the estimated equilibration times 
may not be sufficiently rapid to cause
a saturation in the production of strange hadrons 
before QGP freeze-out.

In particular, assuming low chemical potentials, 
$\mu_d \approx \mu_u \approx 0 = \mu_s$ and a
temperature $T$ higher than the strange quark mass $m_s$, the densities
of all quarks and anti-quarks are nearly the same in the QGP. 
Hence, the probability of forming anti-hyperons by 
combining $\bar{u}, \bar{d}$ and $\bar{s}$ quarks is nearly the same as the
probability of forming strange and non-strange baryons by combining $u, d$ and
$s$ quarks if the freeze out process is rapid and annihilation can
be neglected.

In contrast, the production of an antihyperon-hyperon pair 
produced in nucleon nucleon collisions is greatly suppressed 
by the Schwinger factor \cite{schwinger62a,casher74a} since it is
necessary to tunnel the massive diquark and the strange quark 
through the potential wall in the
chromo-electric field with the string tension 
$\kappa \approx 1$~GeV/fm \cite{wong94a}.
The enhanced production of anti-hyperons
($\bar{\Lambda}, \bar{\Sigma}, \bar{\Xi}$ and $\bar{\Omega}$)
can therefore be used as a QGP signal in the case of zero chemical 
potential \cite{rafelski82a}. 
 
If a QGP is created in heavy ion collisions at AGS or SPS energies,
it will most likely be characterized by nonzero
chemical potentials $\mu_u$ and $\mu_d$. This results in the densities of
$u$ and $d$ quarks being larger than those of the $s$ and $\bar{s}$ quarks,
which in turn are larger than the $\bar{u}$ and $\bar{d}$ densities.
Due to these different abundances the $\bar{s}$ quark is more likely to combine
with a $u$ or $d$ quark to form a $K^+$ or $K^0$ 
(or with two non-strange quarks to form  a $\Lambda$ or $\Sigma$, respectively)
than it is for the $s$
quark to recombine with a $\bar{u}$ or $\bar{d}$ quark thus forming
a $\bar{K^0}$s or $\bar{K^-}$s.
Therefore, in the QGP case 
the $K^+/\pi^+$ ratio in a relativistic heavy ion collision is
different from the $K^-/\pi^-$ ratio \cite{kochp90a}.

The relative abundances of various strange particle species have been
used for the 
determination of relative strangeness equilibration.
To account for incomplete chemical equilibration, a strangeness
fugacity $\gamma_s$ is introduced in a thermo-chemical
approach \cite{rafelski91a,harris96a,heinz94a,sollfrank95a}.
One has also compared the measured ratios and the connected
thermodynamic variables (such as $T, \mu_B$ and the entropy)  
with calculations, either assuming a hadron gas scenario or a 
QGP scenario including some hadronization scheme 
\cite{letessier92a,letessier94a,letessier95a,braun-munzinger95a}.  

There are certain drawbacks to the line of argument presented above:
The strange particle abundances, after freeze out from a QGP, are very close
to those of a fully equilibrated hadron gas at the same entropy 
content \cite{leeks88a}. The reason is \cite{kochp90a,redlich85a,mclerran87a} 
that the volume
of a hadron gas of the same total energy has to be larger
due to the smaller number of available degrees of freedom. Consequently,
one must expect that the abundance of strange quarks is diluted
during the hadronization process. This dilution effect is clearly seen
in hadronization models \cite{kochp86a,barz88a},
where gluons hadronize by conversion
into quark-anti-quark pairs, which predominantly feed the final pion channel.
As a consequence, the $K/\pi$ ratios are significantly reduced.

Furthermore, the computation of particle abundances in the QGP and the 
hadron gas scenario are mostly  based on the assumption 
of chemical and thermal equilibrium (a non-equilibrium calculation
has been published in \cite{barz88a}).
For the hadronic case these assumptions cannot be justified: It
has been shown via rate equations \cite{kochp86a,kochp90a} 
that the strangeness equilibration 
time exceeds the reaction time of a heavy ion collision by at least one
order of magnitude. 

Strangeness production in the hadronic scenario is a
non-equilibrium process. In the early (pre-equilibrium) reaction stages,
typical longitudinal momenta are much higher than in the case
of a thermal momentum distribution.
This leads to enhanced strangeness production \cite{mattiello89a,sorge97a}. 
The system then cools 
down in the course of the reaction. It's final ``equilibrium'' temperature
is therefore only partly connected to the measured 
strange particle yields and spectra.

\paragraph*{Experimental status\\}

An enhancement of the $K/\pi$ ratio has been measured both at the AGS
and at the SPS \cite{kpi_data}. 
At the AGS, $K^+/\pi^+ \approx 0.2$
and $K^-/\pi^- \approx 0.04$.
Furthermore  $K^+/K^-, \bar{\Lambda}/\Lambda$ and $\bar{p}/p$ production
ratios have been measured at the AGS \cite{abbott94a}.
At the SPS,
enhanced production of (anti)hyperons, such as $\bar{\Lambda},
\bar{\Xi}, \Omega$ and $\bar{\Omega}$, has been observed
and ratios of $\bar{\Lambda}/\Lambda, \bar{\Xi}/\Xi, \Xi/\Lambda$
and $\bar{\Xi}/\bar{\Lambda}$ 
have been analyzed by the NA36,WA85 and WA97 collaborations
\cite{anderson92a,anderson94a,abatzis94a,abatzis93a,dibari95a,kinson95a,
helstrup96a,kralik98a}.
The WA94 collaboration has measured antihyperon ratios (i.e. the
$\bar \Xi/\bar\Lambda$ ratio) in pp, pA and AA reactions. They find a
smooth increase in the $\bar \Xi/\bar\Lambda$ value from pp over pA 
to AA \cite{abatzis97a}. 

Recently, very interesting values have been quoted for the 
$\bar{\Lambda}/\bar{p}$
ratio. It has been measured by the NA35, NA49, E866, E878 and
E864 collaborations \cite{alber96a,jones96a,akiba96a,armstrong97a}.
Since it only contains newly produced anti-quarks, it may therefore
represent a rather 
clean measure for the $\bar s/\bar u$ quark ratio in the hot and
dense matter. For pp and pA collisions this ratio is below 0.4, whereas
in AA collisions preliminary analysis give values 
between 3 and 5 \cite{armstrong97a,bormann97a} -- these values are
so high that they could not be obtained in either a hadron gas or QGP 
model with reasonable values for $T, \mu_B$ and $\mu_S$.

The observed strong enhancement of multistrange (anti)hyperons
($\Xi$, $\Omega$, $\bar \Xi$, and $\bar \Omega$) from light to 
heavy collision systems at the CERN/SPS 
\cite{abatzis97a,kralik98a,andersen98a} 
surely constitutes on the experimental side the most intriguing evidence
for a possible non-hadronic enhancement of strangeness.

\paragraph*{Discussion\\}

Hadronic models for particle production
\cite{matsui86b,leeks88a,mattiello89a,brown91a,pang92a} 
work quite well in the case of the observed
$K^+/\pi^+$ enhancement \cite{kpi_data} at the AGS (silicon beam).  
The  reason for strangeness
enhancement in a hadronic scenario is multistep excitation of heavy baryon and
meson resonance states \cite{mattiello89a}.
The AGS value of $K^+/\pi^+\approx 0.2$ is compatible with a 
strangeness equilibrated
hadron gas \cite{braun-munzinger95a}.

AGS data
of $K^+/K^-, \bar{\Lambda}/\Lambda$ and $\bar{p}/p$ ratios
can be fitted with
an equilibrated hadronic fireball with $\mu_s/T = 0.54 \pm 0.11$ and
$\mu_B/T = 3.9 \pm 0.3$ \cite{letessier94a,rafelski94a,braun-munzinger95a}.
However, this does not mean that the system has
always been in the hadronic phase, since an equilibrium state has no
memory on how it has been produced. The system might as well have originated in
the quark phase and then evolve along the phase-boundary, thereby hadronizing
with varying combinations of $T, \mu_B$ and $\mu_S$.
The point is that these ratios provide actually very little information
about the properties of the early time dense system.

The ratios $\bar{\Lambda}/\Lambda,\quad \bar{\Xi}/\Xi,\quad \Xi/\Lambda$
and $\bar{\Xi}/\bar{\Lambda}$ measured by the WA85 and WA97 collaborations 
\cite{abatzis94a,abatzis93a,dibari95a} at the SPS
can be fit in analogy to the ratios at the AGS
by an equilibrium hadron gas model with 
$\gamma_s=0.7$, $\mu_B =0.24$ and $T=180$ MeV \cite{cleymans93a}.
Besides the three parameters $T, \mu_B$ and $\mu_S$ which are used in the  
grand-canonical formalism of statistical mechanics, the additional 
parameter $\gamma_s$ accounts for incomplete 
saturation of strange particles in phase space. However,
data can also be fitted with a hadron
gas model and $\gamma_s \approx 1$ with
$\mu_s/T = 0.24-0.28$ and $\mu_B/T = 1.05$ \cite{braun-munzinger96a},
respectively.
The very same data can also be fit by an instantaneously hadronizing
non-equilibrated QGP with strangeness neutrality and
strangeness saturation $\gamma_s \ge 0.7$ 
\cite{letessier92a,letessier93a,sollfrank94a}.

\begin{figure}[thb]
\begin{minipage}[t]{7cm}
\centerline{\psfig{figure=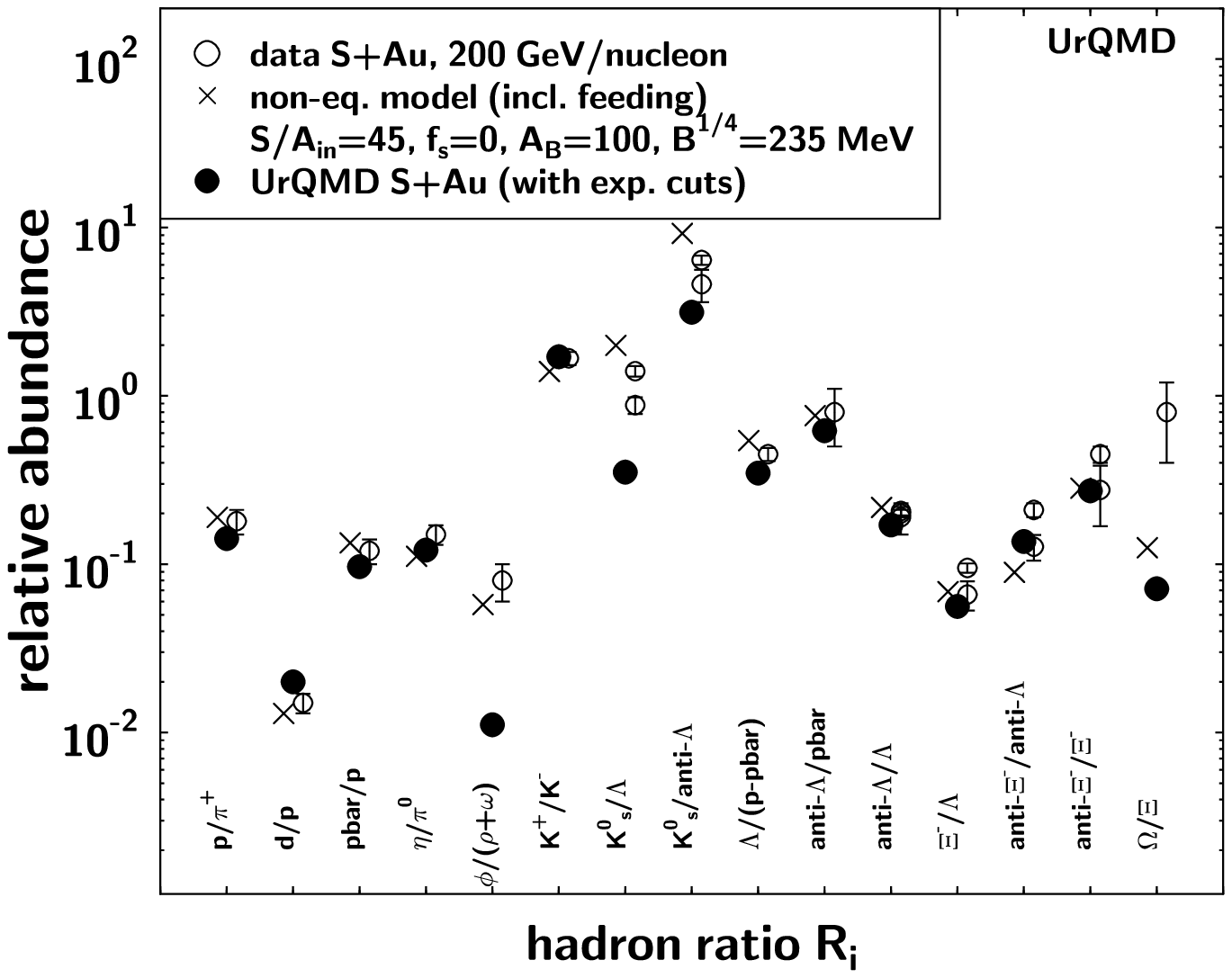,width=7.5cm}}
\end{minipage}
\hfill
\begin{minipage}[t]{6.5cm}
\centerline{\psfig{figure=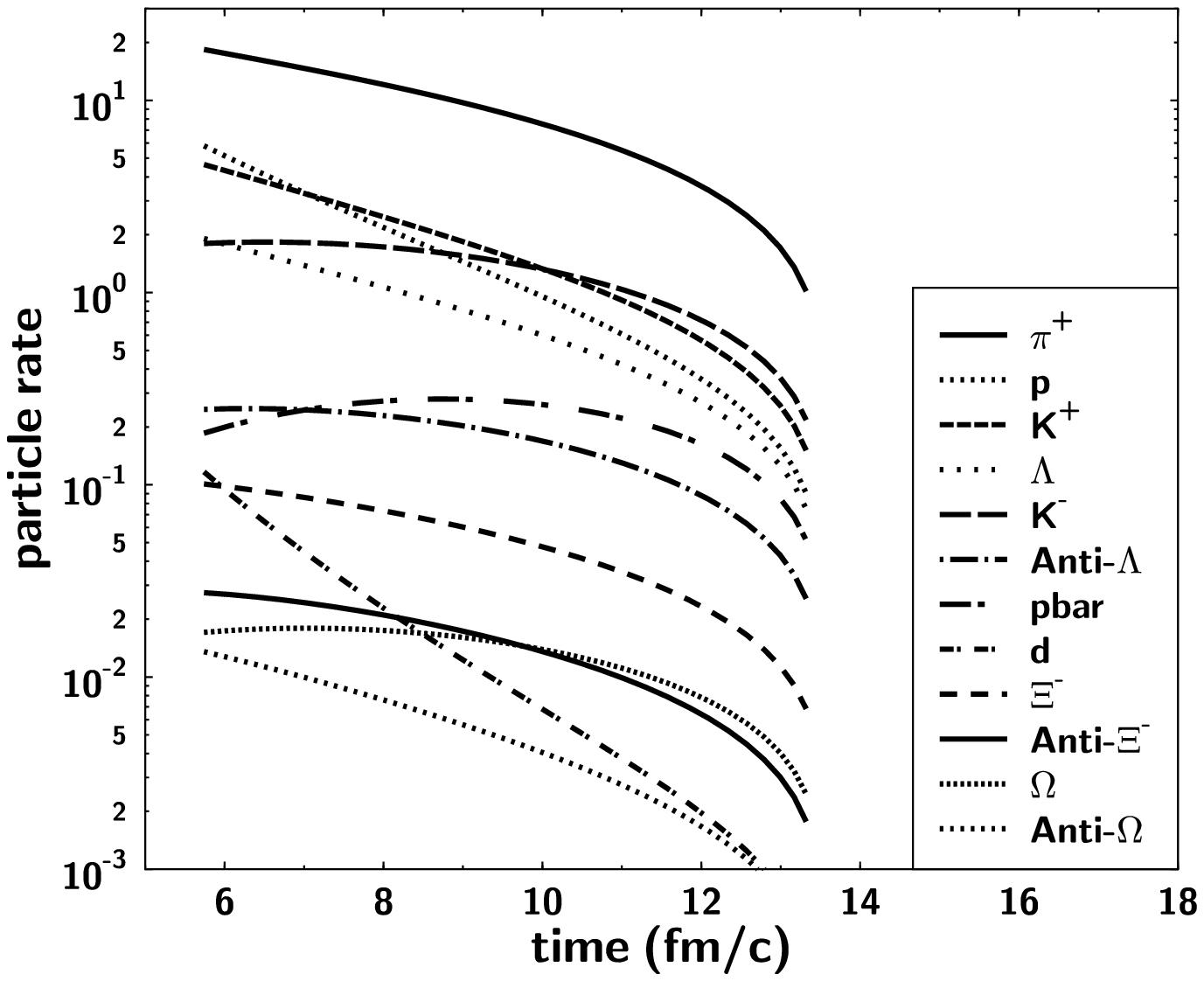,width=7.5cm}}
\end{minipage}
\caption{ \label{ratios-spieles} Left:
final particle ratios computed in UrQMD (full circles) \protect\cite{bass97d} 
and a non-equilibrium hadronization scenario (crosses) 
with initial conditions $A_{\rm B}^{\rm init}=100$, $S/A^{\rm init}=45$, 
$f_s^{\rm init}=0$ and bag constant $B^{1/4}=235$~MeV
\protect\cite{spieles97a}. The data (open circles) 
are taken from various experiments 
as compiled in \protect\cite{braun-munzinger96a}.
Right: corresponding particle production rates as a function of time.
Strong differences in the time-evolution of various particle ratios
are observed.}
\end{figure}

Are the extracted temperatures and chemical potentials really reliable
in view of the simple, static, thermal ansatz? 
Even hadron production in high energy $pp$ and $p{\bar p}$ collisions 
has been calculated by assuming 
thermal and chemical equilibrium and fits the data well
\cite{becattini97a}.  
The fit temperature lies around 130 MeV -- 170 MeV, nearly independent
of the center of mass energy of the incident particles.
A $\gamma_s$ value of $\approx 0.5$ is needed for the fit, indicating
incomplete strangeness saturation already at the pp level.
Does such a model make sense?
The success of the fit can be interpreted as 
hadron production in elementary 
high energy collisions being dominated by phase space rather 
than by microscopic dynamics.

The extrapolation of this conclusion to heavy-ion collisions, however, may
not be valid: Even simple dynamical hadronization schemes \cite{greinerC92a}, 
where thermodynamic 
equilibrium between a quark blob and the hadron layer is imposed, reveal a
more complex picture (see figure~\ref{ratios-spieles}). Particle ratios
can be reproduced nicely with the same number of parameters as in the static
ansatz of a hadron gas in equilibrium, while the space-time evolution 
of the system shows strong changes of
the strange and baryo-chemical potentials due to baryon- and
strangeness-distillery \cite{spieles96a,spieles97a}.
Taking (boost-invariant) longitudinal hydrodynamical expansion into account,
the interplay of the evaporation process and the hydrodynamical expansion
(and vice versa) leads to considerably shorter lifetimes of the 
mixed phase as compared to scenarios without hydrodynamical expansion
\cite{dumitru97b}.
It is very questionable, whether final
particle yields reflect the actual thermodynamic properties of the system at
any one stage of the evolution.

Microscopic transport model calculations are in good agreement with
the measured hadron ratios of the system S+Au at CERN/SPS 
\cite{bass97d,sollfrank98a}. 
They show,
however, that those ratios exhibit a strong 
rapidity dependence. Thus, thermal model fits to data may be distorted
due to varying experimental acceptances for individual ratios.
A thermal model fit to S+Au ratios calculated with the microscopic 
UrQMD transport model (and
extracted within the same range of rapidity for all ratios) 
yields a temperature
of T=145 MeV and a chemical potential of $\mu_B=165$ MeV \cite{bass97d}. 
Hadron ratios for the system Pb+Pb are 
predicted and can be fitted by a thermal model with T=140 MeV and
$\mu_B=210$ MeV. Similar results have been obtained with 
RQMD \cite{sollfrank98a}.
Analyzing the results of a 
non-equilibrium transport model calculations in the framework 
of an equilibrium model may, however, not seem meaningful.

The smooth increase of the  
$\bar \Xi/\bar\Lambda$ ratio from pp via pA to AA reactions 
suggests that production volume
and the degree of thermalization may not be relevant for the production
of antihyperons. Already two overlapping strings (as typically produced
in p+S reactions) are sufficient to yield strong deviations from the 
behavior in pp \cite{sorge97a,miklos208a}. 
Clearly the detailed study of pA reactions
yields important information on the production processes of antihyperons.

However, so far all models fail to describe the recently reported,
unusually high $\bar \Lambda / \bar p$ ratio of $\approx 3-5$ 
(proton-proton collisions yield a ratio of 0.2 - 0.3). One possible
explanation could be that the $\bar \Lambda$ have a far lower annihilation
cross section than the $\bar p$. 
%Assuming that $\bar p$ and $\bar \Lambda$
%are both produced according to a microscopic model, 
This difference in the
annihilation cross section might account for the dramatic $\bar \Lambda$ 
enhancement. A straightforward way to test this hypothesis would be the
measurement of $\bar p$ and $\bar \Lambda$ (anti-)flow. For $\bar p$'s, 
a strong anti-correlation with regard to the ``conventional'' baryon flow
is predicted \cite{jahns94a}. 
This is due to their large annihilation cross section in dense matter.
The same 
would only hold true for the $\bar \Lambda$, if its annihilation cross 
section is correspondingly large. If, however, $\bar \Lambda$ anti-flow 
is not observed,
this would serve as clear indication for a low $\bar \Lambda$ annihilation
cross section. Thus the above explanation for the 
$\bar \Lambda$ enhancement as being due to smaller $\sigma_{ann}$ 
\cite{stoecker97a} would be supported by independent evidence.

Alternatively, the  $\bar \Lambda / \bar p$ enhancement could be 
explained by different medium modifications to the masses of 
non-strange and strange baryons which affect the production 
probabilities. If an attractive
strange scalar condensate lowers the mass of the $\bar \Lambda$ in
hot and dense hadronic matter, even below that of the $\bar p$, 
this could account for the 
$\bar \Lambda / \bar p$ enhancement \cite{stoecker97a,zschiesche97a}.

Thermal model analyses assume constant freeze-out temperatures and
chemical potentials, but at least the more careful ones do not
assume a static source; instead they allow for collective expansion
flow. While the flow does not matter for an analysis of $4\pi$ yields,
it becomes indeed important when comparing the model to data
from limited windows in momentum space. 
Unfortunately, no conclusion is possible unless the freeze-out surface is known.
Most people use too simplistic isochronous ($t=$const.) freeze-out
prescriptions which in fact correspond to a volume freeze-out.
However, it has been shown in the framework of an expanding hadron-gas
that freeze-out is not a state but a reaction stage and that the 
various equilibria (i.e. chemical and thermal equilibrium) necessarily
break down in the final stages \cite{bebie92a}.
Microscopic model calculations
support this picture of a complicated sequential freeze-out depending
on reaction rates and particle species
\cite{sorge95a,weber97a,sorge97a}:
Even if some particles are in thermal and chemical equilibrium 
during the final stages of the
reaction, the problem of how to disentangle the thermal contribution
from the early pre-equilibrium emission would remain. This problem has not
been addressed satisfactorily so far (see section \ref{spectra}).

Hadronic transport models, which are based on a non-equilibrium
scenario, however, are only able to describe the CERN/SPS (anti-)hyperon data
by invoking non-hadronic scenarios such as color-ropes \cite{sorge92a},
breaking of multiple-strings \cite{werner93b} or decay of multi-quark 
droplets \cite{aichelin93a}. Therefore, the (anti-) (strange-) baryon
sector remains a topic of great interest.
Specifically the strong enhancement of multistrange (anti)hyperons
($\Xi$, $\Omega$, $\bar \Xi$, and $\bar \Omega$) heavy collision systems
such as Pb+Pb at the CERN/SPS is of great importance \cite{kralik98a} since
it offers currently the best opportunity to discriminate hadronic from
deconfinement scenarios in the sector of strangeness enhancement.

%%%%%%%%%%%%%%%%%%%%%%%%%%%%%%%%%%%%%%%%%%%%%%%%%%%%%%%%%%%%%%%%%%%%%%%%%%%%%%%
		\subsection{Ashes of the plasma: strangelets and hypermatter}
	\label{strangelets}
%%%%%%%%%%%%%%%%%%%%%%%%%%%%%%%%%%%%%%%%%%%%%%%%%%%%%%%%%%%%%%%%%%%%%%%%%%%%%%%

\paragraph*{Theoretical concepts\\}

The observed abundant production of strange baryons at AGS and SPS energies
led people to speculate about implications for hypermatter
(multi-hyperon clusters or strange quark droplets )
formation \cite{ivanenko65a,bodmer71a,chin79a,farhi84a,witten84a}. 
Speculations about the existence of such multi-strange
objects, with baryon numbers $B > 100$, have been around for decades,
in particular within astrophysics.
Such states are allowed for
by the standard model, although so far their existence
has not been proven in nature, e.g. in the form of strange
neutron stars. 

{\em Quark matter} systems with $A>1$ are unstable,
if they only consist of $u$ and $d$ quarks, due to the large Fermi energy of 
these non-strange quarks. The system's energy may be lowered by
converting some of the $u$ and $d$ quarks into
$s$ quarks (i.e. introducing a new degree of freedom).
The energy gain may
over-compensate the high mass of the $s$ quarks -- thus such strange
quark matter (SQM) may be
absolutely stable \cite{farhi84a}.

Hadrons with $B > 1$ and $S < 0$ have been considered even before 
the advent of QCD \cite{ivanenko65a,bodmer71a}.
However, first the development of the MIT Bag Model \cite{chodos74a} 
allowed to model such states. 
Long hypermatter lifetimes (for hundreds of
quarks and a  strangeness per baryon ratio in the order of 1) 
have been predicted,  up to $10^{-4}$ seconds \cite{chin79a}. 
Further detailed investigations of small pieces of strange quark matter, 
so called {\em strangelets}, reveal possible (meta)stability 
for $B > 6$ \cite{farhi84a,witten84a}.  
The simplest
{\em strangelet} is the $H-$dibaryon with zero charge, $B=2$ and $S=-2$, which
consists of $2u, 2d$ and $2s$ quarks, followed by the 
{\em strange quark--}$\alpha$
with $6u, 6d$ and $6s$ quarks \cite{farhi84a,michel88a}.

For a QGP -- hadron fluid first order phase transition with 
nonzero baryo-chemical potential, a mechanism analogous to associated
kaon-production yields an enriched
population of $s$ quarks in the quark-gluon phase, 
while the $\bar{s}$ quarks drift into the hadron 
phase \cite{greinerc87a,lukacs87a}. 
This strangeness separation
results in the distillation of metastable {\em strangelets}
only if the Bag constants are very small, $B < 180$ MeV/fm$^{-3}$
\cite{greinerc87a}.

Experimentally {\em strangelets} are
distinguishable from normal nuclei
due to their very small or even negative charge to mass ratio. 
The most interesting candidates for long-lived {\em strangelets} are lying
in a valley of stability which starts at the {\em quark--}$\alpha$
and continues by adding one unit of negative charge, i.e.
(A,Z)=(8,-2),(9,-3)\ldots \cite{schaffner96a}. Recent calculations indicate
that positively charged
{\em strangelets} seem only to exist for $A > 12$ and very low bag parameters
\cite{schaffner96a}.

There exist, however, other forms of hypermatter with similar properties
as {\em strangelets}: hyperclusters or MEMO's 
(metastable exotic multihyperon objects) 
consist of  multiple 
$\Lambda, \Sigma$ and $\Xi$ hyperons \cite{schaffner92a}, and  
-- possibly -- also nucleons.
The double-$\Lambda$ hypernucleus ${}_{\Lambda\Lambda}^6$He has
been observed long ago \cite{prowse66a}.
Properties of MEMOs have been estimated using the Relativistic Mean Field
model. MEMOs can contain multiple negatively charged hyperons, therefore
they may also have zero or negative charge-to-mass-ratios.

MEMOs or hyperclusters 
could form a doorway state to {\em strangelet}
production, or vice versa: MEMOs may coalescence in the high
multiplicity region of the reaction. If strangelets are stronger bound than
``conventional'' confined MEMOs, the latter may transform into 
{\em strangelets}.
The cross sections for production of MEMOs in relativistic heavy ion 
collisions rely heavily on
model parameters (e.g. in the in the coalescence model $p_0$ and $r_0$).
The predicted yields are typically $< 10^{-8}$ per event 
\cite{schaffner92a,baltz94a}.

\paragraph*{Experimental status\\}

{\em Strangelet} searches are underway at the AGS 
\cite{kumar95a,rotondo96a,barish96a} 
and SPS \cite{borer94a,dittus95a,klingenberg96a,appelquist96a,pretzl96a}. 
So far no long lived ($\tau > 10^{-7}$ s) {\em strangelets}
have been unambiguously identified 
-- the upper limits for the production cross sections 
established by the experiment are still consistent with 
theoretical predictions for short lived MEMOs since they cannot be
tested in the present long flight path experiments. 
There has been a report of one candidate
with $Z=-1$, $N/Z=7.4$~GeV and $\tau > 85~\mu s$ 
\cite{pretzl96a,klingenberg96a,kabana98a}.
Therefore this exciting topic awaits more experimental effort.

Current experiments are designed to detect {\em strangelets} 
with a small charge-to-mass ratio and 
rather long lifetime ($\tau \ge 12~\mu s$ in
the case of \cite{klingenberg96a,appelquist96a}).
The present experimental setups are hardly sensitive to the most
promising long-lived and negatively charged {\em strangelet} 
candidates beyond the {\em strange quark--}$\alpha$. 
Unfortunately, plans for extending
experiment E864 at the AGS to look for highly charged strangelets
with $B>10$ \cite{sandweiss_priv} cannot be followed because
the AGS fixed target heavy ion  program has been put to rest.

Future experiments at collider energies (STAR at RHIC and ALICE at LHC)
will be sensitive for short-lived metastable hypermatter, too
\cite{harris94a,schukraft96a,spieles96b}.

\paragraph*{Discussion\\}

Due to the possibility of creating  MEMOs in a hadronic scenario and 
their possible subsequent transformation into {\em strangelets}, 
the formation of a QGP is not a necessary
prerequisite for the creation of {\em strangelets}.
The discovery of {\em strangelets} would therefore be no hard proof for
a deconfinement phase transition. 
So far there seems to be no
clear way to distinguish {\em strangelets} from 
MEMOs. Both forms of hypermatter would be extremely interesting to study
and the discovery of one or the other would be worth every effort.
Therefore, experiments should be devoted to the
search for short-lived (anti-) hyperclusters.

Large theoretical uncertainties remain, e.g. how the predicted yields
depend on the model parameters. Both, theory and the current experimental 
results
point towards a future search for {\em strangelets}/hyperclusters 
with rather short lifetimes. Experiments
including a large TPC might be able to observe the decay 
short-lived hyperclusters. Indirect $K^+ K^-$ correlation measurements 
might offer another possibility
of detecting {\em strangelets} or hyperclusters  \cite{soffS97a}.

%%%%%%%%%%%%%%%%%%%%%%%%%%%%%%%%%%%%%%%%%%%%%%%%%%%%%%%%%%%%%%%%%%%%%%%%%%%
\subsection{Radiation of the plasma: direct photons and thermal dileptons}
	\label{photons}
%%%%%%%%%%%%%%%%%%%%%%%%%%%%%%%%%%%%%%%%%%%%%%%%%%%%%%%%%%%%%%%%%%%%%%%%%%%

\paragraph*{Theoretical concepts\\}

The most prominent process for the creation of direct (thermal) photons in a
QGP are $q\bar{q} \rightarrow \gamma g$ (annihilation) and 
$g q \rightarrow \gamma q$ (Compton scattering).
The production rate and the momentum distribution
of the photons depend on the momentum distributions of quarks, anti-quarks
and gluons in the plasma.
Infrared singularities occuring in perturbation theory are softened
by screening effects \cite{kapusta91a,baier92a,baier92b,thoma95a}.
If the plasma is in thermodynamic equilibrium, 
the photons may carry information on this thermodynamic
state at the moment of their production 
\cite{kapusta91a,shuryak78b,sinha83a,hwa85a}.

The main hadronic background processes to compete against are 
pion annihilation $\pi \pi \rightarrow \gamma \rho$ and Compton
scattering $\pi \rho \rightarrow \gamma \pi$ \cite{kapusta91a,song93a}.
The broad $a_1$ resonance may act as an intermediate state in
$\pi \rho$ scattering and thus provide an important 
contribution \cite{xiong92a,song93a} via it's decay into $\gamma \pi$.
In the vicinity of the critical temperature $T_C$ a hadron gas
was shown to
``shine'' as brightly (or even brighter than) a QGP \cite{kapusta91a}.

A finite baryochemical potential yields at 
constant energy density a reduced multiplicity 
of direct photons from a QGP \cite{dumitru93a,traxler95a}. 

Hydrodynamical calculations can be used to compare purely hadronic
scenarios with scenarios involving a first/second order phase transition to a 
QGP. They show a reduction in the temperature of the photon spectrum
in the event of a first order 
phase-transition \cite{alam93a,dumitru95a,neumann95a}.

The rapidity distribution of direct hard photons reflects the initial
rapidity distribution of the produced mesons or directly 
the QGP \cite{dumitru95b}.
It may thus provide insight into the (longitudinal) expansion of
the photon source: If the hot thermal source is initially at rest
and is accelerated by two longitudinal 
rarefaction waves propagating inwards with the velocity of sound,
the photon rapidity distribution is strongly peaked around midrapidity.
In contrast, a Bjorken-like boost-invariant expansion results in a
more or less flat photon rapidity spectrum.

If a very hot plasma is formed (e.g. at RHIC or LHC energies) 
a clear photon signal might be
visible at transverse momenta in the range between 2 and 5 GeV/c
\cite{strickland94a,srivastava92a,chakrabarty92a}. The lower
$p_t$ range (1--2 GeV/c) is dominated by the mixed phase; separated
contributions of the different phases are difficult to see due to transverse
flow effects \cite{alam93a}. These effects, however,
can be important up to transverse
momenta of 5 GeV. Transverse flow effects also destroy the
correlation between the slope and the temperature of the photon
spectrum \cite{neumann95a}. 

Analogously to the formation of a real photon via a quark - anti-quark
annihilation, a virtual photon may be created in the same fashion
which subsequently decays into a $l^+ l^-$ pair (a {\em dilepton}).
Also bremsstrahlung of quarks scattering off gluons can convert into
dileptons.

Dileptons can carry
information on the thermodynamic state of the medium at the moment
of production in the very same manner as the direct photons -- 
since the dileptons interact only electromagnetically
they can leave the hot and dense reaction zone basically undistorted, too.

The main background contributions stem from pion annihilation,
resonance decays \cite{cleymans91a,kochp93a,gale94a,song94a,winckelmann95a}
(two pions can annihilate, forming either a virtual photon
or a rho meson -- both may then decay into a dilepton)
and $\pi-\rho$ interactions \cite{baier97a,murray96a} 
at low dilepton masses and Drell--Yan processes 
\cite{drell70a,spieles97c} at
high masses . Furthermore
meson resonances such as the rho-, omega- or phi- meson may be produced
directly or in the decay of strings and heavier resonances.
As all of those vector mesons carry the same quantum numbers as the photon,
they may decay directly into a dilepton. Resonances can also emit
dileptons via Dalitz decays. The Drell--Yan process describes
the annihilation of a quark
of one hadron with an anti-quark (in proton proton collisions from the
sea of $\bar{q}$) of the other hadron, again resulting in a virtual
photon which decays into a dilepton.
The open charm contribution to the dilepton mass spectrum has been
estimated to be negligible for low dilepton masses \cite{braun-munzinger97a}
at the CERN/SPS. At RHIC and LHC energies, however, charm contributions
dominate the dilepton mass spectrum above 2~GeV \cite{gavin96d}.

Most original calculations on dileptons as signals of a QGP 
at CERN/SPS energies focused
on masses below the rho meson mass
\cite{shuryak78a,domokos81a,kajantie81a,kajantie82a,chin82a,
kochp93a,cleymans86a,cleymans87a,siemens85a,seibert92a}. 
The current understanding of hadronic background 
contributions \cite{cleymans91a,gale94a,song94a,winckelmann95a} shows that 
most probably dileptons originating from a QGP 
are over-shined by hadrons, with the possible exception of
masses around 1 to 1.5 GeV \cite{ruuskanen91a,ruuskanen92a}
where the rates from a plasma (at very high temperatures
around 500 MeV) may suffice  to be visible.
At higher masses, the yield of Drell--Yan processes from first nucleon
nucleon collisions most probably exceeds that of thermal dileptons from a QGP.
Finite baryochemical potential will, at a given 
energy density, reduce the number of dileptons
emitted from a QGP \cite{ko89a,dumitru93b,he95a}, due to the dropping
temperature in that system.

The dependence of the yield of high mass dileptons on the thermalization
time is still a point of open debate \cite{kapusta92a,kampfer92a}. 
The parton cascade  \cite{geiger93b} and other models of
the early equilibration phase \cite{kampfer92a,shuryak93a} 
predict an excess of dileptons
originating from an equilibrating QGP over the Drell--Yan background
in the mass range between 5 and 10 GeV. Then the early thermal evolution
of the deconfined phase could be traced in an almost model independent
fashion \cite{strickland94a}.

The secondary dilepton production via quark-antiquark
annihilation has also been studied on the basis
of a hadronic transport code (UrQMD \cite{bass98a}). Here, one
obtains a realistic collision spectrum of secondary hadrons for SPS energies. 
Using parton distribution  functions and evaluating the contributions of all
individual hadronic collisions one finds  
that meson-baryon interactions
enhance the mass spectrum at mid-rapidity below masses of 3~GeV 
considerably \cite{spieles97c}. 
Preresonance interactions are estimated to enhance this secondary yield
by up to a factor of 5.

\begin{figure}[thb]
\centerline{\psfig{figure=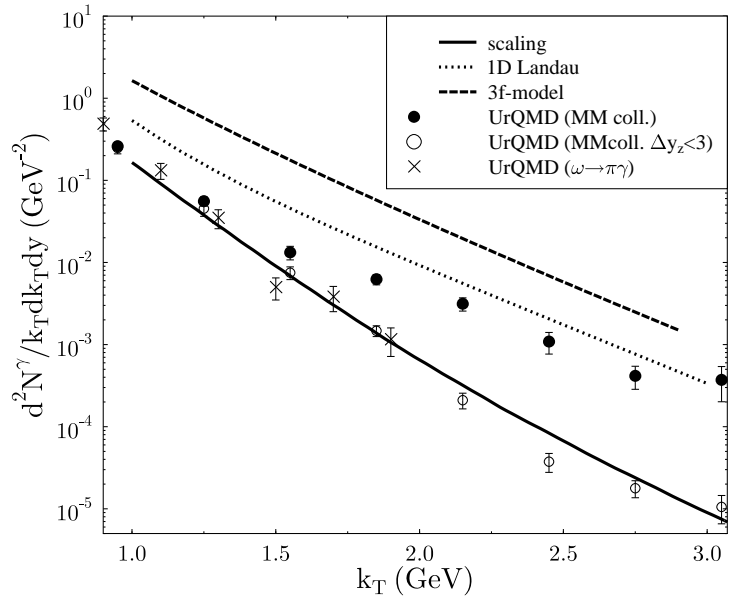,width=10cm}}
\caption{\label{dumiphot}
Transverse momentum spectrum of directly produced photons in
Pb+Pb collisions at 160 GeV/nucleon calculated with the UrQMD model. 
%The contributions of the different processes are shown. 
The resulting spectrum
is compared with different hydrodynamical calculations.
In all models the processes 
$\pi\eta \mapsto \pi \gamma$,  $\pi\rho \mapsto \pi \gamma$
and $\pi\pi \mapsto \rho \gamma$ are considered 
as photon sources. 
The figure has been taken from \protect \cite{dumitru97a}.}
\end{figure}

\paragraph*{Experimental status\\}

Experiments to measure direct photons are carried out at the CERN/SPS
by the WA80/WA98  and the CERES/NA45 collaborations.
Whereas the final analysis of the WA80 collaboration for S+Au indicates
a 5\% photon signal over background (with
a 0.8\% statistical and a 5.8\% systematical error) 
\cite{kampert93a,santo94a,albrecht96a} 
the CERES/NA45 collaboration did not report any direct 
photons, but sets an upper limit of 7\% for the integrated excess
an unconventional photon source might have in central S+Au collisions
\cite{irmscher94a,agakichiev95a,baur96a}. 
The WA80 collaboration has reported upper limits for each measured
$k_t$ bin which yields important information for constraining the
initial temperature of the reaction zone. Within the reported 
systematic errors the results of the WA80 and the CERES/NA45
collaborations are compatible with each other
\cite{awes95a}.

Dileptons can be measured at CERN in form of dimuons 
by the HELIOS3, NA38 and NA50 
\cite{mazzoni94a,masera95a,abreu94a,abreu96a} collaborations
and in form of  electron pairs 
by the CERES collaboration \cite{agakichiev95a}.
Dimuons exhibit an excess in AA collisions 
in the mass range $0.2 < M < 2.5$ GeV/c$^2$ up to  the $J/\Psi$,
as compared to pp and pA collisions. For dielectrons an excess is
observed in the low--mass region $0.2 < M < 1.5$ GeV/c$^2$, 
again relative to pp and pA collisions (c.f. figure \ref{ceresfig1}).

\paragraph*{Discussion\\}

The Pb+Pb analysis on direct photons 
of the WA98 and NA45/CERES collaborations is in progress.
Hydrodynamical
calculations are only compatible with the S+Au data of WA80 
if a phase transition with its cooling
is taken into account 
\cite{srivastava94a,dumitru95a,neumann95a,arbex95a,huovinen98a,huovinen98b}
or if higher mass meson and baryon multiplets are included for
the hadronic EoS.

Microscopic hadronic transport models, however, are not constrained
by the assumption of thermal equilibrium, in particular in the initial stage, 
and yield results compatible
with hydrodynamical calculations without invoking a phase transition scenario 
\cite{dumitru97a}, as can be seen in figure~\ref{dumiphot}.  
They shows that preequilibrium 
contributions dominate the photon spectrum at transverse momenta
above $\approx 1.5$ GeV. The hydrodynamics prediction of a strong
correlation between the temperature and radial expansion velocities
on the one hand and the slope of the transverse momentum distribution
on the other hand thus is not recovered in a microscopic transport model
\cite{dumitru97a}. 

Apart from these ambiguities in the interpretation of the data,
the main problem with regard to the direct photon signal is 
the extremely small cross section in a difficult experimental situation, since
photons from hadronic decays generate a huge hadronic background.
The strong and dedicated effort to improve the
measurements will be continued, also at the more promising collider
energies.

Both, the dielectron as well as the dimuon data 
seem  to be compatible with a hydrodynamic approach
assuming the creation of a thermalized QGP \cite{srivastava96a}.
Hadronic transport calculations are not able to fully reproduce
the observed excess \cite{winckelmann96a,ko96a,bratkovskaya97a}. 
However, at least part of the observed enhancement of lepton pairs
at intermediate and low masses might be either caused by the previously
neglected source of secondary Drell-Yan processes \cite{spieles97a} or
by contributions of heavy mesons, such as the $a_1$ \cite{liGQ97a}.
A detailed discussion
of the dilepton data and its theoretical implications will follow in
conjunction with the discussion on chiral symmetry restoration in
section \ref{dileptons}.

%%%%%%%%%%%%%%%%%%%%%%%%%%%%%%%%%%%%%%%%%%%%%%%%%%%%%%%%%%%%%%%%%%%%%%%%%%%
\subsection{Restoration of chiral symmetry: vector mesons in dense matter}
		\label{dileptons}
%%%%%%%%%%%%%%%%%%%%%%%%%%%%%%%%%%%%%%%%%%%%%%%%%%%%%%%%%%%%%%%%%%%%%%%%%%%

\paragraph*{Theoretical concepts\\}

\begin{figure}[thb]
\centerline{\psfig{figure=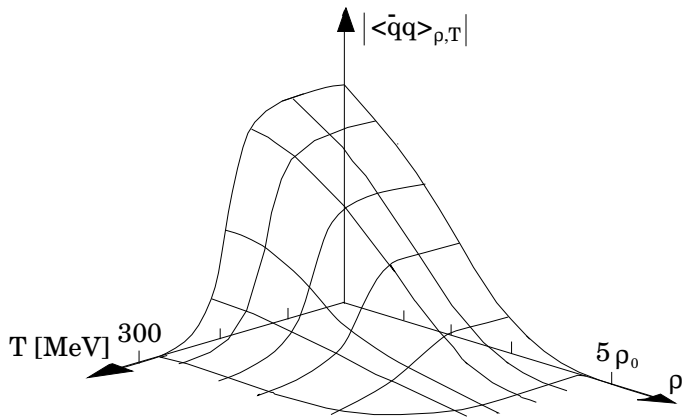,width=9cm}}
\caption{\label{weisebild} quark condensate $-\langle q \bar{q} \rangle$
as a function of temperature $T$ and baryon density $\rho/\rho_0$. The
figure has been adapted from
\protect \cite{weise93a}.}
\end{figure}

The dilepton signal due to the decay of vector mesons, in particular  
from  the rho meson,
is of great  interest: In
conjunction with the chiral symmetry restoration 
\cite{weise93a,birse94a,brown96a,ko97a,koch97a}
the rho, omega and phi mesons (and heavier meson resonances, e.g.
the $a_1$, $a_2$ ...) are
expected to change their spectral function
in the hot, high baryon density medium:
the breaking of the chiral $SU(3)_L \times SU(3)_R$ symmetry 
(an approximative symmetry of QCD) 
results in quark condensates $\langle q \bar{q} \rangle$ in the QCD
vacuum and a ``Goldstone'' boson, i.e. the pion.
The dependence of $\langle q \bar{q} \rangle$ on the temperature
$T$  has been studied in the framework of lattice QCD \cite{bernard92a} and
chiral perturbation theory \cite{gerber89a,gasser89a}. Up to (0.7 - 0.8) $T_C$
$\langle q \bar{q} \rangle$ remains nearly constant and then 
its absolute value decreases rapidly (see figure \ref{weisebild}).
The behavior of the quark condensate at finite baryon densities 
is described in a model independent fashion
by the Feynman-Hellman-theorem \cite{drukarev90a,cohen92a}. 
A model calculation of the  dependence of $\langle q \bar{q} \rangle$ 
on both, the baryon density $\rho/\rho_0$ and temperature
$T$, can be seen in figure \ref{weisebild} -- the drop of 
$\langle q \bar{q} \rangle$ with $\rho$ and $T$ is quite analogous
to the temperature and density dependence of 
the nucleon effective mass in the $\sigma - \omega$ model as noted
in \cite{theis83a}.

The reduction of the absolute value of the 
quark condensate $\langle q \bar{q} \rangle$ in a
hot and dense hadronic environment might reflect itself in reduced masses
of vector mesons 
\cite{pisarski82a,bochkarev84a,gale87a,dosch88a,
furnstahl90a,gale91a,brown91b,seibert92b,karsch95a}.
However, a lowering of the mass of the $\rho$-meson -- most commonly
referred to as ``Brown-Rho scaling'' \cite{brown91b} -- is not 
synonymous with the restoration of chiral symmetry:
It has been shown by employing current algebra as well as PCAC 
that to leading order in temperature, $T^2$, 
the mass of the $\rho$-meson remains nearly constant as a function
of temperature \cite{dey90a}, whereas the chiral condensate is 
reduced \cite{gerber89a}. 

Restoration of chiral may manifest itself in different forms \cite{ko97a}: 
the masses of the $\rho$- and the $a_1$ meson may merge, 
their spectral functions
could mix -- resulting in peaks of similar strength at both masses
(and causing a net-reduction at the $\rho$-peak) -- or both 
spectral functions could be smeared out over the entire mass range.

Dileptons from the in-medium decay of such vector mesons 
with modified masses and spectral functions
would point towards 
the restoration of chiral symmetry at a phase transition.

\paragraph*{Experimental status\\}

At the BEVALAC the DLS collaboration has measured dielectron pairs
in proton induced reactions as well as in d+Ca, He+Ca, C+C and Ca+Ca reactions 
\cite{roche89a,matis95a,porter97a}. Their latest results
\cite{porter97a} for pair masses $M < 0.35$ GeV/c in the Ca+Ca system
show a larger cross section than their previous measurements \cite{matis95a}
and current model calculations \cite{wolf93a,bratkovskaya96a}, suggesting
large contributions from $\pi^0$ and $\eta$ Dalitz decays.  The cross section
$d\sigma / dM$ scales with $A_P \cdot A_T$ up to pair masses of $M = 0.5$
GeV/c. For larger masses the Ca+Ca to C+C cross section ratio is significantly
larger than the ratio of $A_P \cdot A_T$ values.

\begin{figure}[thb]
\begin{minipage}[t]{7cm}
\centerline{\psfig{figure=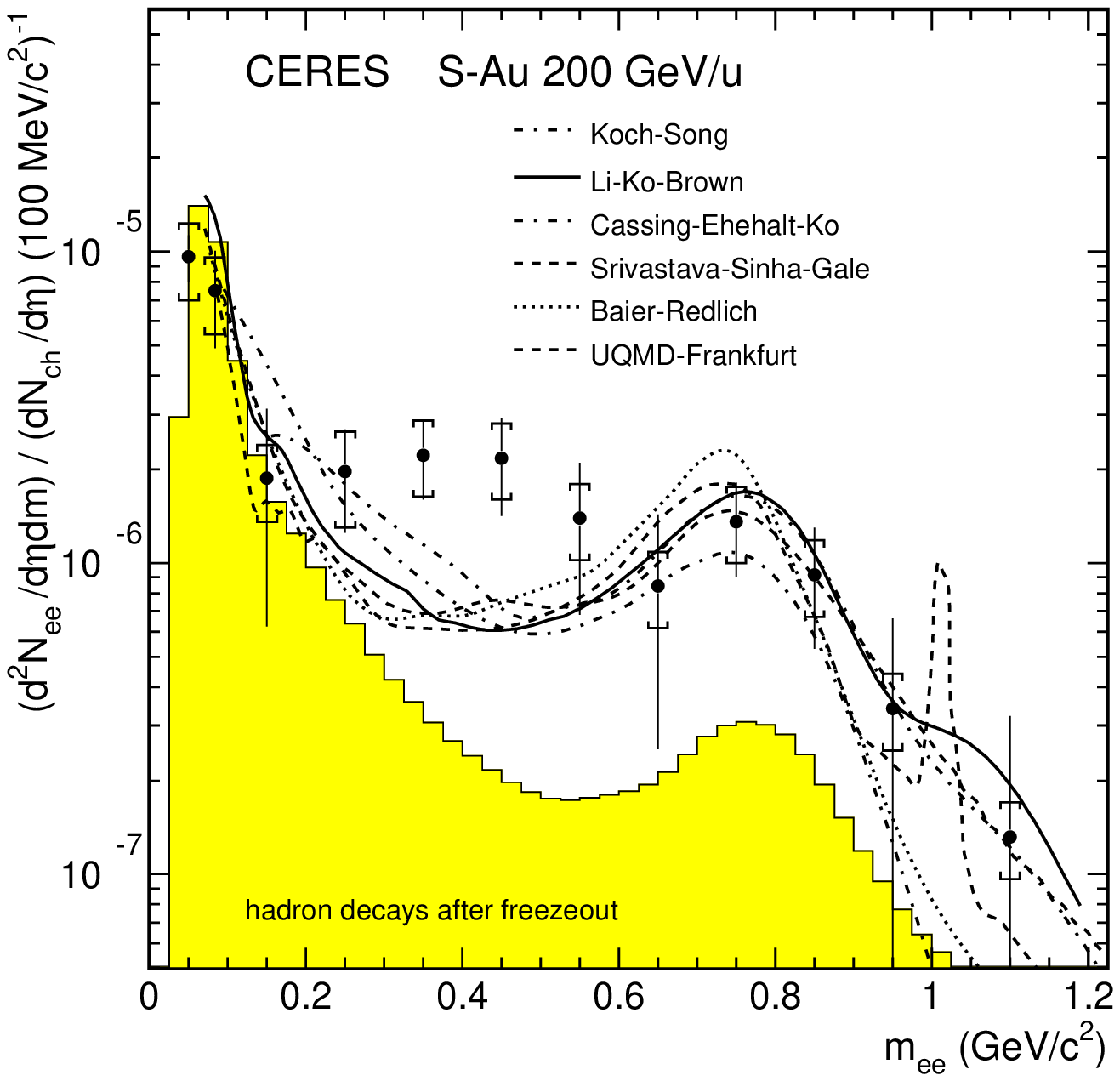,angle=270,width=7cm}}
\end{minipage}
\hfill
\begin{minipage}[t]{7cm}
\centerline{\psfig{figure=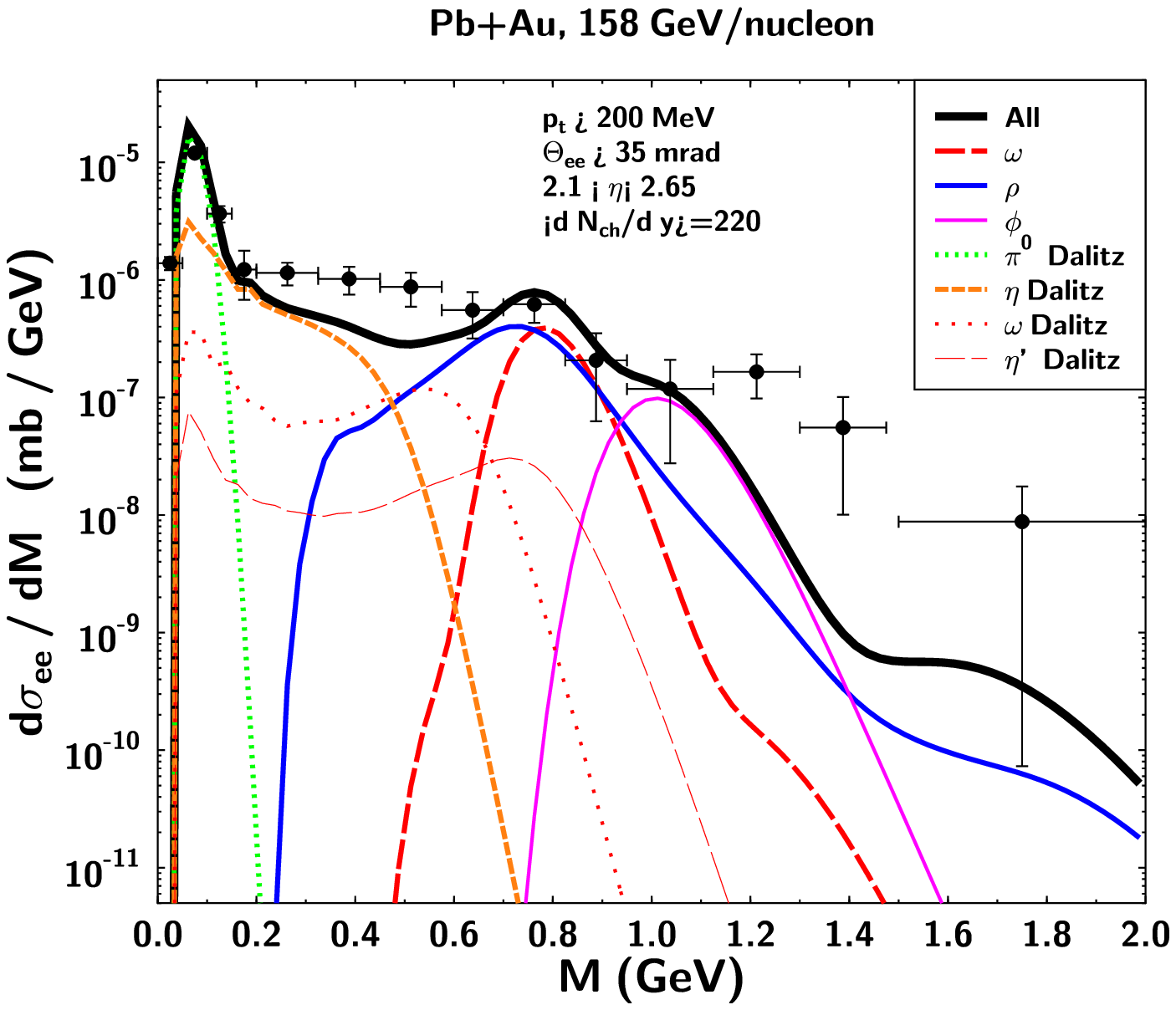,width=7cm}}
\end{minipage}
\caption{\label{ceresfig1}
Left: inclusive $e^+e^-$ mass spectra in 200 GeV/nucleon S+Au collisions 
as measured by the CERES collaboration 
\protect \cite{agakichiev95a}. 
The figures have been taken from \protect \cite{drees96a}.
The shaded area depicts hadronic
contributions from resonance decays. The data are compared
calculations based on a purely hadronic scenario 
\protect \cite{winckelmann96a,srivastava96a,cassing95a,ligq95a,koch96a}. 
Right: UrQMD prediction and data \protect\cite{agakichiev97b}
for Pb+Au at 160 GeV/nucleon.
}
\end{figure}

Unfortunately, there are no experiments capable of measuring dileptons
in A+A collisions at $\sim $ 10 GeV/nucleon.
However, in central sulfur -- gold collisions at the CERN/SPS an enhancement 
has been measured in the invariant mass spectrum of muon pairs  relative
to the normalized proton proton and proton nucleus 
data at 200 GeV/nucleon taken by the
HELIOS/3 and NA38 collaborations:
While the pp and pA data seem well described by measured sources
such as Drell-Yan, open charm and hadronic decays, 
there is an excess in AA observed in the mass range $0.2 < M < 2.5$ GeV/c$^2$
(for the $J/\Psi$, a suppression of the peak relative to 
this background is observed, 
c.f. section \ref{jpsisuppression}) \cite{mazzoni94a,masera95a,abreu94a}.
Similarly, for dielectron pairs in S+Au 
an excess has been observed by the CERES collaboration 
\cite{agakichiev95a}
in the low--mass region $0.2 < M < 1.5$ GeV/c$^2$, 
again relative to pp and pA collisions (c.f. figure \ref{ceresfig1}). 
As in the case of the dimuon excess, the pp and pA data
can be well understood taking known hadronic sources into account.
Data for Pb+Au confirm this low mass dielectron excess 
\cite{ullrich96a,agakichiev97b,jouan98a,braun-munzinger98a}. 

\paragraph*{Discussion\\}

When searching  for chiral symmetry restoration, 
thermal dileptons (see section \ref{photons})
would serve as background. Due to the dominance of hadronic
decays in the mass range up to 1.5 GeV, however, they do not
pose a serious problem for the measurement of vector meson
properties.

As already stated in section \ref{photons},
both, the dielectron as well as the dimuon data 
at the SPS seem  explainable in a hydrodynamic approach
assuming the creation of a thermalized QGP \cite{srivastava96a}.

On the other hand, the same data can be reproduced in the 
framework of microscopic hadronic transport models incorporating
mass shifts of vector mesons 
\cite{cassing95a,cassing96a,ligq95a,ko96a,winckelmann96a}.

However, even bare hadronic transport model calculations, without
any mass shift included,  miss only the data in the 400 to 600 MeV bins 
(by 2 to 3 standard deviations) \cite{winckelmann96a,ko96a}. 

Calculations evaluating in-medium spectral 
functions, due to the coupling of the $\rho$ with nucleon resonances and
particle-hole excitations, also achieve a satisfactory reproduction
of the CERES data \cite{rapp97a}, without requiring a dropping
$\rho$-mass. Recent data on the $p_t$ dependence of this 
phenomenon \cite{agakishiev98a} indicate however that the enhancement
is most pronounced at low $p_t$. This precludes the dominance of 
p-wave baryon-resonance effects.
Since the hadronic transport models did -- up to now -- neglect
contributions, e.g. from
current--current correlation functions \cite{weise96a} and from additional
heavy meson resonances,  it has yet to be determined whether partial
restoration of chiral symmetry or 
Brown-Rho scaling
is the only possible explanation of these
interesting new data. 

On the experimental side the main problems lie with the low signal
to background ratio of 1/10 and limited statistics ($< 1000$ lepton
pairs in Pb+Au). In addition, the shape of the $e^+ e-$ excess at
0.25~GeV~$< m_{ee} <$~0.6~GeV coincides with that of the background.
Currently a strong effort by the CERES collaboration is underway to 
upgrade the experiment with a TPC. The resulting increase in statistics
and resolution should help to verify or falsify some of the conflicting 
hypotheses on the origin of the low mass dilepton enhancement.

The latest DLS data, gives rise to speculations that the observed
enhancement in the mass-range below the free $\rho$ meson mass may be due
to enhanced $\rho$ meson production or a mass shift of the $\rho$ 
in a dense hadronic medium -- without the need for a deconfinement
phase transition. However, all microscopic model calculations which
have addressed this data sofar \cite{ernst98a,bratkovskaya98a}, 
have not been able to provide
a reasonable explanation within the frameworks which work so well
for dilepton production at CERN. 

%%%%%%%%%%%%%%%%%%%%%%%%%%%%%%%%%%%%%%%%%%%%%%%%%%%%%%%%%%%%%%%%%%%%%%%%%%%%%%%%%
                \subsection{Quarkonia Suppression:
Evidence for Deconfinement or Dynamical Ionization}
                \label{jpsisuppression}
%%%%%%%%%%%%%%%%%%%%%%%%%%%%%%%%%%%%%%%%%%%%%%%%%%%%%%%%%%%%%%%%%%%%%%%%%%%%%%%%%

In 1986 Matsui and Satz proposed \cite{matsui86a} that the suppression of
heavy quarkonia-mesons could provide one of the signatures for
deconfinement in QCD at high temperatures. The idea was based on an analogy
with the well known Mott transition in condensed matter systems.  At high
densities, Debye screening in a quark-gluon plasma
reduces the range of the attractive force
between heavy quarks and antiquarks, and above some critical density
screening prevents the formation of bound states.  The larger bound
states are expected to dissolve before the smaller ones as the
temperature of the system increases.  The $\psi'$ and $\chi_c$ states
are thus expected to become unbound just above $T_c$, while
 the smaller $\psi$
state may only dissolve above $\approx 1.2 T_c$.  Heavier $b\bar{b}$
states offer the same features as $c\bar{c}$ states, but require much
shorter screening lengths to dissolve \cite{karsch91a}.  The
$\Upsilon(b\bar{b})$ state may dissolve only around 2.5 $T_c$, while
the larger excited $\Upsilon'$ could also dissolve near $T_c$.

In order to determine the magnitude of suppression, it is obvious that
the initial production mechanism must be well understood.  Charm quark-
anti-quark pairs, $c\bar{c}$, are produced in rare pQCD gluon fusion
processes \cite{baier83a}, ($gg\rightarrow c\bar{c}$) with a cross
section in $pp$ reactions $\sigma_{c\bar{c}}\sim 10\; \mu{\rm b}$ at
$\sqrt s=20$ GeV.  In the rare events when a pair is formed, both the charm
and anti-charm quantum numbers remain approximately conserved, and
either the $c\bar{c}$ emerge from the reaction in hidden charm
quarkonium bound states, $J/\psi(1S_1:3097), \;\psi'(2S_1:3686),\;
\chi_c(1P_{0,1,2}:3500), \; \ldots$, or in continuum open charm states
$D(1869), D^*(2010), \ldots$.  Even though only about 1\% of the
$c\bar{c}$ pairs emerge in $pp$ collisions as $J/\psi$ states, these
vector hidden charm mesons are the easiest to measure because they are
seen as sharp resonances on top of a broad continuum in the invariant
mass spectrum of di-leptons. In contrast, open charm production is
much  harder to measure.  Semi-leptonic open charm decay
contributes to the continuum yield of dileptons mainly below the
$M_\psi$ peak.

Above $M_\psi$, the Drell-Yan (DY) process ($q\bar{q}\rightarrow
\mu\bar{\mu}$) begins to dominate the continuum yield.  The great
importance of DY is that the absence of strong final state
interactions of the produced leptons makes it possible to compute the
absolute DY cross section via pQCD.  The nuclear number dependence of
the cross section is then entirely determined by geometrical (Glauber)
factors, $T_{AB}({\bf b})$ (neglecting small nuclear dependence of the
structure functions). Here, $T_{AB}({\bf b})$ is the number of binary
$NN$ interactions per unit area as a function of the impact parameter.
The measured DY yields thus provide an important constraint on the
impact parameter range associated with specific centrality ($E_T$)
triggers used in the experiment.  The comparison of the centrality
dependence of the $J/\psi$ and DY cross section therefore provides a
calibration tool to determine the magnitude of the suppression factor
of charmonium in nuclear collisions.

Great interest in this
proposed signature arose when NA38 found the first evidence
of suppression in light ion reactions. With the new preliminary
$Pb+Pb$ data of NA50 (see next section)
which have been reported to show
``anomalous'' suppression, it is especially important
to review critically some of the competing dynamical
effects that could forge this quark-gluon plasma signature.

One of the main  problems in the interpretation of the
observed suppression as a signal for deconfinement is that
non-equilibrium dynamical sources of charmonium suppression
have also been clearly discovered in $p+A$ reactions.
The interaction of the
pQCD produced $c\bar{c}$ pair with any QCD medium (confined or not)
decreases significantly the probability of that pair to emerge
in an asymptotic $J/\psi(1S)$ state. 
The observation of $J/\psi$ suppression in $p+A$
is direct proof of this fact since
 the formation of an equilibrated
quark-gluon plasma in such reactions is not expected.

A phenomenological analysis of p+A data yields a dissociation cross
section for both the $J/\psi$ and the $\psi'$ around $7.3 \pm 0.6$ mb 
\cite{gerschel88a,gerschel92a,kharzeev96c,lourenco96a}.
%We note that other estimates of the $J/\psi$ nucleon cross section based
%on the vector dominance model 
%may severely underestimate its value
%\cite{frankfurt88a,frankfurt91a}.
This finding is surprising since
the transverse areas of those two mesons differ by more than a factor
of two. This has led to the 
$c \bar{c}_8$ color octet model interpretation of the (quantum)
formation physics involved.
In this model, it is assumed that 
 the pair is formed in a small octet state accompanied
by a soft gluon that can be  easily stripped off as it propagates 
through a nucleus
\cite{wittmann93a,kharzeev96d} .
This qualitatively accounts for the observed equal nuclear
absorption cross sections since, as a result of  time dilation,
hadronization into the asymptotic $J/\psi$ or $\psi'$  states
is delayed several fm/c at high energies \cite{kharzeev96d,kharzeev96c}.

A recent development is the calculation of the hard contributions
to the charmonium- and bottonium-nucleon cross sections based on
the QCD factorization theorem and the non-relativistic quarkonium
model \cite{gerland98a}. The calculated p+A
cross section agrees well with the data. The non-perturbative contribution
to the charmonium cross section dominates at CERN/SPS energies and
becomes a correction at LHC. The $J/\psi$ production in nucleus-nucleus
collisions at the CERN/SPS can then be reasonably well described by
hard QCD, if the larger absorption cross section of the $\chi$ states
that are predicted by QCD are taken into account.

While the octet model used in conjunction
with the Glauber geometrical model can account for the $pA$ observation,
the corrections to this eikonal picture of nuclear absorption extrapolated
to  $AA$ reactions are, however, not yet under theoretical control.
For example, since the $c \bar{c}_8$ color octet - soft gluon state 
is not an eigenstate of non-perturbative QCD,
its effective hadronic absorption cross section
may vary within the relaxation time. Also possible pileup
of matter and energy loss prior to the gluon fusion event are neglected.
The importance of gaining better theoretical
control of the nuclear absorption process is
underscored by the fact that in central $PbPb$ reactions about one half
of the NA50 observed factor of four suppression of $J/\psi$
is estimated to arise from such non-equilibrium (quantum)
formation physics.

The second major theoretical uncertainty in interpreting charmonium
suppression is distinguishing dynamical ``background'' 
dissociation processes such as $\psi+\rho\rightarrow D\bar{D}$ and
$\psi+\Delta\rightarrow \Lambda_c \bar{D}$ from {\em transient}
partonic dynamical
processes (non-thermal color field fluctuations) and
from the sought after
screening mechanism in  the plasma phase of QCD matter.  

Purely
hadronic dissociation scenarios have been suggested
\cite{neubauer88a,gavin88a,vogt88a,gavin90a,gavin96a,gavin96b}, which could with
suitable parameters account for $J/\psi$ and $\psi'$ suppression
without invoking the concept of deconfinement.  These hadronic
scenarios are referred to as {\em comover models}. Suppression in
excess to that due to preformation nuclear absorption 
is ascribed in such models to interactions of
the charmonium mesons with comoving  mesons and
baryons which are produced copiously in nuclear collisions.  
Unfortunately none of the required absorption cross sections
are experimentally known and estimates are highly model dependent.
A general  criticism of comover suppression estimates is
the use of overly simplified Glauber geometry and idealized boost
invariant expansion dynamics for the produced particles.

Studying the transverse momentum, $p_t$, dependence of $J/\psi$ and $\psi'$
production 
%in general, and 
%comparing the $p_t$ dependence of the $J/\psi$ to that of the $\psi'$
%in particular,
may yield additional information concerning the nature of the 
$J/\psi$-suppression mechanism \cite{blaizot87a,karsch88b}. 
Two common scenarios have been considered:
At sufficiently high $p_t$ final state interactions
 might disappear due to time dilation, while
hadronic absorption effects should be similar for the $J/\psi$ and
$\psi'$. In a deconfinement scenario this idea suggests
that  $J/\psi$ is suppressed  only for low transverse momenta
\cite{satz90a}.
A second scenario assumes that $J/\psi$ and
$\psi'$ acquire  large transverse momenta through multiple elastic
parton-parton collisions. Those multiple collisions, however, are
most likely to occur in the high density QGP region. The consequence
would be that high
$p_t$ $J/\psi$ and $\psi'$ should be even more suppressed than
those with low transverse momenta \cite{kharzeev97a}. A purely
hadronic scenario predicts an increase in the mean transverse momentum
as a function of transverse energy for the heavy Pb+Pb system
\cite{gavin96c}. 

The above discussion emphasizes some of the uncertain theoretical 
elements in the interpretation of charmonium suppression
as a signal for deconfinement. To isolate the final state
interaction effects from initial state nuclear absorption,
it has been proposed 
to combine all available
data using a Glauber model geometric variable, ``$L$''.
Since this is so popular we review below how this variable is defined.

The suppression of the $J/\psi$ production cross section
in $A+B$ collisions can be expressed as
\begin{equation}
\sigma(AB\rightarrow J/\psi)=AB \sigma(pp\rightarrow J/\psi)
e^{-[\sigma^{abs}_{c\bar{c}N} \rho_0 L(A,B,E_T)]}
S_{co}
\;\; .\end{equation}
Here $\sigma^{abs}_{c\bar{c}N}\sim 5-7$ mb is an effective preformation
nucleon absorption cross section, $\rho_0$ is the ground state nuclear
density, and $L(A,B,E_T)$ is a measure of the mean nuclear thickness
evaluated through the Glauber model
\begin{eqnarray}
AB e^{-[\sigma^{abs}_{c\bar{c}N} \rho_0 L(A,B,E_T)]} S_{co}&=&
\int d^2b P(E_T,b)\int dz \int d^2sdz'\nonumber \\  
&\;&\times\rho_A({\bf s},z)\rho_B(|{\bf b-s}|,z') T_{co}({\bf b},{\bf s})
\nonumber \\
&\;& e^{-
\sigma^{abs}_{c\bar{c}N}\int dz'' (\theta(z''-z)\rho_A(b,z'')+
\theta(z'-z'')\rho_B({\bf b-s},z''))} 
\;\; . \label{glaubl}\end{eqnarray}
In the absence of preformation absorption
($\sigma^{abs}_{c\bar{c}N}=0$) and the absence of
comover absorption ($T_{co}=0$),
 the above factor reduces to $AB$ for untriggered data ($P=1$).
As shown in the next section this $AB$ scaling (expected for any hard 
pQCD process without final state interactions)
is observed to hold very well for DY pair production.

In addition to the uncertainties associated with (\ref{glaubl}),
the assumed constancy of the density along the path
in the nuclear overlap region
(neglect of energy loss and density pile-up especially at
moderate SPS energies) and the assumed space-time independence
of the  effective
cross sections of the pre-hadronic $c\bar{c}$ configuration,
other sources of theoretical uncertainties are evident:
A major source of
 model dependence of $L$ enters for triggered data through the
transverse energy impact parameter distribution, $P(E_T,b)$.
The observed transverse energy, $E_T$, 
depends on the details of the experimental 
geometry  and materials and is particularly difficult to simulate
in the multi-target system of NA50. Often this distribution
is simply parameterized such that its integral over
impact parameters reproduces the
the observed global $dN/dE_T$ distribution. The effective length
$L(E_T)$ is computed with the above assumptions by setting $T_{co}=0$.

The main advantage of defining $L$ is that 
 data from different $AB$ systems and $E_T$ triggers
can be  combined in one plot. However, we emphasize that unlike
$E_T$, $L$ is not a measured quantity and is model dependent.
Therefore, interpretations of data plotted as a function of $L$
should be viewed with great caution.
In contrast, it is theoretically much better to plot production cross
sections as a function of $AB$ to combine minimum bias data. 

The comover absorption 
 factor $S_{co}$
 depends sensitively on the magnitude and
time dependence
of the local comoving density of partons or hadrons
as well as their absorption cross section.
In additional,  feed-down processes
associated with  ``charmonium chemistry''
must be taken into account in that calculation.
The final $J/\psi$ include contributions
from radiative decay of higher mass charmonium states.
In one estimate \cite{vogt97}, it was assumed that 
$p({\psi'\rightarrow \psi})\sim 12\%$
 of the observed
$\psi$ arise from radiative $\psi'$ decay, 
and $p(\chi\rightarrow \psi)\sim 30\%$ from $\chi$ decay.
Unlike the small $\psi$, the larger $\psi'$ and $\chi$ states
are expected to have significantly larger absorption cross sections
$\sigma_{co}(nLJ)$. Evidence for  comover absorption
of $\psi'$ has been claimed to be observed
already in $S+U$. Neglecting, for illustration,
nuclear absorption and the impact parameter
variations ,  
the comoving survival factor is thus of the generic form 
\begin{equation}
S_{co}=\sum_{nLJ} p(nLJ) e^{-\sigma_{co}(nLJ)\int_{\tau_0}^{\tau_f}
d\tau \rho_{co}(\tau)}
\; \; .\end{equation}
Even if $\sigma_{co}(\psi(1S_1))=0$, $S_{co}\rightarrow 0.6$
if the  higher mass charmonium states are absorbed.

Often simple scaling assumptions are assumed \cite{vogt97} for the
evolution of 
the comoving density, $\rho_{co} \propto dE_\perp/dy (1/\tau R^2)$.
With reasonable variations of the unknown parameters above,
excellent fits to the NA38 $S+U$ were obtained.
However, even with these parameters fixed, comparisons
for different AB systems require further dynamical assumptions.
Especially important is the assumed A-dependence of the 
the comoving matter density.
Linear Glauber models tend to fail
to reproduce the larger suppression in $PbPb$. However, nonlinear
connections between $\rho_{co}(\tau_0)$ and $E_T$ have been
demonstrated to be also compatible with the data \cite{geiger97b}. 

\paragraph*{Experimental status\\}

\begin{figure}[thb]
\centerline{\psfig{figure=psi1a.eps,width=2.8in}\hfill
            \psfig{figure=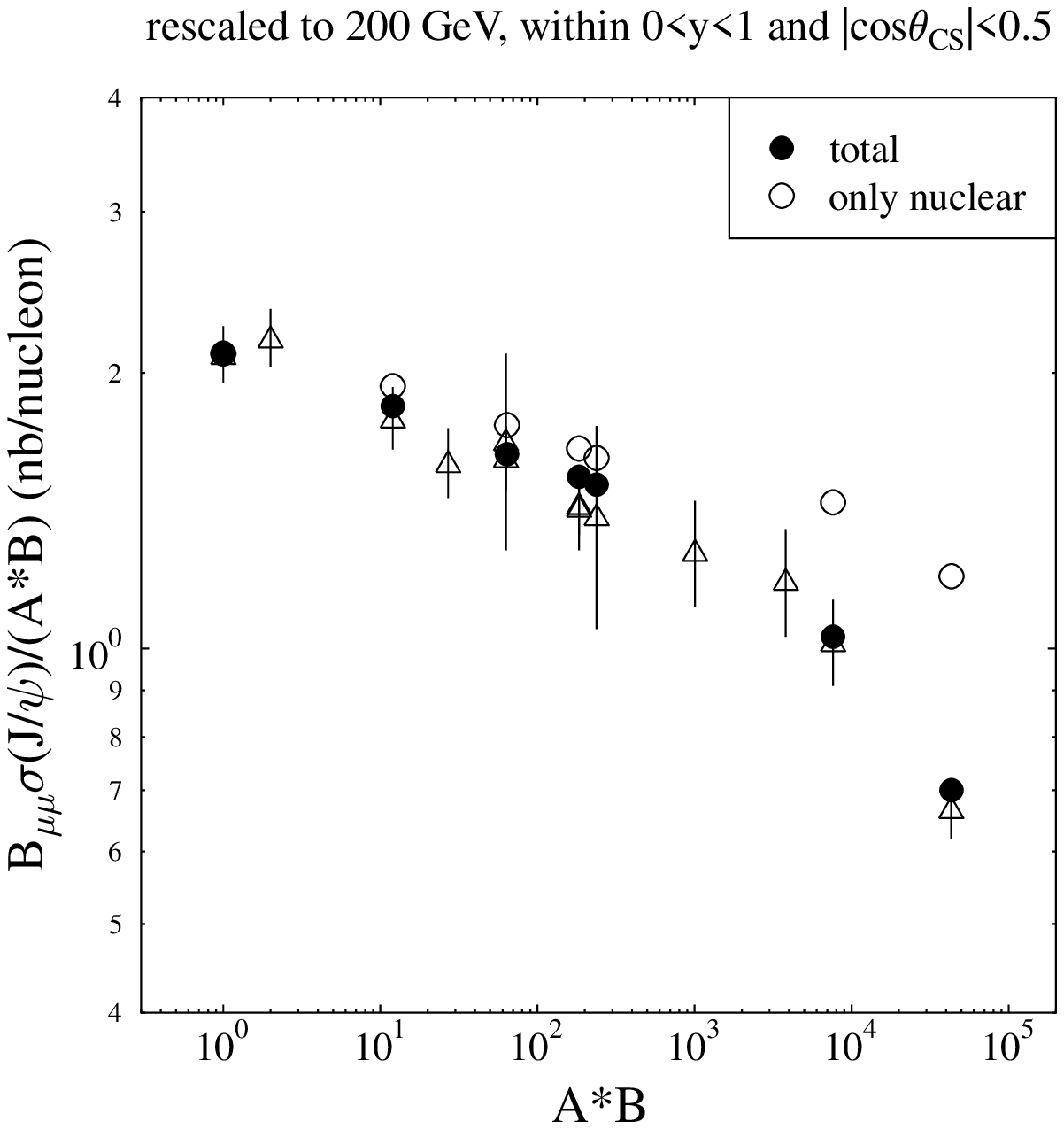,width=3in}}
\centerline{\psfig{figure=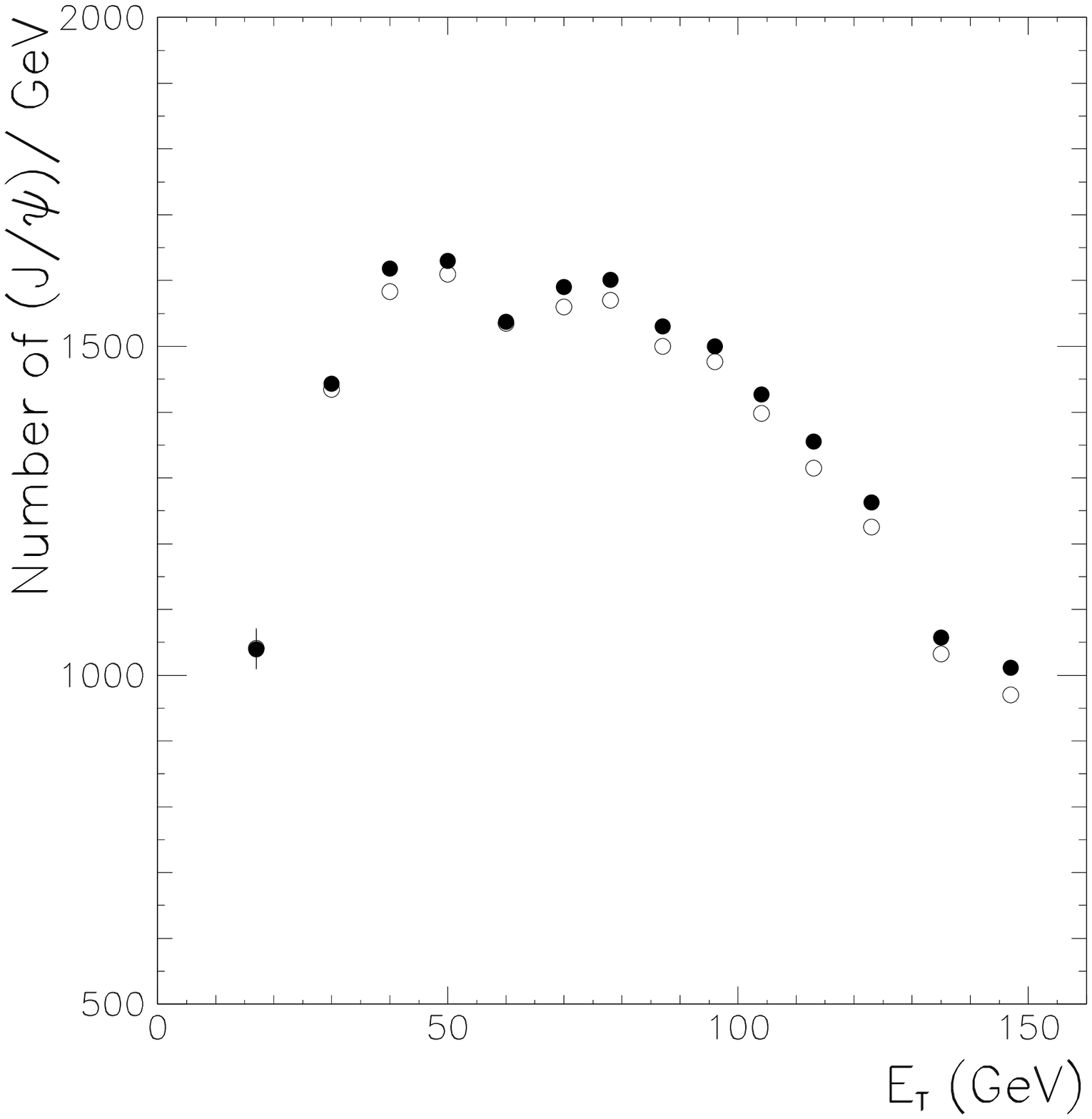,width=2.8in}\hfill
            \psfig{figure=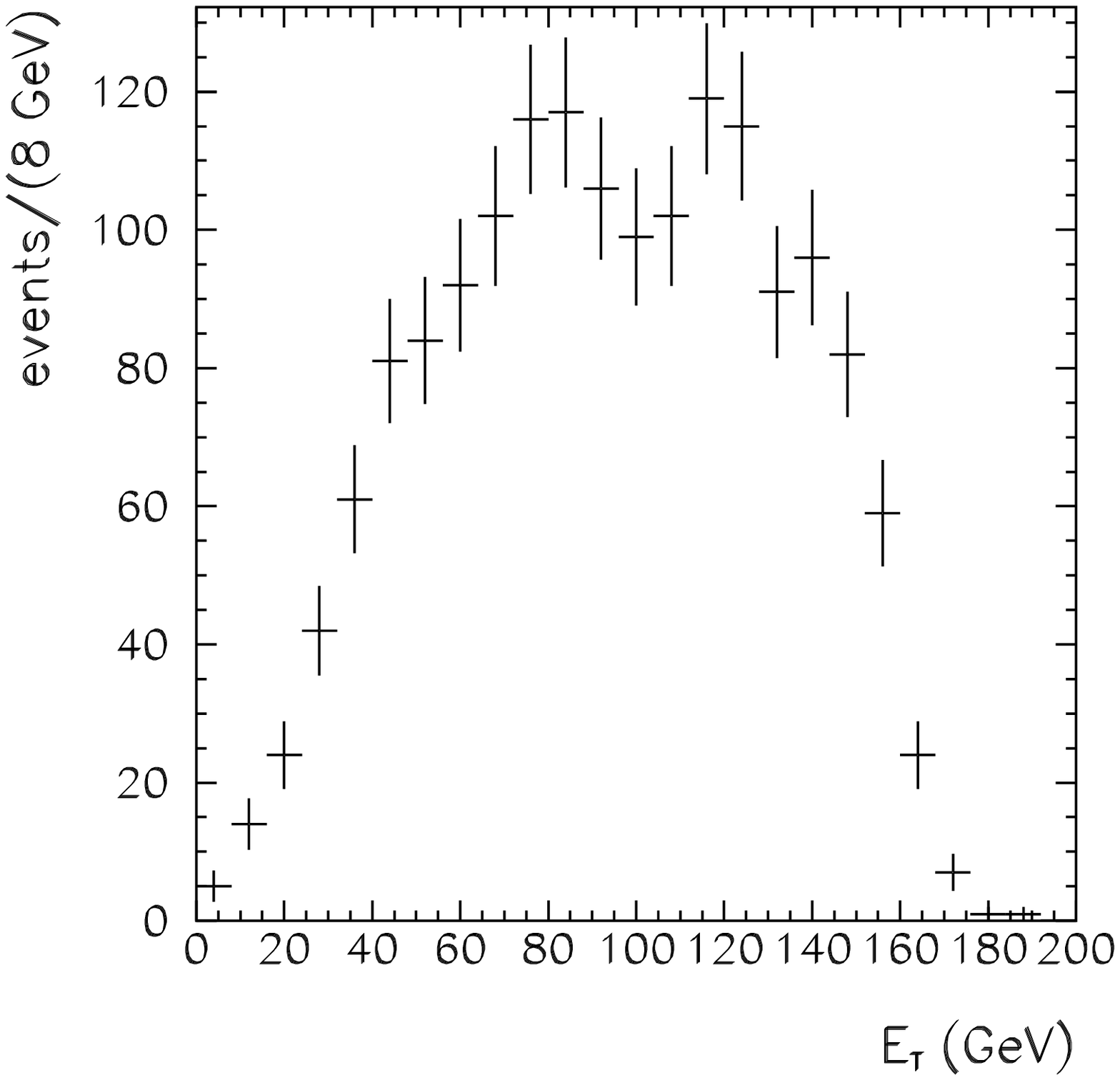,width=2.8in}}
\caption{\label{moriond} Preliminary $J/\psi$ to Drell-Yan data
from NA50 \protect\cite{moriond98,ramello98a}. 
The top left frame shows that Drell Yan scale with $A\times
B$ over five orders of magnitude. Note that the value of the $K$ is actually
1.8 and not 2.3 which refers to GRV LV.
The top right shows 
a suppression of $J/\psi$ reaching almost a factor of 4
in $Pb+Pb$ collisions (open triangles). The additional
25\% suppression of $J/\psi$ in Pb+Pb relative to
extrapolation of $p+A$ and $S+U$ is referred to as anomalous.
The full circles depict a UrQMD calculation \protect\cite{spieles98b} including
comoving mesons whereas the open circles show the same calculation with
only nuclear absorption. 
In the bottom left and right frames the number of $\psi$ and DY pairs
observed as a function of the uncorrected NA50 transverse energy
are shown. The kink in the bottom left frame 
at $E_T\sim 40-50$ GeV shows the rapid onset of
anomalous suppression in the impact parameter range estimated to be $b\sim 8$
fm.}

\end{figure}

Systematic measurements of $J/\psi$, $\psi'$ and 2 -- 5 GeV
continuum processes  (Drell-Yan  processes, open charm decay, etc.)
have been performed by the NA38 collaboration
using proton, oxygen and sulfur beams at CERN
\cite{baglin89a,baglin90a,baglin91a,drapier92a}.
The first preliminary data on $Pb+Pb\rightarrow J/\psi$
was reported by the NA50 collaboration in 1996. 
The data analysis is still not complete
but in a recent conference \cite{moriond98} all the data have been combined
as a function of $AB$ as shown in Fig.~\ref{moriond}.
The first striking result in the top left frame is that the Drell-Yan
yield of dileptons with mass $>2.9$ GeV scale within
20\% as $\sigma^{th}_{DY}\propto AB$
 over five orders of magnitude in that variable.
The so called $K\sim 2$ factor depends on the choice
of the proton structure functions.
In the top right frame the deviation of the $J/\psi$ production cross section
from this simple scaling is obvious (data are depicted with open triangles).
After rescaling data from different energies to 200 AGeV,
all but the PbPb data lie on what appears to be 
a universal curve, $(AB)^{0.92}$.
The fact that the $p+A$ data and $S+U$ data lie on the same curve
suggests the common preformation physics interpretation
discussed before.
The minimum bias $PbPb$ data point is about 25\% below the
curve extrapolated up to $208^2$.
This is the so called anomalous suppression. 

\begin{figure}[thb]
\centerline{\psfig{figure=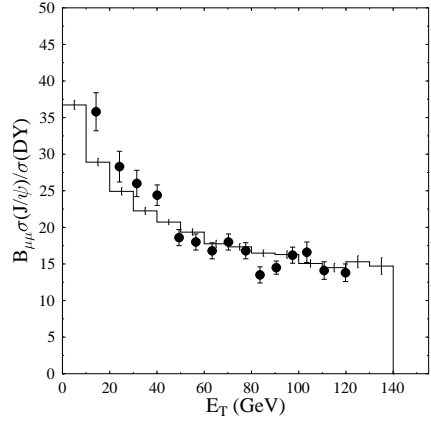,width=7cm}}
\caption{\label{jpsifig} 
Ratio of $J/\psi$ to Drell-Yan as a function of
the experimentally measured
transverse energy for the system Pb+Pb at 160 GeV/nucleon. The data
are taken from \protect\cite{moriond98}, the histogram is a UrQMD calculation
\protect\cite{spieles98b}. Both, data and calculation, have been rescaled
to 200 GeV/nucleon.}
\end{figure}

\begin{figure}[thb]
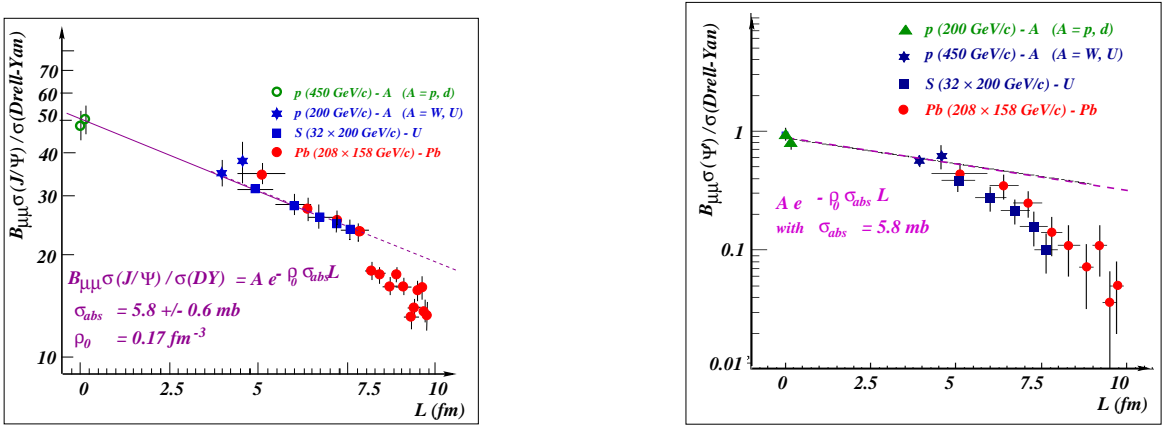

\centerline{
\psfig{figure=psi2a.eps,width=2.5in}\hfill
\psfig{figure=psi2b.eps,width=2.5in}
}
\caption{\label{lplot} Preliminary data on   $J/\psi/$DY and  $\psi'/$DY
plotted as a function of the Glauber model nuclear thickness
parameter, L, discussed in text. This plot suggests a sudden extra 
suppression of $J/\psi$ when the effective nuclear depth increases to 
about 8 fm.
}
\end{figure}

In the lower left and right frames the centrality $E_T$ dependence
of the $J/\psi$ and DY yields are compared.
(Note that recently the $E_T$ scale has been recalibrated
to be $\sim 10-20$ smaller than shown here, but no final publication
has appeared as of this writing).
The low $E_T<40$ GeV are more peripheral collisions
with geometries comparable  to central $S+U$.
For $E_T>50$ GeV corresponding to $b<8$ fm
central collisions, the shape of the DY and $J/\psi$
distributions appear different.
It should be noted that the $J/\Psi$ and Drell-Yan distributions are
affected by systematical errors which do not affect the ratio 
between the two variables and therefore the direct comparison of theory
to these distributions may be misleading.
The ratio of 
$J/\psi$ to Drell-Yan yields for Pb+Pb  
as a function of $E_t$ is displayed in figure \ref{jpsifig} (full circles).  
Within the Pb+Pb sample a discontinuity has been hinted at around
$E_t=50$~GeV. New minimum bias data with highly improved statistics
show a rather smooth increase with decreasing $E_T$. However, the data
situation remains unclear for both, the low and the high $E_T$ limit.
The $\langle p_t^2 \rangle$ does not increase anymore
for $E_t \ge 100$~GeV in Pb+Pb collisions.

Finally in Fig.~\ref{lplot} the provocative ``L'' plot \cite{moriond98}
is shown that suggests that a sudden increase in $\psi$ absorption occurs
for effective nuclear depths above $\sim 7-8$ fm,
while the larger $\psi'$ state is similarly absorbed in $S+U$ and
and $Pb+Pb$.

\paragraph*{Discussion\\}

The initial plasma interpretations of the
$J/\psi$ suppression in light nuclear beam data
in 1987 have been reformulated as a result
of extensive $p+A$ data proving the importance of pre-formation
absorption phenomena in confined QCD matter.
In addition, the similar suppression of $\psi$ and $\psi'$
as a function of $A$ have allowed the determination
of the octet $c\bar{c}$ effective cross section.
The light ion suppression pattern and the $p+A$ data
are now commonly agreed to
be consistent  with hadronic (confined matter) dissociation scenarios
%involving $c \bar{c}_8 - g$ states and to a lesser extend
%$J/\psi$ and $\psi'$ states 
\cite{gerschel88a,gavin88a,vogt88a,kharzeev96c}. 
As pointed out , however, the pre-resonance $c \bar{c}_8 - g$ state
is not an eigenstate of QCD and the use of an effective
cross section for it in Glauber
models is not without ambiguity. 

The data by the NA50 collaboration with the Pb+Pb 
experiment has given rise to renewed speculation 
on an additional suppression and
on the
possible creation of a deconfined phase 
\cite{kharzeev96b,wong96a,blaizot96a}.
Plotting the total $J/\psi$ over Drell-Yan cross section ratio as 
a function of collision system mass ($A_P \cdot A_T$) an additional
suppression in  the order of 20\% is observed between S+U and Pb+Pb.
However, when studying the cross section as a function of $E_T$,
a discontinuity at $E_T > 50$ GeV may only be inferred when
comparing the experimental $J/\psi$ over Drell-Yan ratio vs. $E_T$ with
a Glauber-type calculation (without QGP formation) employing the same 
{\em comover}-density as in the S+U case.
The same Glauber-type hadronic absorption models \cite{gavin96b}
are also capable of reproducing
the lead data of NA50 if a higher {\em comover}-density is employed for
Pb+Pb than for S+U \cite{gavin_priv}. 
For low transverse energies current hadronic {\em  comover}-models
have difficulties in fitting the data.
Therefore the issue whether
simple {\em comover} models  describe the lead data is not yet fully 
settled.

Hadronic transport
model calculations incorporate the full collision dynamics and
go far beyond the commonly used simplified version of the Glauber theory  
\cite{cassing96b,gerland97a,geiss98a,spieles98b}. 
These transport model calculations
are very sensitive to certain input
parameters such as the formation time of the $J/\psi$ and the comovers. 
In a first round of early calculations
the
HSD transport model \cite{cassing96b} was fully able reproduce the NA50 lead
data while assuming a fixed formation time of 0.7--0.8 fm/c for both,
$J/\psi$ and comovers. However, the hadron density which causes
the suppression in the model may be unreasonably high.  The UrQMD
model \cite{bass98a}, however, uses for the comovers 
a variable formation time  emerging from
the Lund string fragmentation formalism (here the formation time 
depends on the hadron mass) and zero formation time for the $J/\psi$.
The assumption of zero formation time is valid if the $J/\psi$ is
considered as pre-resonance $c \bar{c}_8 - g$ state with a hadronic
dissociation cross section of 7 mb.
However, in this mode the UrQMD model did not reproduce the 
additional suppression of the Pb+Pb experiment \cite{gerland97a}.  
The question of formation time might be a central issue since in the
color octet model the dissociation cross section is actually higher during
the lifetime of the pre-resonance $c \bar{c}_8 - g$ state 
\cite{braaten96a,kharzeev96b} than after
hadronization. Furthermore the amount of comover - charmonium interaction
will crucially depend on the formation time (and thus the cross section)
of the comovers and whether the comovers are allowed to interact
within their formation time as ``pre-formed'' states (analogously to
the charmonium ``pre-resonance'' states). 

Recently, a new HSD calculation studied the influence of strings on the 
$J/\psi$ suppression. The $c \bar c$ pairs were produced perturbatively 
and the influence of dissociation of those pairs by strings was taken 
into account by regarding strings as longitudinal geometrical 
objects with a specific transverse radius \cite{geiss98a}.
Good agreement with the data from NA38 and NA50 was found for a string
radius of $R_s \approx 0.2 - 0.3$~fm.
A new UrQMD calculation employs a microscopic Glauber simulation for
$J/\psi$ production and the full microscopic transport calculation 
for nuclear and comover dynamics as well as for rescattering \cite{spieles98b}.
The dissociation cross sections are calculated using the QCD factorization
theorem \cite{gerland98a}, feeding from $\psi'$ (5\%) and $\chi$  states (40\%)
is taken into account and the $c \bar c$ dissociation cross sections
increase linearly with time  during the formation time of the charmonium
state. Using only nuclear dissociation yields, a far smaller suppression
than seen in the data is achieved. 
However, if comovers are taken into account
($\sigma_{meson} \approx 2/3 \sigma_{nucleon}$, the agreement between theory and
data is impressive (see figure~\ref{jpsifig}). 
The strong dependence of these results on details, such as 
the treatment of the formation time or the time dependent dissociation
cross section, remain to be studied further.

Quantum effects such as energy dependent formation and coherence lengths
must be taken into account \cite{huefner96a}
before definite statements can be made with regard to the nature
of the $J/\psi$ suppression.

Whereas there exist techniques to calculate the $J/\psi$ nucleon cross
section without using the vector dominance model \cite{brodsky94a},
the size of the $c \bar c$-{\em comover} interaction is still unclear.
TJNAF and HERMES experiments may be able to
address the question of interaction of spatially small configurations
with the nuclear medium. Such a research will help to understand the
interaction of the $\bar c c$ wave packet with {\em comovers} at
comparatively low energies \cite{frankfurt94a}.

Interpretations of the data based on plasma scenarios
are also increasingly evolving away from the original Mott
transition analog. For example, $J/\psi$ suppression due
to large coherent color fields (strings/ropes) have been 
proposed \cite{geiss98a}.
Percolation of longitudinal strings of transverse radius $\sim 0.25$ fm
have been proposed to explain the possible sudden drop
of the $J/\psi$ yield at moderate impact parameters (low $E_T$)
\cite{satz98a}. It is clear that much work needs to be
done theoretically to sort through the many competing dynamical
models of charmonium absorption. The observed effects
are among the most striking results found in heavy ion reactions
and deserve the intense attention they now receive.

On the experimental side, it would be especially important
to map out carefully the $A$ dependence in the intermediate mass
range $30 < A < 200$ to confirm if there is a discontinuity
or rapid change in the mechanism for moderate nuclear depths
as was hinted at by the Pb data. In addition the functional
dependence of the $J/\Psi$ yield as a function of $E_T$ needs
to be clarified for the low ($E_T < 40$~GeV) and high $E_T$ ($E_T > 100$~GeV) 
limit. The beam energy dependence
would be valuable to know, given the very rapid
variation in $pp$ at present SPS energies and to study the suggested
expansion of small wave packets \cite{gerland98a}.
Finally, an independent experimental confirmation of the results
is essential. Most likely only at RHIC will
there be an independent experiment, PHENIX, capable
of addressing this very important observable.
While nominally running at $\surd s=200$ AGeV,
RHIC will be able to approach SPS conditions down to
$\sim 30$ AGeV.

Of course one of the most important additional
experimental checks would be the search for
discontinuities in other observables such as the strangeness fraction
and HBT radii for less central reactions with
$E_T$ in the region suggested by NA50 where new physics
may arise. Thusfar, the other experiments have concentrated on more
central collisions and the observables have varied smoothly
as a function of control parameters, $A, E, $ ...

In summary, a striking pattern of charmonium
suppression has been discovered by NA38/NA50 at the SPS.
The theoretical debate on its interpretation
is far from settled, but great
strides have been made in the past decade to refine concepts and
models. The rapid change in the suppression could be 
the smoking gun of deconfinement, but it is not likely
to be due to simple Debye screening effect originally
hoped for.  Rather novel QCD dynamics in a non-equilibrium plasma 
(e.g. enhanced color field fluctuations in moderate frequency
$\omega\sim 0.5-1.0$ GeV) may emerge as the final culprit. 
A  goal of further theoretical work
will not be to continue to try to rule out 
more ``conventional'' explanations, but to give positive proof of
additional suppression by QCD-based calculations which actually
{\em predict} the $E_T$-dependence of the conjectured signature.
Consistency tests and a
detailed simultaneous analysis of 
all the other measured observables are needed, if at least the same
standards as for the present calculations involving other signatures
are to be held up.

%%%%%%%%%%%%%%%%%%%%%%%%%%%%%%%%%%%%%%%%%%%%%%%%%%%%%%%%%%%%%%%%%%%%%%%%%%%%%
        \section{Summary and Outlook}
%%%%%%%%%%%%%%%%%%%%%%%%%%%%%%%%%%%%%%%%%%%%%%%%%%%%%%%%%%%%%%%%%%%%%%%%%%%%

\paragraph*{AGS and SPS Milestones\\}

In the  last two years the heavy ion research  at Brookhaven
and CERN  have succeeded to achieve the measurement
of a wide spectrum of observables with truly
heavy ion beams $Au+Au$ and $Pb+Pb$.
As these programs continue to measure with greater
precision  the beam energy, nuclear size, and centrality dependence
of those observables, it is important to recognize
the major milestones past thusfar in that work.
Experiments have conclusively demonstrated the existence of
strong nuclear $A$ dependence of
\begin{itemize}
\item {\em baryon stopping power} \cite{videbaek95a,wienold96a}
\item {\em hadronic resonance} production \cite{barrette95a}
\item {\em collective (transverse, directed, and elliptic) flow} 
of baryons and mesons both at AGS and SPS energies 
\cite{barrette96a,nishimura98a,appelshaeuser98a}
\item {\em strangeness enhancement} \cite{anderson92a,
abatzis94a,alber96a,bormann97a}
\item {\em meson interferometric source-radii} 
\cite{beker94a,sullivan93a,     
        miskowiec96a,franz96a,kadija96a,rosselet96a,roehrich97a}
\item {\em dilepton-enhancement} below the $\rho$ mass 
        \cite{agakichiev95a,ullrich96a}
\item {\em anomalous $J/\psi$ and $\psi'$ suppression} 
        \cite{gonin96a,ramello98a,moriond98} 
\end{itemize}

These observables prove that high energy- and baryon density
matter has been  created in nuclear collisions.
The global multiplicity and transverse energy measurements
prove that substantially more entropy is produced
in $A+A$ than simple superposition of $A\times$pp
would imply. Multiple initial and final state
interactions play a critical role in all observables.
The high midrapidity baryon density and the
observed collective radial and directed flow patterns
constitute one of the strongest
evidence for the existence
of an extended period ($\Delta \tau\sim 10$ fm/c)
of strong final state interactions.
The enhanced $\psi'$ suppression in $S+U$ relative to $p+A$
again attests to this fact. The anomalous low mass dilepton
enhancement shows that substantial in-medium modifications
of multiple collision dynamics exists, possibly related
to mass shifts or in-medium broadening of vector mesons.
The systematics of the strangeness (and anti-strangeness)
quantum number production shows that novel non-equilibrium
production processes arise in these reactions.
Finally, the centrality dependence
of $J/\psi$ absorption in $Pb+Pb$ collisions 
hints towards the non-equilibrium nature of such reactions, but
can also be seen -- in the case of an anomalous centrality dependence --
as indication that high frequency gluon modes
may be excited in such reactions. Is this the sought after
quark-gluon plasma that thusfar has only existed
as a binary array of predictions  inside
teraflop computers? It is too early to tell.
Theoretically there are still too many ``scenarios''
and idealizations to provide a satisfactory answer.
And there are experimental gaps such as
lack of intermediate mass $A\sim 100$ data
and the limited number of beam energies studied thusfar.
The field is at the doorstep of the next milestone:
$A+A$ at $\surd s=30-200$ AGeV due to begin at RHIC/BNL in 1999.

At the AGS, where 
particle spectra already have transverse slopes $T > T_C$,
the highest chances for the discovery of
partonic degrees of freedom lie in the measurement of the collective flow 
excitation function and the search for novel
{\em strangelet} configurations. 
The investigation of the physics of high baryon
density (i.e. partial restoration of chiral symmetry via properties of
vector mesons), 
for which the gold beam at the AGS would be ideal,
are unfortunately not accessible due to 
the lack of experimental setups capable of
measuring electro-magnetic probes in AA collisions.  

\begin{figure}[thb]
\centerline{\psfig{figure=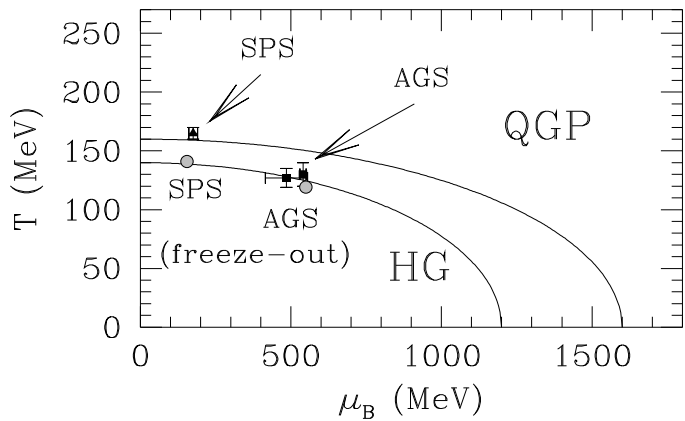,width=9cm}}
\caption{\label{phtrans1fig}
Phase diagram of nuclear matter, adapted from 
\protect \cite{harris96a,annrevcp}. 
The two dashed lines mark the location of the expected phase
boundary at its level of uncertainty. The data points mark
thermal freeze out parameters deduced from AGS and SPS data, taking
flow into account. The arrows indicate how the freeze out conditions
may be reached during the expansion of the fireball. The two gray
points show thermal model fits to UrQMD predictions for heavy collision
systems.}
\end{figure}

At the CERN/SPS new data on 
electro-magnetic probes, strange particle yields (most importantly 
multistrange (anti-)hyperons) and heavy quarkonia will be interesting to
follow closely. 
Energy densities estimated from 
rapidity distributions and temperatures extracted
from particle spectra indicate that initial conditions
should be near or just above
the domain of deconfinement and chiral symmetry restoration. 
With respect to
HBT (meson interferometry) source radii, the matter 
is not yet adequately  resolved (extensive
and precise comparisons with hadronic and deconfinement model
calculations have yet to be performed). Directed flow has been observed --
a flow excitation function, filling the gap between 10 AGeV (AGS)
and 160 AGeV (SPS),  would be extremely interesting
to look for the softest point of the QCD
equation of state. An effort to perform experiments in this energy region
at the SPS is underway. However, dedicated runs would be mandatory to really
explore these intriguing effects in the excitation function. It is
questionable, whether this key program will actually get support
at the SPS.
Also the excitation function of particle yield ratios ($\pi/p, d/p, 
K/\pi ...$) and, in particular, multistrange (anti-)hyperon yields, 
may be a sensitive
probe of changes in the physics of the EoS.
Most intriguing, however, would be the search for novel,
unexpected, forms of $SU(3)$
matter, e.g. MEMOS, {\em strangelets} or even {\em charmlets}.
Such exotic QCD mesonic and nuclear configurations
would extend the present periodic table of elements
into hitherto unexplored flavor dimensions.
A strong experimental effort should continue  in that direction.
The current status concerning the phase-diagram of nuclear matter
is depicted in figure~\ref{phtrans1fig}.

We note that exotic non-equilibrium phenomena such as Disoriented
Chiral Condensates (DCC) that could effect very small transverse
momentum pion spectra and charge/neutral meson fluctuations
should also be continued to look for. These are highly speculative
and scenario dependent phenomena, but worth to be searched for along
with the above exotic mesonic and nuclear states.

%The experimental 
%programs at AGS and SPS with the gold and lead beams are 
%still underway. Further data, 
%especially from the NA49, WA98 and CERES collaborations will
%help to resolve remaining ambiguities of interpretation.

Another intriguing possibility of observing the transition from
nuclear matter to deconfined quark matter may lie in the timing
structure of Pulsar spin-downs \cite{glendenning97a}:
Pulsars contain a huge amount of 
angular momentum and rotational energy. The emission of electro-magnetic
radiation and electron-positron pairs (over a time-span of millions
of years) causes a reduction in the angular velocity and thus also
a reduction in the centrifugal deformation of the Pulsar. Consequently
the interior density of the Pulsar increases and may rise from below
the critical density for a phase transition to above the critical value. 
In the resulting phase-transition a conversion from rather
incompressible nuclear matter to highly compressible quark matter
will take place (starting at the center of the Pulsar) which reduces
the radius of the Pulsar and causes and anomalous decrease of the
moment of inertia with decreasing angular velocity. This is superposed
on the normal reduction of angular velocity due to radiation loss.
In order to conserve angular momentum, the deceleration of the angular
velocity may decrease or even change sign resulting in a Pulsar
spin-up. The time-span in which this effect may be observable is estimated
to be in the order of $10^5$ years. Since the mean life of Pulsars
is around $10^7$ years, 1\% of the 700 currently known Pulsars 
may  currently be undergoing this phase-transition.

\paragraph*{New facilities: RHIC and LHC\\}

RHIC will begin operation in 1999 with
four  detectors:
two medium scale ones, BRAHMS and PHOBOS, as well as two large scale  %$
detectors, PHENIX and STAR. BRAHMS (Broad RAnge Hadron Magnetic Spectrometer)
is a conventional spectrometer (adapted from the AGS program)
with particle ID, covering
the cm rapidity range 0 to 4. 
PHOBOS is a two arm magnetic spectrometer which
will be able to measure low $p_t$ charged hadrons 
{\em and} leptons at selected
solid angles. 

PHENIX is a large solenoid with a variety of multi-purpose  detector arrays; 
its goal is the multiple detection of  phase transition signatures via 
the measurement of hadrons, leptons and photons in the same central
rapidity bin \cite{gregory92a}. 
Apart from the QGP signatures which are already discussed, 
the PHENIX experiment will also
search for a disoriented chiral condensate (DCC)
\footnote{A first preliminary result was published on the DCC search
at CERN/SPS by the WA98 collaboration \cite{wyslouch96a}: 
At a 90\% confidence level they rule out a
DCC admixture of greater than 20\%} 
\cite{pisarski84a,bjorken92a,rajagopal95a}.

The main emphasis of the STAR 
(Solenoidal Tracker At Rhic) detector
will be the correlation of many (predominantly hadronic) observables 
on an event-by-event basis\footnote{
Recently, the NA49 collaboration presented first measurements of event-by-event
fluctuations at the CERN/SPS, which may have significant implications for the
issue of thermalization and critical fluctuations near a phase transition \cite{stock_pc}. 
The measured disappearance of strong dynamical fluctuations in
$\langle p_t \rangle$ suggest a high degree of equilibration or at least 
rescattering. The absence of non-Gaussian fluctuations furthermore may exclude
the possibility that the system has been close to a phase-transition.}
\cite{harris94a}.

The great energy range and beam target range accessible
with RHIC will allow a dedicated systematic search
for the quark-gluon phase matter at energy density
an order of magnitude above the transition domain.
This occurs not only because the rapidity density
of hadrons  is expected to be 2-4 times larger than in central SPS
collision, but also because   $pQCD$ dominated mini-jet
initial conditions are finally reached at collider ($\surd s\sim 200$ AGeV)
energy range. A whole class of new signatures involving hard pQCD
probes (high $p_T$ and jets) becomes available.

At yet higher energies at LHC, 
QGP research efforts and planning are centered around the 
ALICE detector. Its design is similar to that of STAR. However,
also dimuon arms (like in PHENIX) are planned. 
ALICE will be the only large scale heavy ion detector setup at LHC.
At $\surd s \sim 5$ ATeV even bottom quarkonia are copiously produced
and transverse momenta twice as high ($p_\perp \sim 60$ GeV/c)
will be readily measurable to probe even deeper into the multiparticle
dynamics in a QGP.

For applications to nuclear collision observables,
an extension of the QGP concept to non-equilibrium conditions
is required. The popular use of simple fireball
models may provide convenient parameterizations
of large bodies of data, but they will never provide
a convincing proof of new physics. 
Microscopic transport models are required that can address
simultaneously all the observables
and account for experimental acceptance and trigger configurations.
Present work in parton cascade dynamics
is based largely on analogy of transport phenomena in
known abelian QED plasmas. A significant new feature of QCD plasmas
is its ultrarelativistic nature and
the dominance of (gluon) radiative  transport. These greatly complicate
the equations.
The role of quantum coherence phenomena beyond classical
transport theories has only recently been established
within idealized models. Much further work will be required
in this connection. The outstanding theoretical task
will be the development of practical (vs. formal)
tools to compute  quantum non-equilibrium
multiple collision dynamics in QCD. 
Recent work \cite{tdlee}
 along the lines of non-compact lattice formulations of
gauge theories may provide one of the most promising  avenues in
that direction. As yet unrealized techniques utilizing
supersymmetry and string theory should also be explored.

Experiments and
data on ultra-relativistic
collisions are essential in order to motivate,
guide, and constrain such theoretical developments.
They provide the only terrestrial probes of
non-perturbative aspects of QCD and its dynamical vacuum.
The understanding of confinement and chiral symmetry
remains one of the key questions at the end of this millennium.

\ack
We wish to express our thanks to our colleagues 
Adrian Dumitru, Jens Konopka, Christian Spieles, Markus Bleicher,
Sven Soff and Lars Gerland for many fruitful discussions and suggestions.
We are very grateful to Lonya Frankfurt, Carlos Lourenco and 
Dieter R\"ohrich for their
comments and improvements to the manuscript.
S.A.B., M.G. and H.S. thank the Institute for Nuclear Theory at the
University of Washington for its hospitality and the Department of
Energy for partial support during the completion of this work.
Finally, S.A.B. thanks Berndt M\"uller for many enlightening
discussions.

This work was supported by DFG, GSI, BMBF and in part by DOE grant 
DE-FG02-96ER40945 and the A. v. Humboldt Foundation.

%%%%%%%%%%%%%%%%%%%%%%%%%%%%%%%%%%%%%%%%%%%%%%%%%%%%%%%%%%%%%%%%%%%%%%%%%

\end{document}